\RequirePackage{fix-cm}
\documentclass[natbib,smallextended]{svjour3}       
\smartqed  
\usepackage[dvipsnames]{xcolor}
\usepackage{graphicx}
\usepackage{amssymb,amsmath,mathrsfs}
\usepackage{comment}
\usepackage{hyperref}
\usepackage[normalem]{ulem}
\hypersetup{
    colorlinks,%
    citecolor=blue,%
    linkcolor=blue,%
    urlcolor=blue
}

 \journalname{my petit journal}
\newcommand{\wdchap}{White Dwarf Chapter}

\newcommand{\ratechap}{Rates Chapter}
\newcommand{\disrupchap}{Disruption Chapter}
\newcommand{\flowchap}{Formation of the Accretion Flow Chapter} 
\newcommand{\diskchap}{Accretion Disc Chapter}
\newcommand{\simchap}{Simulation Methods Chapter}
\newcommand{\emischap}{Emission Mechanisms Chapter}

\newcommand{\binchap}{Binaries Chapter}

\begin{document}

\title{The Process of Stellar Tidal Disruption by Supermassive Black Holes}
\subtitle{The first pericenter passage}
\titlerunning{Stellar Tidal Disruption by SMBH}        

\author{E.M.~Rossi \and N.C.~Stone \and J.~A.~P.~Law-Smith \and M.~MacLeod \and G.~Lodato \and J.~L.~Dai \and I.~Mandel}

\authorrunning{Rossi et al.} 

\institute{E.M. Rossi \at
              Leiden Observatory\\
              Leiden University\\
              PO Box 9513\\
              2300 RA, Leiden (the Netherlands)\\
              \email{emr@strw.leidenuniv.nl} 
              \and 
              N.C. Stone \at
              Racah Institute of Physics \\
              The Hebrew University\\
              Jerusalem, 91904 (Israel) \\
             \email{nicholas.stone@mail.huji.ac.il}
              \and 
              J. A. P. Law-Smith \at
              Department of Astronomy and Astrophysics\\
              University of California, Santa Cruz\\
              Santa Cruz, CA 95064 (USA)\\ \email{lawsmith@ucsc.edu} 
              \and
              M. Macleod \at
              Harvard-Smithsonian Center for Astrophysics \\ 
              60 Garden Street, Cambridge, MA, 02138, (USA) \\ \email{morgan.macleod@cfa.harvard.edu}
              \and
              G. Lodato \at
              Dipartimento di Fisica\\ 
              Unversit\`a degli Studi di Milano\\ Via Celoria 16, Milano (Italy) \\  \email{giuseppe.lodato@unimi.it}
              \and    
              J. L. Dai \at
              Department of Physics\\ 
              The University of Hong Kong\\ Pokfulam Road, Hong Kong (HK) \\  \email{lixindai@hku.hk} 
              \and 
              I. Mandel \at
              School of Physics and Astronomy\\
              Monash University \\
              Clayton, Victoria 3800 (Australia) \\
              The ARC Centre of Excellence for Gravitational Wave Discovery -- OzGrav \\
              Birmingham Institute for Gravitational Wave Astronomy, School of Physics and Astronomy, \\
              University of Birmingham, \\
              Birmingham, B15 2TT (United Kingdom) \\ \email{ilya.mandel@monash.edu}
}

\date{Received: date / Accepted: date}

\maketitle

\begin{abstract}
Tidal disruption events (TDEs) are among the brightest transients in the optical, ultraviolet, and X-ray sky. These flares are set into motion when a star is torn apart by the tidal field of a massive black hole, triggering a chain of events which is -- so far -- incompletely understood. However, the disruption process has been studied extensively for almost half a century, and unlike the later stages of a TDE, our understanding of the disruption itself is reasonably well converged. In this Chapter, we review both analytical and numerical models for stellar tidal disruption.  Starting with relatively simple, order-of-magnitude physics, we review models of increasing sophistication, the semi-analytic ``affine formalism,'' hydrodynamic simulations of the disruption of polytropic stars, and the most recent hydrodynamic results concerning the disruption of realistic stellar models. Our review surveys the immediate aftermath of disruption in both typical and more unusual TDEs, exploring how the fate of the tidal debris changes if one considers non-main sequence stars, deeply penetrating tidal encounters, binary star systems, and sub-parabolic orbits. The stellar tidal disruption process provides the initial conditions needed to model the formation of accretion flows around quiescent massive black holes, and in some cases may also lead to directly observable emission, for example via shock breakout, gravitational waves or runaway nuclear fusion in deeply plunging TDEs.
\end{abstract}

\section{Introduction}
\label{sec:intro}
The process of tidal disruption of a star by a supermassive black hole (SMBH) was originally studied by \citet{Hills75} as a mechanism to fuel active galactic nuclei, whose emission had recently been associated to SMBH gas accretion by \citet{lynden-bell69}. Later, however, it became clear that the stellar disruption rate may not be sufficient \cite[e.g.][see also the \disrupchap]{Frank&Rees76} for producing the copious ($\sim 10 M_\odot$ yr$^{-1}$) and steady accretion flows needed to explain bright quasars. Rather, \citet{Rees88} suggested that tidal disruption events (TDEs) could be used to identify the presence of {\em quiescent} SMBHs in nearby galaxies, with the distinctive signature of an accretion-powered flare lasting up to a few years. By the first decade of the 21st century, the ubiquity of SMBHs in galactic nuclei was established, with an overwhelming majority of low-redshift SMBHs being quiescent \cite[e.g.][]{FerrareseMerritt2000}. Thus, TDEs are currently regarded as a unique tool to deliver a census of SMBH properties, including mass, spin and occupation fraction up to redshifts of a few. This is vital information to unravel the galaxy formation process, which is tightly linked to cosmological evolution of SMBHs. Beside black hole demographics, the time-dependent emission of TDE flares can be exploited to understand the physics of accretion and jet launching through different accretion regimes and/or states, similar to the goal of X-ray binary observations.

This Chapter describes theoretical efforts and progress over the last 45 years to understand the (magneto-hydro) dynamics of the stellar tidal disruption process. The tidal destruction of a self-gravitating body by a denser companion is a venerable problem in astrophysics, dating back to the 19th century work of Roche.  For most of its history, the problem was studied primarily in the circular-orbit limit.  While stars can approach SMBHs on quasi-circular orbits, the resulting tidal mass transfer is generally stable, and therefore far less luminous than the TDEs which are our primary subject here. TDEs differ from standard Roche-lobe overflowing systems through the orbits of the disrupted stars, which are generally parabolic, or nearly so.  Consequently, the entire star can be destroyed in a single pericenter passage, faster even than unstable mass transfer in the circular-orbit Roche problem. Alternatively, if the star's pericenter only grazes the tidal sphere, it may suffer limited stripping of its outer envelope: a partial disruption.

While the disruption process itself is not expected to be highly luminous (although there exist some notable, albeit speculative, possibilities for observing the disruption, which we discuss later on), the dynamics of partial and full disruption set the stage for later events in a tidal disruption flare.  The efficiency and the qualitative manner in which an accretion flow is formed, and the resulting light curve of the TDE, are all dictated by the rate at which tidal debris falls back to the SMBH after disruption.  These later stages in the evolution of TDEs are, at the time of this writing, quite incompletely understood, and large open questions exist about the hydrodynamics of accretion disc formation and the emission mechanisms operating during TDEs.  In comparison, the actual process of tidal disruption is itself reasonably well-understood.  The focus of this Chapter is limited in scope\footnote{The one exception to this is the evolution of the star's debris which is dynamically unbound during the disruption process; because it does not participate in later stages of the bound debris evolution, we cover its evolution here.} to events occurring in the immediate vicinity of the tidal sphere; the subsequent evolution of dynamically bound tidal debris is picked up in the \flowchap, \diskchap, and \emischap.

Here we review theoretical models of the tidal disruption process.  In \S \ref{sec:dyn_basics}, we present a general theoretical framework for tides, in both the Newtonian and general relativistic regimes.  We then overview both analytic and semi-analytic models for the disruption process and the dynamical properties of the stellar debris as it exits the tidal sphere.  In \S \ref{subsec:hydro}, we survey the substantial literature of numerical hydrodynamic simulations of full tidal disruptions (see also the \simchap\ for a more detailed discussion of the numerical techniques).  \S \ref{sec:partial_TDE} likewise surveys past numerical hydrodynamic simulations of partial tidal disruption.  In \S \ref{sec:stellar_types}, we explore how the disruption process depends on the detailed stellar type being examined.  This section goes beyond the primarily polytropic disruption simulations of the prior sections to examine realistic models for both main sequence and giant-branch stars.  In \S \ref{sec:high_beta} we discuss the subset of highly penetrating TDEs as opposed to more common grazing disruptions, examining three as-yet unobserved signatures of deeply penetrating encounters: shock breakout, gravitational wave emission, and thermonuclear fusion. \S \ref{sec:unbound} reviews the fate of the $\approx 50\%$ of the star that is dynamically unbound from the black hole, and does not participate in later stages of the bound debris evolution.  In \S \ref{sec:variations}, we explore ``unusual'' sub-types of TDEs, such as stable but extreme-mass-ratio Roche lobe overflow, disruption of stars on non-parabolic orbits, tidal disruption of binary stars, and repeated partial disruptions.  Finally, we conclude in \S \ref{sec:conclusions}.

\section{Analytical modelling of the process of tidal disruption}
\label{sec:dyn_basics}

In Newtonian gravity, tides are a differential acceleration between two (initially) nearby points, objects, or fluid elements.  If we focus on the tidal forces exerted by a massive black hole, with mass $M_{\rm BH}$, on an object at distance $r$ away, then the ``tidal approximation'' will apply if the object's physical size $R_\star \ll r$.   In this limit, we may Taylor expand the Newtonian gravitational field around the finite size of the object, which leads to an approximate tidal acceleration $a_{\rm t} \sim GM_{\rm BH}R_\star / r^3$, where $G$ is the Newtonian gravitational constant.  The order unity numerical prefactor on $a_{\rm t}$ varies depending on which region of the object we are concerned with, but from this approximation alone, it is straightforward to define a tidal disruption radius: the distance interior to which objects are torn apart by tides from a black hole.  If specifically we consider a self-gravitating star of mass $M_\star$ and radius $R_\star$, this tidal radius will be, approximately \citep{Hills75},
\begin{equation}
    R_{\rm t} = R_\star \left( \frac{M_{\rm BH}}{M_\star} \right)^{1/3}. \label{eq:rt}
\end{equation}
This equation is approximate in that it neglects a variety of effects: the internal structure of the star, the finite duration over which tides strongly perturb a star on a parabolic orbit, the positional variation of $a_{\rm t}$ across the star's surface, the stellar spin and general relativistic corrections.  Ultimately, the true order unity prefactor on Eq. \ref{eq:rt} can only be computed through (relativistic) hydrodynamic simulations of the disruption process.  Fortunately, however, most of these effects are subsumed into the cube root, and Eq. \ref{eq:rt} is therefore reasonably accurate.  

We may make the above dynamical arguments more mathematically rigorous by computing the exact tidal tensor $C_{ij}$, which describes differential accelerations experienced in a rest frame centered on the victim object.  Our presentation of $C_{ij}$ will be brief, but a more thorough treatment can be found in \citet{Brassart&Luminet08}.
Working once more in the tidal approximation ($R_\star \ll r$), the tidal acceleration may be computed via the second derivatives of the Newtonian gravitational potential $\Phi(\vec{r}) = -GM_{\rm BH}/r$.  If the star's position, in a lab frame centered on the SMBH, is $\vec{r}$, then
\begin{equation}
    C_{ij} = -\partial_{r_i} \partial_{r_j} \Phi(\vec{r})= \frac{GM_{\rm BH}}{r^3} \left(-\delta_{ij} + \frac{3r_i r_j}{r^2} \right).
\end{equation}
Here $\delta_{ij}$ is the Kronecker delta.  The tensor $C_{ij}$ has been defined so that, in the {\it tidal} reference frame centered on the star's center of mass, the acceleration of a test particle with position $\vec{x}$ will be
\begin{equation}
    \ddot{x}_i = x_j C_{ij}(\vec{r}). \label{eq:tidalAccelerationN}
\end{equation}
Throughout the notation in this section, repeated indices denote summation.  Because Newtonian orbits about a point mass are planar, we may specialize to a lab frame coordinate system where one of our reference axes is orthogonal to the stellar orbit, so that
\begin{align} C(\vec{r}) =&
\begin{bmatrix}
-1 + \frac{3r_1 ^2}{r^2} & \frac{3r_1 r_2}{r^2} & 0 \\
\frac{3r_1 r_2}{r^2} & -1 + \frac{3r_2 ^2}{r^2} & 0 \\
0 & 0 & -1
\end{bmatrix} \notag \\
=&
\begin{bmatrix}
-1 + 3\cos^2f & 3\cos f \sin f & 0 \\
3\cos f \sin f & -1 + 3\sin^2f & 0 \\
0 & 0 & -1
\end{bmatrix}. \label{eq:tidalTensorN}
\end{align}
Here, $r_1$ and $r_2$ represent positions along rectilinear coordinate axes in the orbital plane; the tidal tensor is independent of the third, orthogonal direction, $r_3$.  In the second line of Eq. \ref{eq:tidalTensorN}, we have replaced these coordinates with the Keplerian true anomaly (azimuthal angle) $f$.  The tidal tensor has three eigenvalues,
\begin{align}
    \lambda_1 =& \frac{2GM_{\rm BH}}{r^3} \notag  \\
    \lambda_2 =& -\frac{GM_{\rm BH}}{r^3}\\
    \lambda_3 =& -\frac{GM_{\rm BH}}{r^3} \notag,
\end{align}
which encode the tidal accelerations along the three principal axes (eigenvectors) of the problem, $\vec{u}_1$, $\vec{u}_2$, and $\vec{u}_3$.  The first two of these eigenvectors lie within the orbital plane: $\vec{u}_1 \parallel \vec{r}$, and $\vec{u}_2 \perp \vec{u}_1$.  These two eigenvectors will, therefore, rotate as the star moves along any non-radial orbit.  The vector $\vec{u}_3$ is orthogonal to the orbital plane, and remains fixed in direction.  Notably, $\lambda_1>0$, implying a ``stretching'' acceleration, while the negative values of $\lambda_2$ and $\lambda_3$ imply a ``compressional'' type of acceleration. Since in the plane the star is stretched in the radial direction, a rigorous but ``generous" tidal radius could be defined by equating $\lambda_1$ to $GM_\star / R_\star^2$, i.e. $R_{\rm t} = R_\star (2M_{\rm BH}/M_\star)^{1/3}$.  This is the largest radius at which any fluid elements of the star will be unable to resist the tidal pull of the black hole through self-gravity.

A similar type of estimate may be made to account for the fully general relativistic tidal field of a Schwarzschild or Kerr metric SMBH.  By constructing a locally orthonormal coordinate tetrad that is parallel-propagated along the star's center of mass geodesic (a Fermi Normal Coordinate system), it is possible to create a local tidal tensor, $\Gamma_{ij}$, very analogous to the Netwonian $C_{ij}$ we have just discussed \citep{Marck+83}.  Specifically, by projecting the Riemann curvature tensor\footnote{In general relativity, as in Newtonian gravity, we may view tides as differential gravitational forces.  The geometric nature of general relativity allows for a second, equivalent, interpretation, where tides reflect the local curvature of spacetime.  This is most easily seen in the geodesic deviation equation.} onto this coordinate tetrad, we obtain a local but relativistic tidal tensor, which describes the accelerations of particles in a small radius around the star's center of mass:
\begin{equation}
    \ddot{x}_i = x_j \Gamma_{ij}(\vec{r}).
\end{equation}
For equatorial motion in the Kerr spacetime (with a SMBH of spin $a_{\rm BH}$),
\begin{equation} \Gamma(\vec{r}) =
\begin{bmatrix}
-1 + 3(1+K/r^2)\cos^2\Psi & 3(1+K/r^2)\cos \Psi \sin \Psi & 0 \\
3(1+K/r^2)\cos \Psi \sin \Psi & -1 + 3(1+K/r^2)\sin^2\Psi & 0 \\
0 & 0 & -1 + 3K/r^2 \label{eq:tidalTensorGR}
\end{bmatrix}.
\end{equation}
The similarities with $C_{ij}$ are self-evident, allowing for straightforward continuity of our Newtonian intuition.  The azimuthal angle $\Psi$ is analogous to, though distinct from, the Keplerian true anomaly $f$ (both angles are $0$ at pericenter).  The primary difference between the two tensors is the presence of $K\equiv (L_{\rm z} -a_{\rm BH} \varepsilon)^2 + Q$, a combination of the Kerr constants of motion: relativistic energy $\varepsilon$ ($\varepsilon=1$ for a parabolic orbit), $z$-component of angular momentum $L_{\rm z}$, and Carter constant $Q$.  In the Schwarzschild limit, $K$ is just the total orbital angular momentum.  For inclined orbits in the Kerr geometry, $\Gamma_{ij}$ becomes considerably more complicated, and Lense-Thirring precession ``mixes up'' the tensor's eigenvalues\footnote{Speaking more rigorously, inclined Kerr orbits are no longer planar, meaning that the off-diagonal terms in $\Gamma_{ij}$ are no longer equal to $0$; as we shall see shortly, this complicates the dynamics of disruption.} \citep{Luminet&Marck85}.

In the equatorial Kerr limit, however, we may once again compute a simple tidal radius by examining the positive eigenvalue of the tidal tensor \citep{Beloborodov+92, Kesden12}.  The eigenvalues of $\Gamma_{ij}$ are
\begin{align}
    \Lambda_1 =& \frac{2GM_{\rm BH}}{r^3}(1+3K/r^2) \notag  \\
    \Lambda_2 =& -\frac{GM_{\rm BH}}{r^3}(1+3K/r^2)\\
    \Lambda_3 =& -\frac{GM_{\rm BH}}{r^3} \notag,
\end{align}
and therefore the effective tidal disruption radius\footnote{Note that $R_{\rm t}^{\rm GR}$ reduces in the Newtonian limit to the ``generous'' tidal radius derived from $C_{ij}$: $R_\star(2M_{\rm BH}/M_\star)^{1/3}$, rather than to Eq. \ref{eq:rt}.  As mentioned before, the order unity prefactor on the tidal radius is uncertain, sensitive to hydrodynamic and self-gravitational effects, and best calibrated through hydrodynamical simulations.} is \citep{Kesden12}
\begin{equation}
    R_{\rm t}^{\rm GR} = R_\star\left(\frac{\Lambda_1}{GM_{\rm BH}/r^3} \frac{M_{\rm BH}}{M_\star} \right)^{1/3}. \label{eq:rtGR}
\end{equation}
The process of tidal disruption for misaligned orbits has received little analytic study, so for now we will focus mainly on the Newtonian and (to a more limited extent) planar general relativistic regimes.

As a star enters the tidal disruption radius, fluid and self-gravitational forces become subdominant to the tides from the SMBH.  The process of tidal disruption can be understood through various levels of approximation.  At the simplest level, we may postulate that at the moment of disruption (usually assumed to be the first moment when $r=R_{\rm t}$), the star impulsively ``shatters'' to pieces, with internal forces becoming negligible and each fluid element free-falling along a Keplerian trajectory (in Newtonian gravity) or timelike geodesic (in general relativity).  This assumption is simplistic, but allows for exact solutions to the future evolution of the star, and provides important physical insights.  Historically, analytic TDE theory based around this assumption were often referred to as ``freezing'' or ``frozen-in'' models, due to the assumption that the debris immediately freezes in to a fixed set of ballistic or geodesic trajectories; in this text, instead, we will refer to this as the ``impulsive disruption'' approximation.

The semi-analytic ``affine models'' study the disruption process with a greater degree of realism, at the cost of exact analytic solutions.  These models, first developed by \citet{Carter&Luminet83}, couple the tensor virial theorem with strong assumptions on the geometry of the disrupting star.  As long as these geometrical assumptions remain valid, the interplay between SMBH tides and weaker internal forces can be studied, and more sophisticated variants of the original affine model provide sometimes surprising degrees of physical accuracy.  Of course, the greatest degree of physical realism will come from numerical hydrodynamic simulations of the disruption process, which are discussed later in this Chapter (and in the \simchap).  For the remainder of this section, we discuss the analytic and semi-analytic insights provided by impulsive and affine models for the disruption process.

\subsection{Tidal compression and the affine model}
Although the semi-analytic affine model is in some ways more sophisticated than purely analytic impulse-approximation solutions, we present it first for two reasons.  Most obviously, it was the earliest approach developed to studying tidal disruption, predating impulsive models by five years \citep{Carter&Luminet83, Rees88}.  Second, it is focused on providing an accurate picture of the early details of disruption, while the impulse approximation is more concerned with accounting for the aftermath.

The early affine models considered the tidal disruption of a star in Newtonian gravity, and assumed that throughout the disruption process, the stellar geometry would follow nested, coaxial ellipsoids of deformable axis ratios, interacting in a way that satisfies the tensor virial theorem \citep{Carter&Luminet83}.  More specifically, the affine model (in its simplest form) can be visualized as treating nested ellipsoids of gas that evolve due to combinations of self-gravity, external (tidal) gravity, and internal pressure.  Fluid elements inside the affine star have positions
\begin{equation}
    R_i = q_{ij} \hat{\vec{R}}_j,
\end{equation}
where $\hat{\vec{R}}$ is the initial position of a fluid element in the unperturbed star (note that the hat notation indicates a unit vector), and $q_{ij}$ is a deformation matrix describing the warping and rotation of the star's principal axes under tidal stress.  For now, we follow the earlier implementations of the affine model and assume that $q_{ij}$ is independent of $\vec{R}$.

The power of the affine approximation comes from the fact that, at lowest order, Newtonian tides induce quadrupolar deformations in a spherical star \citep{Press&Teukolsky77}, making the ellipsoidal approximation very good for weak (non-disruptive) tidal encounters, and reasonable for the early stages of a TDE.  Under these assumptions, \citet{Carter&Luminet83} derive a Lagrangian formulation for the process of tidal disruption, with equations of motion given by:
\begin{align}
    \dot{P}_i=& -M_\star \frac{\partial \Phi(\vec{r})}{\partial r_i} + \frac{\mathcal{M}_\star}{2}q_{lk}q_{jk} \frac{\partial C_{lk}(\vec{r})}{\partial r_i} \label{eq:affineExternal} \\
    \dot{p}_{ij}=& \mathcal{M}_\star C_{ik}(\vec{r})q_{kj}+\Pi q_{ij}^{-1} + \Omega_{ik}q_{jk}^{-1}. \label{eq:affineInternal}
\end{align}
Here $\vec{r}$ is the position vector of the stellar center of mass relative to the SMBH, $P_i = M_\star \dot{r}_i$ is its total (``external'') momentum, and $p_{ij}=\mathcal{M}_\star \dot{q}_{ij}$ is an ``internal'' momentum tensor.  In other words, the first of these equations describes the motion of the stellar center of mass in its orbit about the SMBH, while the second describes internal deformations of the star, which are encoded in $q_{ij}$.  From the definitions of $P_i$ and $p_{ij}$, we can re-express Eqs. \ref{eq:affineInternal} and \ref{eq:affineExternal} as sets of second-order ordinary differential equations (ODEs) for the evolution of $q_{ij}$.  Overall, we have 12 coupled second-order ODEs, supplemented by the scalar, quadrupolar moment of inertia (evaluated for the original, unperturbed star)
\begin{equation}
    \mathcal{M}_\star = \frac{1}{3}\iint \hat{R}_i\hat{R}_j{\rm d}M,
\end{equation}
the gravitational self-energy tensor
\begin{equation}
    \Omega_{ij}=-\frac{G}{2}\iint \frac{(R_i-R_i')(R_j-R_j')}{|\vec{R}-\vec{R}'|^3}{\rm d}M{\rm d}M',
\end{equation}
and the volume integral of the local pressure $P$,
\begin{equation}
    \Pi = \int \frac{P}{\rho}{\rm d}M.
\end{equation}
In the definitions of $\Pi$ and $\Omega_{ij}$, it is useful to remember that $q_{ij}$ can be used to relate (Lagrangian) mass coordinates to the original positions of stellar gas parcels.  Finally, an equation of state is needed to relate local pressures $P$ to local densities $\rho$.  These definitions and equations of motion have been presented without proof or elaboration; the reader interested in a more rigorous mathematical treatment of the affine model should consult \citet{Carter&Luminet85}; it is also covered more thoroughly in the \simchap.

\begin{figure}
\begin{center}
\includegraphics[width=0.8\textwidth]{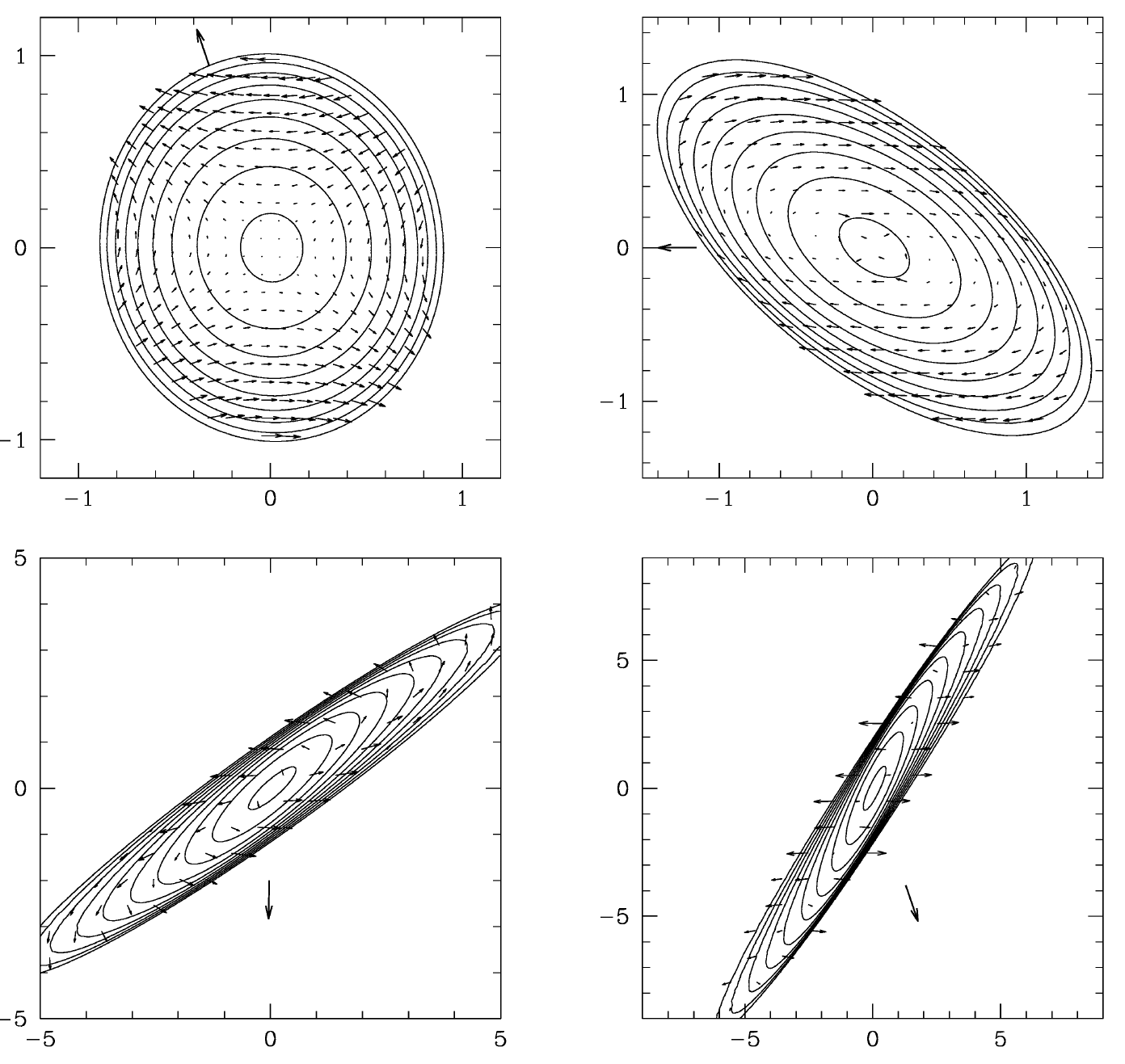}\caption{The internal geometry of a star being disrupted in the framework of the extended \citep{Ivanov&Novikov01}, general relativistic \citep{Ivanov+03} affine model.  Panels show different moments in time in the evolution of a $\beta=0.72$ encounter.  The top left panel shows the star during its approach to pericenter, with $r=1.5R_{\rm p}$.  The top right panel shows the star at pericenter.  The bottom left and bottom right panels show the star after pericenter passage, at distances $r=1.9R_{\rm p}$ and $r=2.73R_{\rm p}$, respectively.  In this figure, we can see that the extended affine model permits different axis ratios and orientations in different internal mass shells.  Vectors inside the star denote internal motions, while the vector outside the star points to the SMBH.  Taken from \citet{Ivanov+03}, their figure 5.\label{fig:extendedAffine} }
\end{center}
\end{figure}

So far, we have written the simplest version of the affine model, and many generalizations exist that incorporate additional physical effects or, alternatively, loosen the underlying assumptions.  By adding additional terms to the underlying Lagrangian, it is possible to model the effect of viscosity, other sources of internal dissipation, and internal rotation \citep{Carter&Luminet85, Luminet&Carter86}.  By replacing Eq. \ref{eq:affineExternal} with the geodesic equations and $C_{ij}$ with $\Gamma_{ij}$, the model can be made general relativistic \citep{Luminet&Marck85}.  The addition of heating terms and a nuclear reaction network enable the study of nuclear fusion reactions triggered by tidal compression \citep{Luminet&Pichon89}.  More recent generalizations of the affine model have generalized the underlying geometry, specifically by allowing the ellipsoidal orientations and axis ratios (i.e. $q_{ij}$) to vary at a single moment in time as one moves from inner mass shells to outer ones \citep{Ivanov&Novikov01}.  This generalized affine model was derived in a Newtonian context, but it has also been applied to the general relativistic tidal problem \citep{Ivanov+03, Ivanov&Chernyakova06}.  Fig. \ref{fig:extendedAffine} illustrates results and geometrical assumptions in the extended affine model.

The most prominent application of the affine model has been to the study of tidal compression during the star's destruction.  As was first noted in \citet{Carter&Luminet82}, the decoupling of vertical from in-plane acceleration in Eq. \ref{eq:tidalAccelerationN} leads to a homologous collapse of the star in the direction orthogonal to the fixed orbital plane (we will refer to this as the ``vertical'' or $z$ direction).  The vertical deformation of the star is extremely pronounced because of the coherent effect of tidal acceleration: in Newtonian gravity\footnote{This statement also holds true in the Schwarzschild spacetime, and in the Kerr equatorial plane.}, $\ddot{z}$ is always negative (positive) for $z>0$ ($z<0$), so the star is uniformly compressed by tides in this direction.  This evolution is markedly different from tidal acceleration within the orbital plane, where the eigenvectors of the tidal tensor $C_{ij}$ must rotate to follow the Keplerian trajectory of the stellar center of mass.  In the star's reference frame, an in-plane direction that is getting stretched at one moment in time will be squeezed at a later one, and therefore the degree of in-plane deformation during the disruption process does not exceed factors of order unity\footnote{Note that after the stellar debris leaves the tidal sphere, it is completely deconfined along the direction of motion, and its ballistic expansion elongates the debris into a very narrow, spaghettified stream.}. 

The degree of vertical compression is thus severe, and turns out to depend strongly on the penetration factor
\begin{equation}
    \beta = R_{\rm t}/R_{\rm p},
\end{equation}
a dimensionless inverse pericenter.  Analytic arguments \citep{Carter&Luminet82} suggest that the vertical collapse velocity achieved during the star's passage through the tidal sphere is $w_{\rm c} \sim \beta V_\star$, where we have made use of the star's natural velocity ($V_\star = \sqrt{GM_\star / R_\star}$).  If the star were made of test particles, and were collapsing uniformly everywhere, it would compress into an infinitely thin pancake somewhere in the vicinity of pericenter. However, this compression will be reversed by the buildup of internal gas pressure.  If we assume the pressure increase is adiabatic, then the internal energy of the star at peak tidal compression is
\begin{equation}
    U_{\rm c} \sim \beta^2 U_\star,\label{eq:affineU}
\end{equation}
its peak density is
\begin{equation}
    \rho_{\rm c} \sim \beta^{2/(\gamma -1)} \rho_\star,\label{eq:affineRho}
\end{equation}
and the duration of the peak compression is 
\begin{equation}
    T_{\rm c} \sim \beta^{-(\gamma+1)/(\gamma -1)} T_\star. \label{eq:affineTau}
\end{equation}
Here we have assumed a polytropic equation of state (adiabtic index $\gamma$) and made use of other ``natural'' stellar variables, namely $\rho_\star = M_\star/R_\star^3$, $U_\star =GM_\star^2R_\star^{-1}/(5\gamma-5) $, and $T_\star =1/\sqrt{G\rho_\star}$.  Since most of the distortion of the stellar shape happens along one axis, the height at peak compression is
\begin{equation}
z_{\rm c} \sim R_\star \rho_\star / \rho_{\rm c} \sim \beta^{-2/(\gamma-1)} R_\star.  \label{eq:zc}
\end{equation}
These scaling relations are simple, but match numerical integrations of the affine model \citep{Luminet&Carter86} as well as 1-dimensional hydrodynamic simulations of collapsing stellar columns \citep{Brassart&Luminet08}.  Their validity has not been explored across a wide parameter space of 3-dimensional hydrodynamic simulations.

\begin{figure}
\begin{center}
\includegraphics[width=0.8\textwidth]{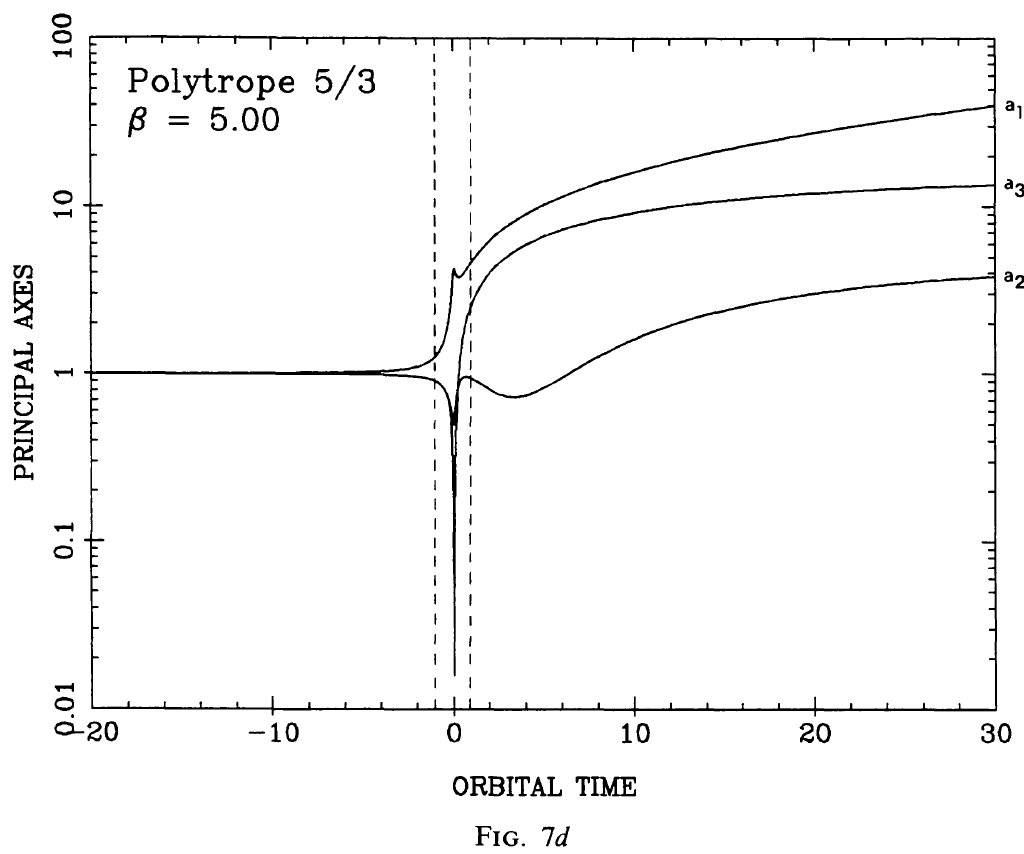}\caption{The evolution of the principal axes of a star, as simulated by the simple affine model.  The principal axes are shown in units of $R_\star$, and are plotted against a dimensionless time $t$ (in units of the orbital time at $r=R_{\rm t}$).  The disruption simulated is that of a $\gamma=5/3$ star with $\beta=5$, in Newtonian gravity.  The duration of the encounter inside the tidal sphere is shown with vertical dashed lines.  Severe compression of the second principal axis (the vertical direction) is visible at $t \approx 0$.  Taken from \citet{Luminet&Carter86}, their figure 7d.\label{fig:affineResults} }
\end{center}
\end{figure}

The degree of vertical compression in a high-$\beta$ TDE can be severe: if one assumes $\gamma=5/3$, then $\rho_{\rm c}\propto \beta^{3}$.  Under a softer equation of state, such as $\gamma=4/3$, the adiabatic compression is even more violent ($\rho_{\rm c} \propto \beta^{6}$).  This phase of stellar pancaking reverses itself rapidly, in an intense burst of hydrodynamic acceleration: for $\gamma=5/3$ ($\gamma=4/3$), the time of peak compression $T_{\rm c} \propto \beta^{-4}$ ($T_{\rm c} \propto \beta^{-7}$).  The vertical compression of a $\gamma=5/3$ star in a $\beta=5$ TDE is illustrated in Fig. \ref{fig:affineResults}.  Under such violent conditions, additional physics may come into play, such as shock heating or thermonuclear reactions.  While these effects have been incorporated in approximate ways into the affine model \citep{Luminet&Carter86, Luminet&Pichon89}, they are sensitive to the internal structure of the collapsing star, and are in principle more accurately treated in hydrodynamical simulations with sufficient spatial resolution.  It should be noted that strong shock heating or thermonuclear detonation will cause the stellar collapse to become non-adiabatic, invalidating the assumptions behind the analytic scaling relations in Eqs. \ref{eq:affineU}, \ref{eq:affineRho}, \ref{eq:affineTau}. A description of physical phenomena caused by stellar vertical collapse around the time of pericenter passage can be found in Section~\ref{sec:high_beta}.

The extended affine model of \citet{Ivanov&Novikov01} has an additional application, which is to determine the amount of mass lost from stars in partial TDEs, with $\beta \lesssim 1$.  By decoupling individual mass shells from each other, the extended affine model allows the outer shells of the star to achieve positive total energy, at which point they are treated as unbound.  As we will show in \S \ref{sec:partial_TDE}, the predictions of the extended affine model are in fairly impressive agreement with three-dimensional hydrodynamic simulations of partial disruption.

\subsection{Impulsive disruption approximation}
\label{sec:impulse}
One of the earliest and most robust results in the context of tidal disruption events is without a doubt the ``$t^{-5/3}$'' decay of the fall-back rate after the total disruption of a star. The result is so fundamental that it 
is often viewed as the classical signature of TDEs, although it is non-trivial to observe directly. It was initially derived by \citet{Rees88}, although in the original paper the result was quoted with an incorrect $t^{-5/2}$ exponent, later corrected to $t^{-5/3}$ by \citet{Phinney89} (interestingly, the $t^{-5/3}$ law had been independently discovered just one week after \citealt{Rees88} in another Nature paper, by \citealt{Curtis88}, to describe disc formation from supernova fall-back\footnote{We thank Sterl Phinney for pointing out this paper during a conference.}). The basics of the argument is very simple and is based only on Kepler's third law. Consider a star of mass $M_\star$ and radius $R_\star$, on a parabolic orbit around a black hole of mass $M_{\rm BH}$, with a pericenter distance $R_{\rm p}=(1/\beta) R_{\rm t}$, where $R_{\rm t}$ is the tidal radius. We will assume for now that the penetration factor $\beta=1$. 
The argument by \citet{Rees88} assumes that the star is almost unperturbed until it reaches pericenter, where it has an impulsive interaction with the black hole and gets torn apart. This is clearly an approximation, but, as we shall see, is not a bad one and deviations from it can be incorporated in the theory. It is straightforward, in this approximation, to compute the spread in the specific orbital energies of the debris as being due to the different depths in the potential well of the black hole across the stellar radius,
\begin{equation}
\Delta E=\frac{GM}{R_{\rm t}^2}R_\star,
\label{eq:energy_spread}
\end{equation}
which corresponds to velocities of the order of
\begin{equation}
v_{\rm ej} \sim \left(\frac{M_{\rm BH}}{M_\star}\right)^{1/6} V_\star, 
\label{eq:vej}
\end{equation}
for the highest-energy ejecta. 
For a typical mass ratio between the black hole and the star of $10^6$, the debris can reach 10 times the stellar escape velocity $V_\star$. Interestingly, tidal forces also induce a spin in the debris, but this can only accelerate the debris up to the escape velocity (see, for example, \citealt{Sacchi19} for a recent description). The argument by \citet{Rees88} then continues by assuming that the orbital energy $E$ distribution is flat among the debris:
\begin{equation}
    \frac{\mbox{d}M}{\mbox{d}E}=\frac{M_\star}{2\Delta E},
    \label{eq:distr}
\end{equation}
and that the return time of the debris to pericenter simply follows from Keplerian dynamics: 
\begin{equation}
    T^2=\frac{4\pi^2}{GM_{\rm BH}}\left(\frac{GM_{\rm BH}}{2|E|}\right)^3.
    \label{eq:kepler}
\end{equation}
From the above, one can calculate the return time of the most bound debris, by setting $E=\Delta E$ in equation (\ref{eq:kepler}):
\begin{align} \label{eq:t_min}
    t_{\rm min}=&\frac{\pi}{\sqrt{2}}\left(\frac{R_\star^3}{GM_\star}\right)^{1/2}\left(\frac{M_{\rm BH}}{M_\star}\right)^{1/2}\\
    \approx& 40\, \mbox{days}~ \left(\frac{M_{\rm BH}}{10^6 M_\odot} \right)^{1/2} \left(\frac{M_\star}{M_\odot} \right)^{-1} \left(\frac{R_\star}{R_\odot} \right)^{3/2}. \notag
\end{align}
One can also obtain the distribution of return times (that is the fall-back rate) as:
\begin{equation} \label{eq:mdot}
    \dot{M}=\frac{\mbox{d}M}{\mbox{d}t}=\frac{\mbox{d}M}{\mbox{d}E}\frac{\mbox{d}E}{\mbox{d}t}=\frac{M_\star}{3t_{\rm min}}\left(\frac{t}{t_{\rm min}}\right)^{-5/3},
\end{equation}
where in the last equation we have used Eqs.~(\ref{eq:distr}) and (\ref{eq:kepler}). For the case of a $1M_\odot$ star disrupted by a $10^6M_\odot$ black hole, the peak fall-back rate $M_\star/(3t_{\rm min})$ 
corresponds to roughly 100 times the Eddington rate.  

\citet{Lodato09} refined this calculation further by estimating the differential distribution of debris mass with respect to specific energy, ${\rm d}M/{\rm d}E$.  This may be visualized as a ``salami slicing'' of the star at the moment of breakup: each infinitesimally thin cylindrical slice of star will have the same $\epsilon$ value.  More specifically, for a spherically symmetric star with internal mass density profile $\hat{\rho}=\rho(x)/\rho_\star$, where $0 \le x=R/R_\star \le 1$,
\begin{equation}
\frac{{\rm d}M}{{\rm d}E} = 2\pi \frac{M_\star}{\Delta E} \int_x^1 \hat{\rho}(x')x' {\rm d}x',
 \label{eq:salami}
\end{equation}
The cylindrical slabs of the star we integrate over are axisymmetric about a vector connecting the SMBH to the star's center of mass, and the center of each cylindrical slab is a distance $R$ from the center of the star.  Each cylinder, at the beginning of tidal free fall, freezes in to its specific orbital energy $E = x\Delta E$ (note that the approximation of constant energy across the cylindrical slab requires $R_\star \ll R_{\rm t}$).  In this way, we can evaluate the distribution of debris energies that accounts for the nontrivial internal structure of the star.

It is important to note that the impulse approximation yields an accurate spread of specific energy when applied at the tidal disruption radius  $r=R_{\rm t}$ as in Eq. \ref{eq:energy_spread} and in Eq. \ref{eq:salami}, rather than at periapsis $r=R_{\rm p}$.  For high-$\beta$ disruptions, $GM_{\rm BH}R_\star / R_{\rm p}^2$ may over-estimate the specific energy spread by one to two orders of magnitude  \citep{Guillochon&RamirezRuiz13, stone+2013}.  The primary reason for this is the relatively short duration the star spends at radii much less than $R_{\rm t}$, which limits the amount of work internal forces (self-gravity and hydrodynamic pressure) can do to alter the energy spread that exists during the crossing of the tidal sphere.  The ability of internal forces to do work on the debris is further reduced by their near-cancellation during the star's entry into the tidal sphere (when the star is not so far from hydrostatic equilibrium); later on, self-gravity will be further reduced in importance by the increasing physical size of the star.

A further elaboration of the impulse approximation was provided by \cite{Stone+13}, who used the free solutions to the parabolic Hill equations \citep{Sari+10} to write explicit orbital elements for every individual fluid element of the disrupted star (once again, under the assumption of instantaneous, impulsive freeze-in to ballistic motion once $r=R_{\rm t}$).  Each of the six solutions represents small perturbations of the Keplerian orbital elements around the star's parabolic center-of-mass trajectory, and the ballistic, post-disruption orbits of the stellar debris are linear combinations of the six free solutions.  One notable feature of the parabolic free solutions, already evident in Eq. \ref{eq:tidalTensorN}, is the decoupling of motion within and orthogonal to the orbital plane.  It is also possible to derive somewhat more complicated free solutions that allow for internal motions at the time of disruption \citep{Stone+13}, accounting for the effects of e.g. stellar rotation; for the sake of brevity we reproduce neither set here.   

This geometrical picture of the disruption process is exact under the (strong) assumption of impulsive disruption and subsequent ballistic motion, and allows one to compute several quantities of interest.  After entering the tidal sphere, at a true anomaly (i.e. azimuthal angle) $f_{\rm t}=-\arccos(2/\beta -1)$, the star will undergo a homologous vertical collapse.  If it were made of test particles, peak vertical compression of the star would occur shortly after pericenter passage, at a true anomaly 
\begin{equation}
f_{\rm c}= \arctan(1/\sqrt{\beta -1}).
\end{equation}
At this point in the orbit, each component of the star will be vertically free-falling with a speed
\begin{equation}
    w_{\rm c} = -\beta \frac{z_0}{R_\star} \sqrt{\frac{GM_\star}{2R_\star}}\left( \sqrt{1-\beta^{-1}} +1\right),
\end{equation}
and the in-plane principal axes of the deformed star will have lengths
\begin{align}
    \tilde{r}_{\rm long} \approx& \frac{4}{5}\beta^{1/2}+\frac{22}{5}\beta^{-1/2} \label{eq:freeLong} \\
    \tilde{r}_{\rm short} \approx& 2\beta^{-1/2} - \frac{23}{2}\beta^{-3/2}. \label{eq:freeShort}
\end{align}
If pressure gradients during peak compression are unable to accelerate significant motions within the orbital plane, the energy spread will not change during the compression process, and the frozen-in specific energy for each fluid element will remain
\begin{equation}
    E = \Delta E \left(\frac{x_0}{R_\star}(1-2\beta^{-1}) + 2\frac{y_0}{R_\star}\sqrt{\beta^{-1} - \beta^{-2}} \right). \label{eq:freeEnergy}
\end{equation}
Here we have denoted initial positions of fluid elements inside the star as $x_0$, $y_0$, and $z_0$, with an origin at the stellar center of mass. By combining the free solutions with a simple approximation for the hydrodynamics of the bounce, \citet{Stone+13} argue that the frozen-in energy spread is unlikely to be altered by even severe degrees of tidal compression.  This conclusion stems from the homology of the star's vertical compression, and could be altered if some source of asymmetry (e.g. a substantial stellar or SMBH spin component that is misaligned with the orbital angular momentum) breaks the homology of collapse, and increases the magnitude of in-plane pressure gradients at peak compression.

A general relativistic version of the impulse approximation was developed by \citet{Kesden12b}, who numerically computed the spread of geodesics that stellar debris would find itself on, assuming disruption of stars on equatorial orbits in the Kerr spacetime.  As with other efforts in this subsection, this work assumed that a static star shatters to pieces at $r=R_{\rm t}$, although here the tidal radius in question is the general relativistic one provided by Eq. \ref{eq:rtGR}.  \citet{Kesden12b} finds that the specific energy spread is not altered greatly by the relativistic disruption process, although for orbits where the gravitational radius $R_{\rm g}=GM_{\rm BH}/c^2$ is comparable to $R_{\rm p}$, $\Delta E$ may change at the factor of two level.

\section{Numerical simulations of the disruption process} \label{subsec:hydro}
The analytical picture outlined above has been confirmed numerically by various works, starting with the early simulations of \citet{Evans89}, who used Smoothed Particle Hydrodynamics (SPH) to demonstrate that the fall-back rate is indeed proportional to $t^{-5/3}$, and that the energy distribution of the debris is approximately flat\footnote{It is worth noting that the energy distribution presented by \citet{Evans89} and some others is on a logarithmic scale, which artificially ``flattens'' it to the eye, but it is correct that at late times, the material falling back from a full disruption is sampled from a flat part of the ${\rm d}M/{\rm d}E$ curve.}. More recently, \citet{Lodato09} considered the effects of changing the internal structure of the star on the fall-back rate, both analytically (as described above) and in numerical hydrodynamics simulations. While \citet{Evans89} had modelled the star as a polytropic sphere with an index $\gamma=5/3$, \citet{Lodato09} consider instead a range of indices, finding that the mass fallback rate can depend significantly on stellar structure.

\begin{figure}
\begin{center}
\includegraphics[width=0.8\textwidth]{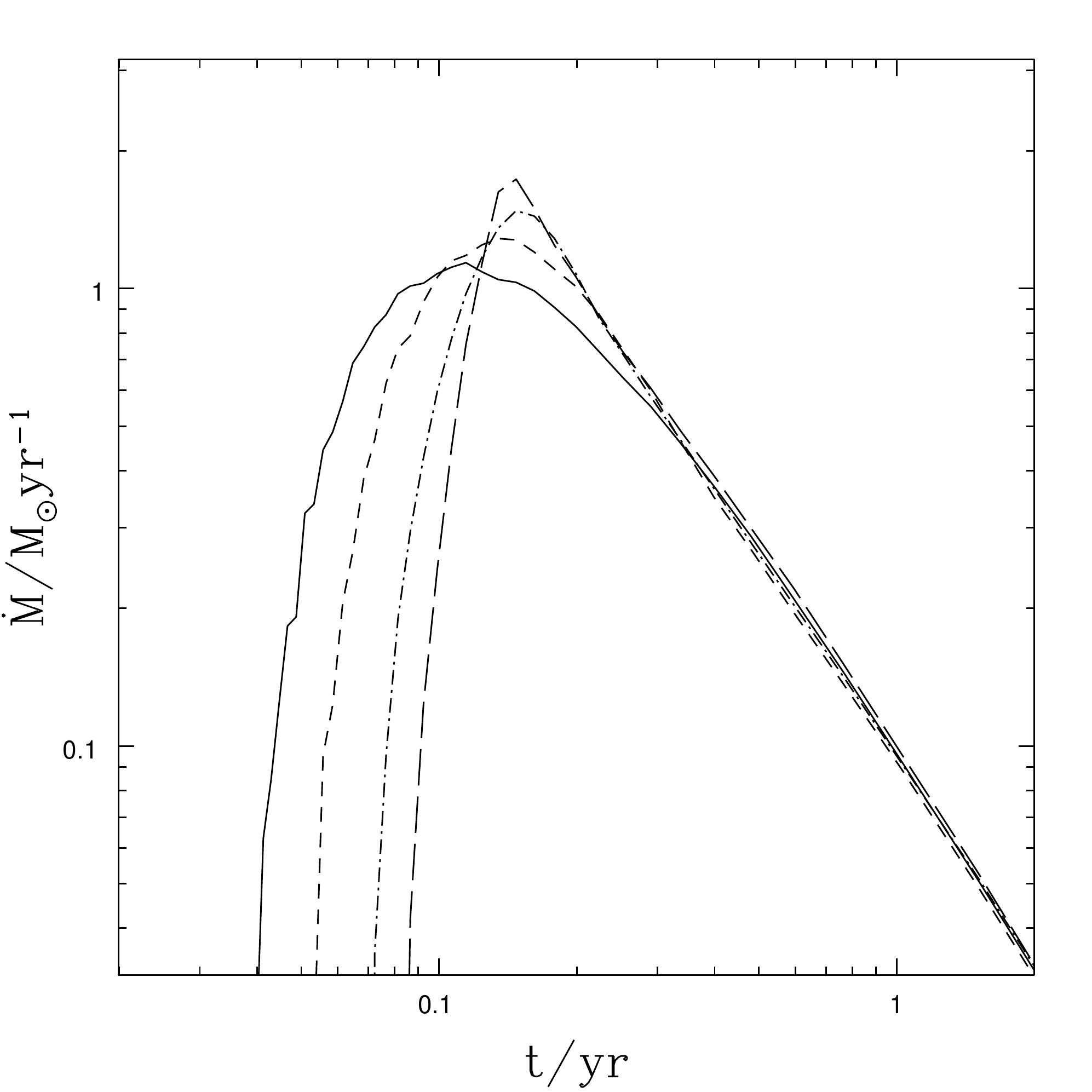}
\caption{Fall-back rate as a function of time from the numerical simulations of \citet{Lodato09}, for different polytropic models of the star: $\gamma=1.4$ (solid line), $\gamma=1.5$ (short-dashed line), $\gamma=5/3$ (dot-dashed line), $\gamma=1.8$ (long-dashed line).  At late times, the returning mass is always drawn from the flat, central portion of the ${\rm d}M/{\rm d}E$ curve, and the fallback rate asymptotes to a $t^{-5/3}$ solution regardless of internal structure. Figure taken from \citet{Lodato09}, their figure 10, left panel.}
\label{fig:fallbackrate}
\end{center}
\end{figure}

Firstly, the energy distribution of the debris should depend on the internal structure of the star, and in particular, more centrally concentrated (less compressible) stars should have a steeper energy distribution, resulting in a slower rise to the peak of the fall-back rate. This result follows from a more precise determination of the energy distribution of the debris (as is expressed analytically in Eq. \ref{eq:salami}), rather than the simple flat distribution of Eq. (\ref{eq:distr}). Note, however, that for any reasonable stellar density profile, the energy distribution of the least bound debris, which originates near the stellar center of mass and determines the late fallback rate, should be indeed characterized by a flat ${\rm d}M/{\rm d}E$. Secondly, the analytical model for ${\rm d}M/{\rm d}E$ is tested numerically with Smoothed Particle Hydrodynamics, and gives good qualitative agreement (see Fig. \ref{fig:fallbackrate} for numerical results). However, since the star is perturbed before
reaching pericenter, quantitative deviations from the analytical models appear. \citet{Lodato09} show that such deviations can be accounted for in the impulse approximation by allowing for homologous expansion of the star at pericenter, due to the reduced effective gravity. A subsequent analysis by \citet{Guillochon&RamirezRuiz13} using a grid-based code has closely confirmed this picture. Recently, \citet{Law-Smith+2019} and \citet{Ryu+20a} have also studied the disruption of realistic stellar models, as opposed to simple polytropes, and discuss the differences in the resulting fall-back rates (see Section \ref{sec:stellar_types}).

The above discussion considered the case $R_{\rm p}=R_{\rm t}$, or $\beta=1$. What happens for more, or less, penetrating events? For more penetrating TDEs, the main result of numerical hydrodynamics is in agreement with the analytical arguments of \S \ref{sec:impulse}: one can still use the impulse approximation to estimate $\Delta E$, but the energy spread needs to be evaluated at the tidal radius rather than at pericenter. This was first seen numerically in the work of \citet{Guillochon&RamirezRuiz13}, and has been investigated in greater detail more recently by \citet{steinberg+19}, who simulate $\beta=5, 6, 7$ disruptions of polytropic stars in Newtonian gravity.  For this set of highly penetrating TDEs, there is little variation in the final ${\rm d}M/{\rm d}E$ for a given unperturbed stellar structure.  For less concentrated $n=3/2$ polytropes (representative of lower main sequence stars), the assumptions of the impulse approximation work reasonably well at all $\beta$, and the final $\Delta E$ is within tens of percent of analytic predictions.  For more highly concentrated $n=3$ polytrope models, internal forces are seen to do substantially more work for the portion of the orbit where $r < R_{\rm t}$, and the final energy spread is enhanced by a factor of a few over the analytic $\Delta E$ estimate. 
For less penetrating encounters the disruption is only partial, as we discuss in section \ref{sec:partial_TDE} below. 

Until very recently, most simulations have neglected the effect of initial stellar rotation on the tidal disruption process. This is mostly due to the fact that tidal torques massively spin up the star immediately prior to disruption, so that pre-existing rotation will only be important 
for initial stellar spins close to break-up. The effect of stellar rotation has been studied recently by \citet{Golightly19} and by \citet{Sacchi19}, who find that prograde stellar rotation (with respect to the orbital axis) enhances the rate of mass fallback (possibly leading to a faster and more luminous flare), while the opposite occurs for stars whose spin is retrograde with the orbital plane. If the star is rapidly spinning in a retrograde sense, tidal disruption might be completely inhibited so that the outer stripped layers of the star re-accrete onto the star rather than onto the black hole, possibly giving rise to a fainter X-ray flare \citep{Sacchi19}. 

General relativistic effects on the disruption process have been considered analytically by \citet{Kesden12b} (see above), but also by a number of recent hydrodynamical simulations. Relativistic simulations of tidal disruptions by spinning black holes have been performed by \citet{Haas12} in the case of white dwarf disruption by intermediate black holes, by \citet{Evans15} for stellar disruptions and more recently by  \citet{Tejeda+17, Gafton19, Liptai19}.  
While in general, the frozen-in energy spread agrees with the Newtonian limit at the factor of $\approx 2$ level, there are some cases where large general relativistic enhancements to the energy spread are seen for deeply plunging ($\beta \gtrsim 10$) disruptions.  This has been attributed both to shock-heating during the vertical compression of the star \citep{Tejeda+17} and also to prompt self-intersection of debris streams before they leave the region of pericenter \citep{Evans15}.

During the disruption phase, any internal magnetic field in the star could in principle be amplified. This effect has been studied by \citet{Guillochon17} and by \citet{Bonnerot17}. The magnetic field can be significantly amplified by at least an order of magnitude, but does not generally have a strong dynamical effect or modify the fall-back rate.  However, the presence of a strong magnetic field can have implications for the resulting accretion flow.

Finally, the fall-back rate can be strongly affected if the stellar disruption is due to a black hole that is a member of a close binary system. The first studies of this process were by \citet{Liu09}, who used N-body simulations to predict that in this case the fall-back rate would suffer several, almost periodic interruptions. This was used by \citet{Liu14} to argue for the presence of a hidden black hole binary system based on the lightcurve of an observed TDE.  A large set of hydrodynamical simulations of this process have been performed by \citet{Coughlin17}, while a more systematic exploration of the parameter space has been provided by \citet{Vigneron18}. While in general the interruptions are very sharp, in some cases, especially if the binary orbit is perpendicular to the stellar orbit, the interruption can result in a relatively gentle decrease in the fall-back rate, which might resemble the lightcurve observed in ASASSN-15lh \citep{Coughlin18}. More details on disruption by SMBH binaries are provided in the dedicated Chapter \binchap~within this book.

\section{Partial tidal disruptions}
\label{sec:partial_TDE}

\begin{figure}
\begin{center}
\includegraphics[width=0.8\textwidth]{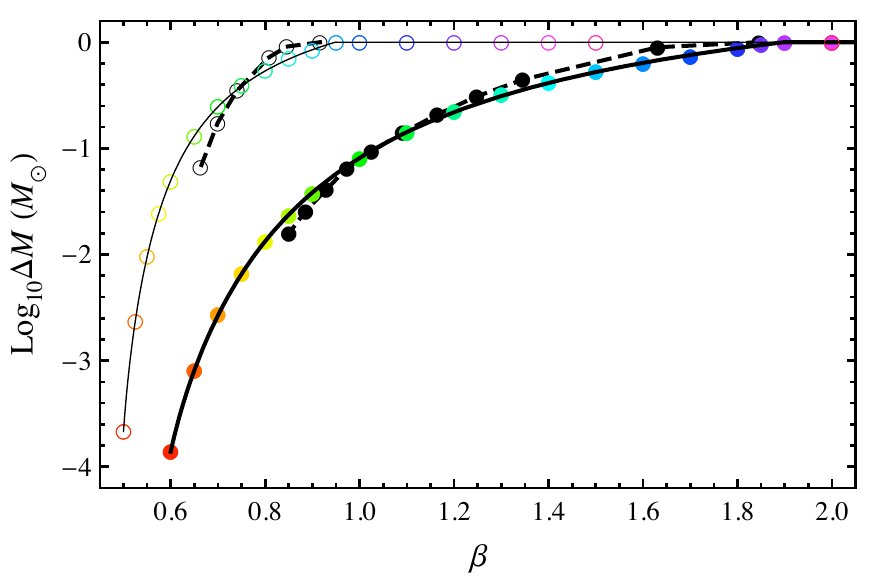}
\caption{The amount of mass lost from a single pericenter passage of a Solar-type star around a $10^6 M_\odot$ SMBH with penetration parameter $\beta$. Thick curves and solid circles show $\gamma=4/3$ models of stellar structure, while thin curves and open circles show $\gamma=5/3$ models.  The solid lines and colored data points show results from three-dimensional hydrodynamical simulations; dashed lines and black data points show results from the extended affine model of \citet{Ivanov&Novikov01}, which is generally in good agreement.  The critical $\beta$ demarcating full from partial disruption is roughly $\beta=1.875$ for $\gamma=4/3$ ($\beta=0.975$ for $\gamma=5/3$). This is figure 4 from \citet{Guillochon&RamirezRuiz13}.}
\label{fig:partialDisruptions}
\end{center}
\end{figure}

In star--black hole encounters with periapsis radii significantly greater than the tidal radius, non-disruptive tides can act on the star as it passes through pericenter. In this case, oscillatory motions of the star's envelope are excited by the tide and continue after the star has passed periapsis \citep[e.g.][]{Press&Teukolsky77}. Deeper encounters, meaning those with higher $\beta$, lead to distortions and subsequent oscillations of progressively larger amplitude in the star. At a critical impact parameter, with $\beta$ of the order of unity, a fraction of the stellar material is unbound from the star, in a partial disruption. In still-deeper encounters, the entire star is disrupted and no self-bound remnant survives. 

This section focuses on the phenomenology of {\em partial} tidal disruptions, in which only the external layers are peeled off the star. 
Partial tidal disruptions occur because stars have differentiated interiors. The simple definition of the tidal radius states that the density enclosed by the tidal sphere at periapsis is equal to the mean density of the star, i.e. $M_{\rm BH}/R_{\rm t}^3 = M_\star / R_\star^3$, or $\rho_{\rm t} = \rho_\star$. If we imagine a star with a constant density interior (an $n\rightarrow 0$ polytrope), the entire interior experiences an equal ratio of tidal gravitational force to self-binding force in a given encounter. For any realistic star with a stratified interior, this statement is no longer true: the stellar density $\rho(R)$ decreases towards the surface, and consequently external layers have larger effective tidal radii, and one can define a radius-dependent tidal radius, $R_{\rm t}(R)>R_{\rm t}$ (see e.g. \citealt{Ryu+20c} for a more detailed example of this). Therefore, in grazing encounters with $R_{\rm p} = R_{\rm t}(R) > R_{\rm t}$, the stellar core within roughly a radius $R$ of the center remains bound by self-gravity and proceeds on its orbit away from the black hole, stripped of its outer, more tenuous layers. Clearly, the way the density changes within the star determines the mass of the surviving core as a function of $\beta$.

More quantitative versions of these statements in the literature have relied on semi-analytic models as well as on numerical simulations. In the following, we review these efforts in chronological order. The first hydrodynamical simulations of the partial disruption process were performed by \citet{Diener+97}, in Eulerian simulations that made use of a relativistic tidal tensor (the inclined generalization of Eq. \ref{eq:tidalTensorGR}).  While these simulations were the first to resolve the survival of a self-bound core following tidal stripping of its envelope, the computational expense limited their coverage of parameter space. Further progress originated from the semi-analytic, nested-affine model of \citet{Ivanov&Novikov01}. By assuming shells are lost when they gain positive energy, these authors were able to estimate not just the degree of nonlinearity imparted by tides, but also fractional mass losses. 

The first detailed sampling of the parameter space of partial disruptions with hydrodynamical simulations were performed in Newtonian gravity by \citet{Guillochon13}. Figure \ref{fig:partialDisruptions} presents these results, showing also a comparison to the semi-analytic predictions of \citet{Ivanov&Novikov01}. As a function of $\beta$, this figure shows the fraction of the star unbound in the encounter ($\Delta M/M_\star=1$ implies a complete disruption). Figure \ref{fig:partialDisruptions} demonstrates important differences that occur for polytropes of differing internal structure. The $\gamma = 1+ 1/n = 4/3$ models transition from partial mass removal near $\beta \sim 0.6$ to full disruption near $\beta\sim 2$. The $\gamma=5/3$ models, by contrast, are partially disrupted in a narrower range of $\beta \sim 0.5$ to $\beta \sim 0.9$. This numerical result aligns with our qualitative discussion above: the $\gamma=4/3$ polytropic star has a wider range of internal densities and self-binding forces than the $\gamma=5/3$ star, which is less centrally condensed. 

\begin{figure}
\begin{center}
\includegraphics[width=0.8\textwidth]{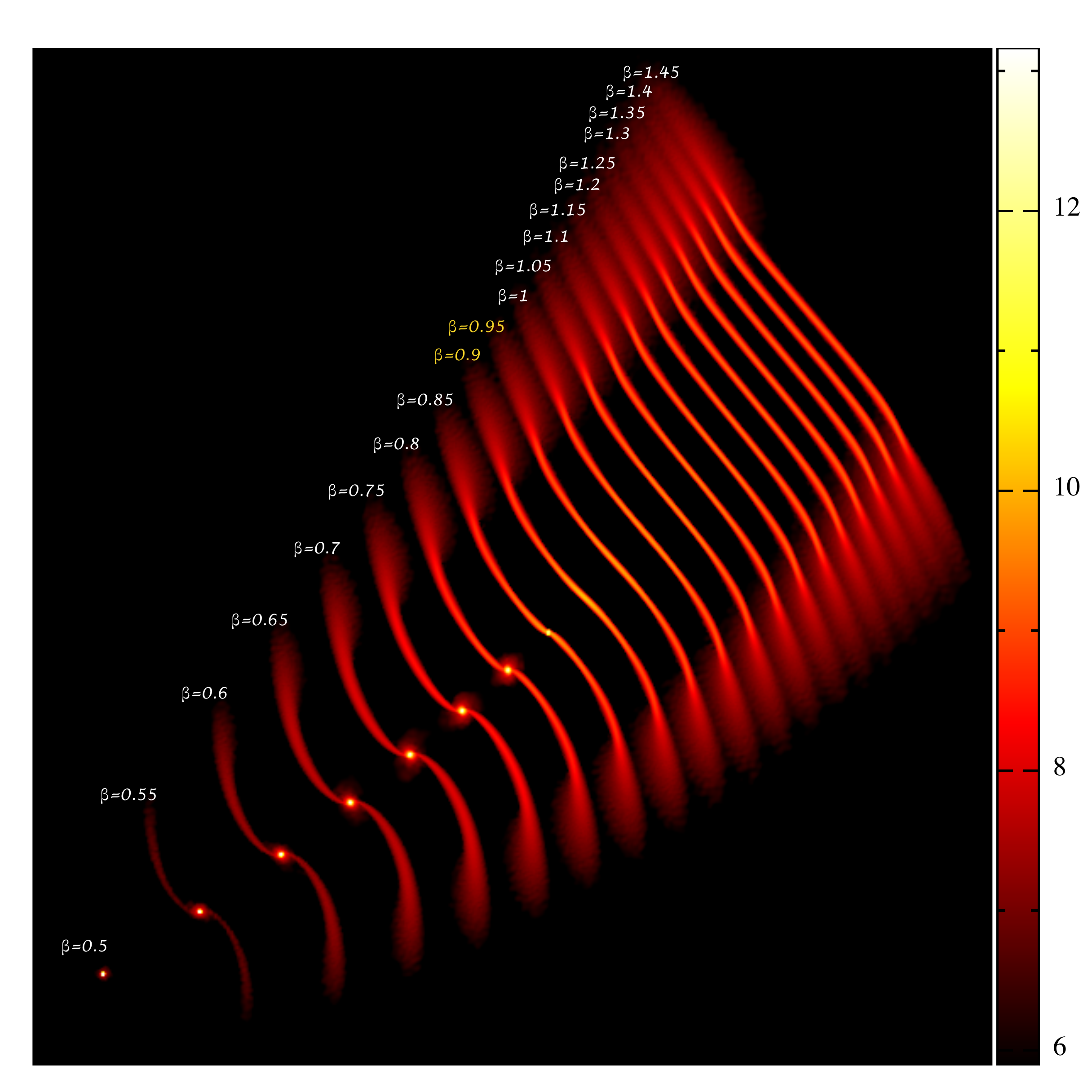}
\caption{Snapshots of $\gamma=5/3$ polytropes following encounters with differing impact parameter $\beta$. These simulations are computed using the smoothed-particle hydrodynamic code GADGET2. This is figure 2 in \citet{Mainetti+17}.}
\label{fig:hydropartial}
\end{center}
\end{figure}

\citet{Mainetti+17} re-examined the quantitative outcomes of partial disruption by using multiple numerical hydrodynamic methods to study the precise impact parameter that differentiats full disruption from partial disruption for $\gamma=4/3$ and $\gamma=5/3$ polytropes. Figure \ref{fig:hydropartial} shows snapshots 
for the $\gamma=5/3$ stellar models. As $\beta$ increases, the surviving core gets smaller and smaller while the mass and extent of the tidal tails grows. At $\beta \approx 0.9$, the star is entirely disrupted and no self-bound core remains. The results seen in this paper with both discrete-mass and discrete-volume techniques are quantitatively close to each other and those in \citet{Guillochon13}, indicating a converged understanding of polytropic stellar disruption in Newtonian gravity.

\begin{figure}
\begin{center}
\includegraphics[width=\textwidth]{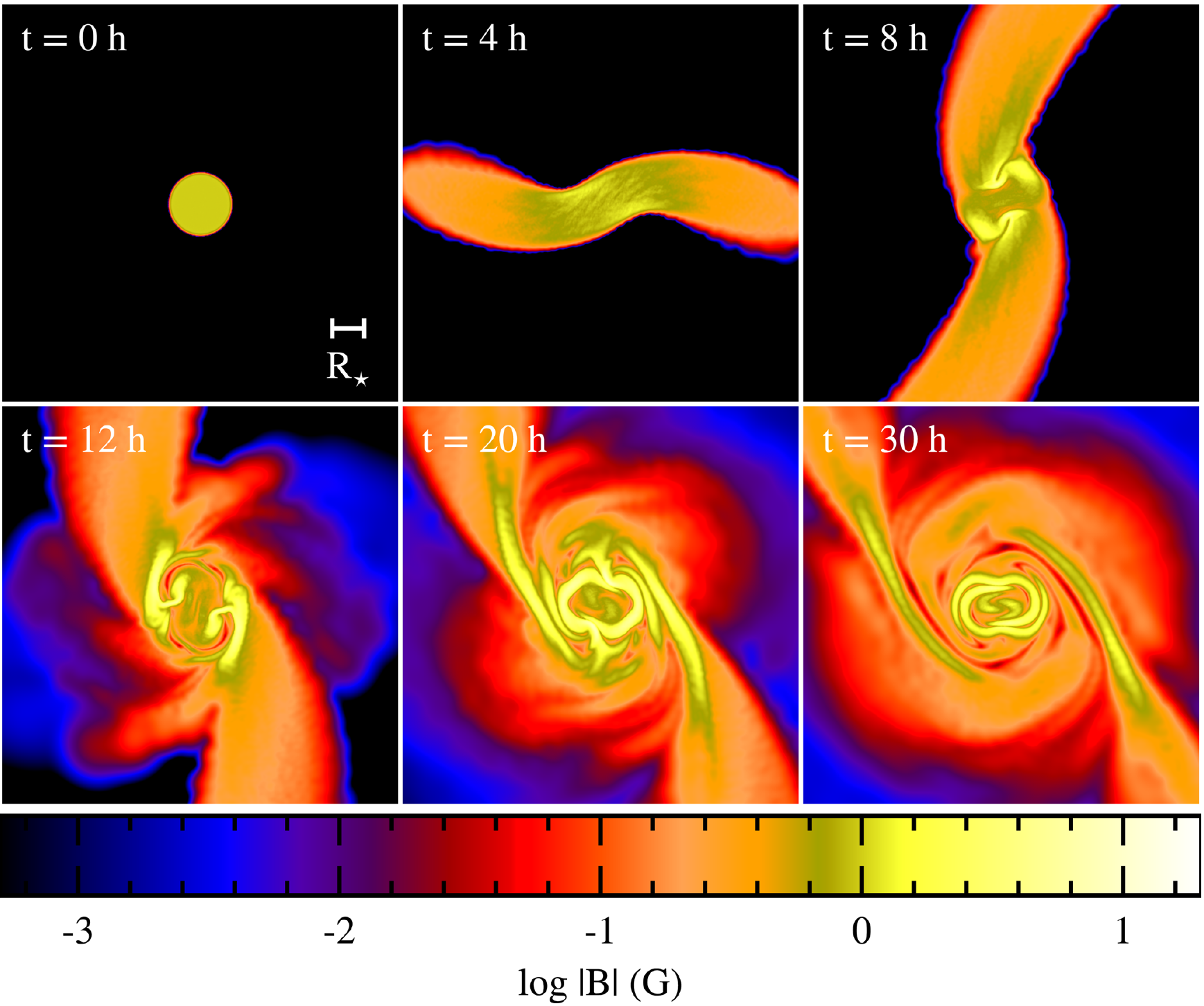}
\caption{The configuration of magnetic field strength (shown color-coded, with a logarithmic color scale at the bottom of the figure)  in a simulated partial tidal disruption event at different snapshots in time.  The top left snapshot shows the star before disruption, the top middle snapshot shows the first episode of re-accretion of the stripped gas, while later snapshots (with hours after disruption shown as white text labels) show the formation of vortices in the surviving core. Figure reproduced from \citet{Bonnerot17}, their figure 5}
\label{fig:mhdpartial}
\end{center}
\end{figure}

Hydrodynamic simulations of partial tidal disruptions have revealed the morphology and dynamics of the gas around the surviving stellar core.
Of the material stretched and distorted into the tidal debris streams, some remains bound to the core. The core itself is distorted by tides and may emerge from the encounter oscillating non-radially (typically dominated by an $l=m=2$ fundamental mode).  The re-accretion of bound material from the debris streams creates spiral shocks and vortices within the surviving core. These stages are clearly visualized by the magnetohydrodynamic simulations of \citet{Bonnerot17} and \citet{Guillochon17}, these former of which are reproduced in Figure \ref{fig:mhdpartial}.  Both sets of magnetohydrodynamic simulations find that the initial magnetic field strength is amplified by a factor $f_{\rm amp} \approx 10$ from vortex-driven dynamo activity, although in neither case does a self-sustaining dynamo emerge.  The final degree of amplification in each study is resolution-dependent and unconverged, suggesting that these results may be lower limits on the true magnetic field strength present in a surviving core following partial disruption.  \citet{Bonnerot17} and \citet{Guillochon17} both highlight the importance of repeated partial disruptions, which arise naturally for stars deep in the empty loss cone regime.  For example, if a star undergoes $N$ partial disruptions before a terminal full disruption, its initial field will be amplified by a factor $\sim f_{\rm amp}^N$, possibly producing enough magnetic flux for the final disruption to power a strong, relativistic jet.

A partial disruption produces a distinctive temporal 
behaviour of the fall-back rate. \citet{Guillochon&RamirezRuiz13} find numerically that it asymptotically approaches a power-law with index $n_\infty \approx -9/4$. Interestingly, this result can also be derived and understood in the context of the impulse approximation. \citet{Coughlin19} show that the ``frozen-in'' spread in debris energies (e.g. Eqs. \ref{eq:distr}, \ref{eq:salami}) will be modified by the gravitational influence of a surviving core embedded within the stream.  Solving the Lagrangian equation of motion for the combined stream-core-SMBH system, they find a late-time fallback rate that is a power law with index $n_\infty\approx -1 - (\sqrt{73}-1)/6 \approx -2.257$, and which is at leading order independent of the mass of the surviving core.

We close by noting that the processes of partial tidal disruption that we have described is quite sensitive to the interior structure of the star. The discussion above has focused on polytropic models. In the following section, we explore more closely how the basic principles of both full and partial disruption apply to realistically evolving stars with a more complex internal structure.

\section{Exploring different stellar types} 
\label{sec:stellar_types}

The stellar clusters that surround galactic center black holes are believed to consist of a wide spectrum of stellar masses and evolutionary types. In our own Galactic Centre's nuclear star cluster, we observe light from giant-branch stars, substructures of young, massive stars, and diffuse light from an old population of main sequence stars \citep[e.g.][]{2007A&A...469..125S}. In extragalactic nuclear clusters, there is also evidence for a diversity of stellar ages and types (see, for example, the work on NGC 404 of 
\citealt{2010ApJ...714..713S}, or the wide range of star formation histories seen in the nuclear star cluster sample of \citealt{Georgiev&Boker14}). Each of these types of stars can be scattered into orbits that lead to their disruption. In this section, we review the spectrum of possible stellar disruptions and the characteristics that relate their unique stellar evolutionary state to the outcome of a close passage by the black hole. 

Differences in the disruption processes of different stellar types stem first of all from their different tidal radii. The smallest pericenter for a non-plunging parabolic orbit around a BH is the innermost bound spherical orbit (IBSO). This latter depends on the BH mass, spin, and orbital inclination, is $4GM_{\rm BH}/c^2$ for a non-spinning SMBH, and can be as small as $GM_{\rm BH}/c^2$ (for prograde orbits in the equatorial plane of a maximally-spinning SMBH). Setting the tidal radius equal to this distance gives an approximate estimate of the maximum black hole mass for tidal disruption to occur.  This upper limit is usually called the Hills mass \citep{Hills75}, and for a non-spinning black hole is
\begin{equation}
    M_{\rm H} \sim \frac{R_\star^{3/2}}{M_\star^{1/2}} \left(\frac{c^2}{4G} \right)^{3/2} = 4\times 10^7 M_\odot \left(\frac{R_\star}{R_\odot} \right)^{3/2} \left(\frac{M_\star}{M_\odot} \right)^{-1/2} \propto \rho_\star^{-1/2}.
    \label{eq:Mmax}
\end{equation}
for a non-spinning black hole\footnote{Many calculations in the literature equate the tidal radius to the Schwarzschild horizon radius, $2GM_{\rm BH}/c^2$, rather than the IBSO radius (e.g. \citealt{Hills75}).  For the reasons described above, no quasi-Newtonian calculation of the Hills mass is more trustworthy than a factor of a few, and it is better to use fully relativistic results.}. While the scalings in Eq. \ref{eq:Mmax} are accurate, the prefactor can only be trusted to within a factor of a few, because (i) general relativistic tides differ from Newtonian tides in their strength, and (ii) physical radii such as the IBSO are coordinate-dependent quantities in general relativity.  For these reasons, it is more accurate to perform fully general relativistic calculations using the Kerr metric tidal tensor (Eq. \ref{eq:tidalTensorGR}).  This can be done analytically (see \citealt{Kesden12}, and also the discussion in the \ratechap) or with hydrodynamical simulations performed in a Kerr metric tidal field (e.g. \citealt{Ryu+20d}). 

Eq. \ref{eq:Mmax} shows that a variety of stellar types are needed in order to probe the full range of SMBH masses. Exact relativistic calculations indicate that tidal disruptions of main sequence (MS) stars by Schwarzschild black holes\footnote{We note, however, that $M_{\rm H}$ is a strong function of SMBH spin $\chi_{\rm BH}$ \citep{Kesden12}, and a favorably oriented orbit around a $\chi_{\rm BH}\approx 1$ SMBH can increase the MS Hills mass almost to $10^9 M_\odot$, as in \citet{Leloudas+16}.} only happen for $M_{\rm BH} \lesssim 10^8 M_\odot$ (\citealt{Kesden12}), while evolved stars can be disrupted by SMBHs with $M_{\rm BH} \gtrsim 10^8 M_\odot$. Typical white dwarfs (WDs) can only be disrupted when $M_{\rm BH} \lesssim 10^5 M_\odot$, although low-mass helium WDs with extended hydrogen envelopes can be partially disrupted so lon gas $M_{\rm BH} \lesssim 10^7 M_\odot$. For the same stellar type, high-$\beta$ events can only occur when $M_{\rm BH} \ll M_{\rm H}$. The $\beta-M_{\rm BH}$ parameter space where tidal disruptions can occur is often visualized with a ``TDE triangle'' diagram (see e.g. Fig. 1 in \citealt{Luminet&Pichon89b}, or Fig. 1 in \citealt{Stone+2019}).

\begin{figure}
\begin{center}
\includegraphics[width=0.8\textwidth]{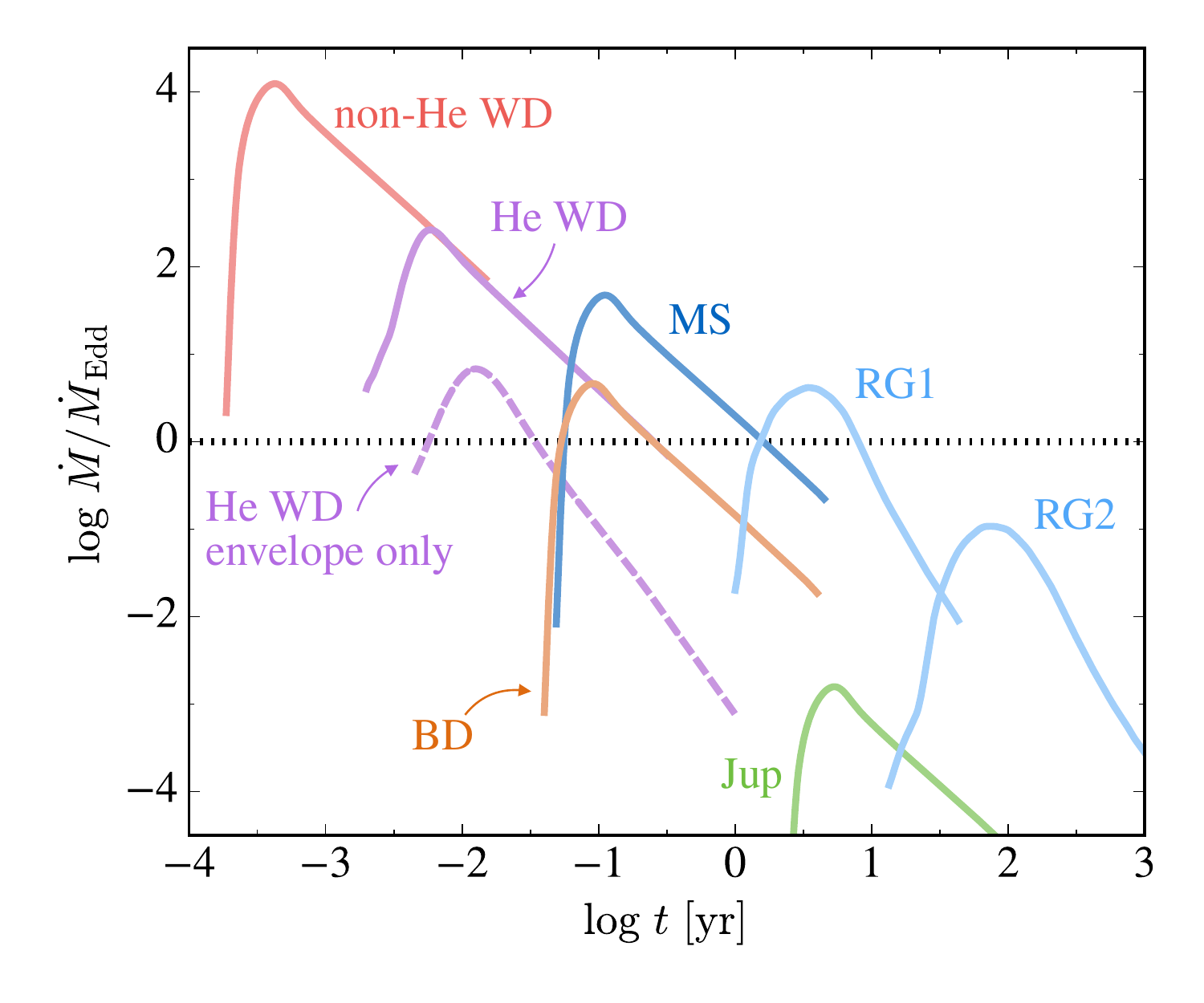}
\caption{Mass fallback rates $\dot{M}$ for six representative objects scaled to a single SMBH mass---$M_{\rm BH}=10^6 M_\odot$ is chosen for comparison, but note that a non-He WD cannot be disrupted outside the horizon of a Schwarzschild black hole for this mass. 
We show a $0.6 M_\odot$ non-He WD in red, a 0.17 $M_\odot$ He WD in purple, a 0.6 $M_\odot$ MS star in blue, a 50 $M_{\rm Jup}$ brown dwarf in brown, a 1 $M_{\rm Jup}$ planet in green, and a 1.4 $M_\odot$ red giant at 10 $R_\odot$ (RG1) and at 100 $R_\odot$ (RG2) in light blue. We show a $\beta=0.9$ encounter (full disruption) for the non-He WD, MS star, BD, and planet, and a $\beta=1.5$ encounter for the giant stars. For the He WD, we show two curves for comparison: the solid line shows a full disruption and the dashed line shows an envelope-stripping encounter. Time is relative to pericenter passage. Figure from \citet{LawSmith+17b}, their figure 12.}
\label{fig:menu_mdots}
\end{center}
\end{figure}

Figure~\ref{fig:menu_mdots} plots mass fallback rates against time seen in hydrodynamical simulations \citep{MacLeod+12,Guillochon&RamirezRuiz13,LawSmith+17b} for several representative objects: a main-sequence star, a red giant at two different evolutionary stages, white dwarfs (He and CO/ONe), a brown dwarf, and a Jupiter-mass planet. The peak fallback rates and timescales span several orders of magnitude. One can understand this at the order-of-magnitude level without the need for hydrodynamical simulations, as the characteristic timescale and fallback rate for a tidal disruption scale with the BH mass, stellar mass, and stellar radius (Equations \ref{eq:t_min} and \ref{eq:mdot}). The fall-back rate $\dot{M}$ can range from highly super-Eddington to highly sub-Eddington, with peak timescales ranging from less than $1$ day to more than $100$ years.
The way that the potential energy implicit in these mass return rates is converted into radiation is still highly debated (see the \flowchap, \diskchap, and \emischap for more details). In several---though not all---current models of accretion flow formation, the efficiency with which $\dot{M}$ is converted into radiation is a strong function of the dimensionless parameter $R_{\rm p}/R_{\rm g}$.  In these models, luminous flares will arise predominantly from encounters with $R_{\rm p} \lesssim 10R_{\rm g}$, biasing observations towards only finding TDEs from $\{M_\star, M_{\rm BH}\}$ pairs where the SMBH mass is within a factor $\approx 10$ of the Hills mass \citep{StoneMetzger16}; see Fig. 1 of \citet{LawSmith+17b} for this phase space.  

\subsection{Main Sequence Stars}
The internal structure of MS stars changes with stellar mass: more massive MS stars are more centrally concentrated. The effect of stellar structure on the disruption process was first thoroughly studied by \citet{Lodato09} and \citet{Guillochon&RamirezRuiz13}, using polytropic stellar structure models (see Section~\ref{subsec:hydro} for discussion).  While the polytropic approximation has some validity, particularly at the two extremes of the zero-age main sequence mass spectrum, it does not self-consistently account for important components of stellar physics (such as the changing structure of the star as it evolves in the MS) and has difficulty modeling stars with $0.5M_\odot \lesssim M_\star \lesssim 1.0 M_\odot$.  Compared to realistic stellar models, polytropic stars will have slightly different thresholds (in $\beta$) for the onset of partial mass stripping, and also for the transition between partial and full TDEs.

Aside from ``bulk'' questions related to total mass loss, there are more subtle (but nonetheless observationally testable) predictions that can only be made with realistic stellar models.  As a star evolves along the MS, its composition profile changes, and this is reflected in the composition of the debris returning to the SMBH. Since the late-time fallback rate is dominated by material from the stellar core, and the early-time fallback rate is dominated by material from the stellar envelope, the chemical composition of fallback material (which may be reflected in emission line equivalent widths) will change over time if the progenitor star is chemically differentiated \citep{2016MNRAS.458..127K}.  Using a semi-analytic fallback framework \citep[][see also Eq. \ref{eq:salami}]{Lodato09}, \citet{2018ApJ...857..109G} calculated the time evolution of the composition of the fallback material for MS stars of varying mass and age. For most stars, they predict an enhancement in helium and nitrogen and a depletion in carbon (relative to solar) with time. The strength and timing of these abundance variations in the mass fallback depend on the mass and age of the star, and can thus help determine the properties of the victim star in an observed TDE.

\begin{figure}
\begin{center}
\includegraphics[width=0.6\textwidth]{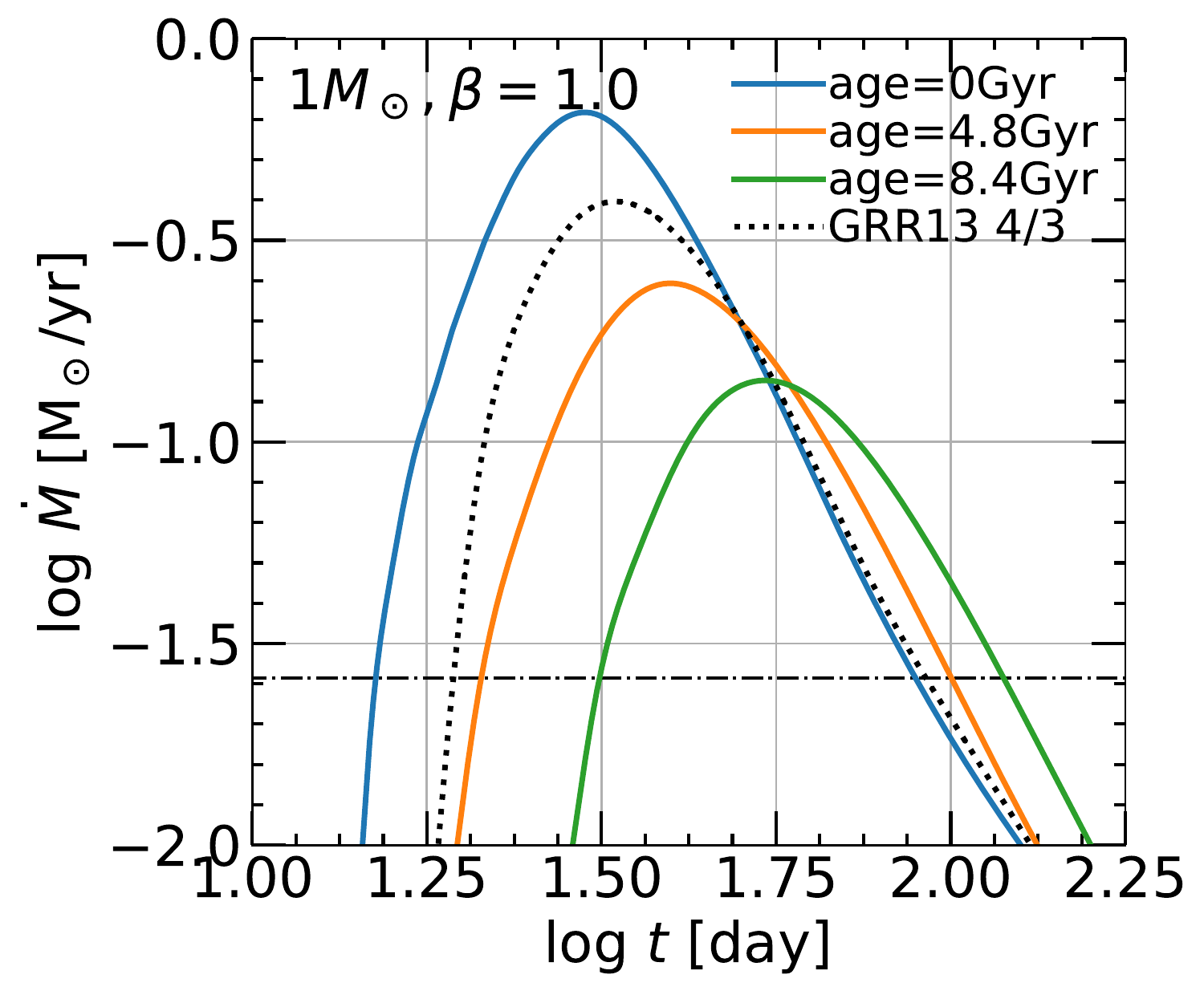}
\caption{Mass fall-back rate to the BH as a function of time for the disruption of a 1$M_\odot$ star at three different ages for a $\beta=1$ encounter with a $10^6 M_\odot$ BH. The result for a $\gamma=4/3$ polytrope from \citet{Guillochon&RamirezRuiz13}, scaled to the radius of a ZAMS 1$M_\odot$ star, is in dotted black. The Eddington limit for this BH is shown by the dot-dashed line. Figure adapted from figure 3 in \citet{Law-Smith+2019}.}
\label{fig:ms_mdots_beta1}
\end{center}
\end{figure}

\citet{Law-Smith+2019} developed a simulation framework in which stars built using \texttt{MESA} are used as inputs for tidal disruption calculations in the 3D adaptive-mesh code \texttt{FLASH} \citep{2000ApJS..131..273F} with the Helmholtz EOS. This framework uses accurate stellar density profiles and tracks the chemical abundance of the debris for 49 elements. \citet{Law-Smith+2019} studied the tidal disruption of 1$M_\odot$ and 3$M_\odot$ stars at zero-age main sequence (ZAMS), middle-age, and terminal-age main sequence (TAMS). 
They find that the initial density structure of the star leads to different susceptibilities to disruption: e.g. for a ZAMS star a $\beta=2$ encounter is a full disruption, whereas for a TAMS star this is a grazing partial disruption. In addition, significant differences in the fallback rate curves for a given stellar age and mass have been found compared to results for polytropes.  This is illustrated in Figure~\ref{fig:ms_mdots_beta1}. In terms of the composition of the fallback material, the authors found that abundance anomalies in nitrogen, carbon, and helium are present before the time of peak fallback rate for some disruptions of MS stars.

Deviations between fallback curves for polytropic and realistic stellar models were also found in the work of \citet{Golightly+2019b}, which simulated the disruption of $0.3 M_\odot, 1 M_\odot$ and $3 M_\odot$ stars at three different ages (for $\beta=3$ encounters). The authors find qualitative differences with polytropic TDEs, and use this comparisons to argue that determinations of SMBH mass from TDE light-curve fitting using models with polytropic stellar structures can be incorrect by as much as a factor of 5.

\citet{Goicovic19} performed disruption simulations of a $1 M_\odot$ ZAMS star constructed in the 1D stellar evolution code \texttt{MESA} \citep{2011ApJS..192....3P} for a range of $\beta$'s, using the 3D moving-mesh code \texttt{AREPO}. Their $\Delta M$ vs. $\beta$ and fallback-rate results agree relatively well with the $\gamma=4/3$ polytrope model from \citet{Guillochon&RamirezRuiz13}, which is expected as a ZAMS $1 M_\odot$ star is reasonably well approximated by a $4/3$ polytrope. \citet{Goicovic19} also studied the internal dynamics of the stellar remnant following a partial disruption.

\citet{Ryu+20a} very recently introduced fully relativistic simulations for a grid of stellar masses, impact parameters, and SMBH masses, at a single stellar age (halfway through the MS for each star, with models taken from \texttt{MESA}). They provide fitting formulae to describe the trends they find in several disruption parameters, such as the critical pericenter distance for full disruption, the mass of the remnant, and the spread in the debris energy distribution. For all partial disruptions, they find that mass-loss continues for many stellar dynamical times after pericenter.

\subsection{Giant Stars}
\label{sec:giant}

As stars evolve off the main sequence, their radii grow by factors of tens to hundreds. Giant stars, therefore, are particularly vulnerable to tidal forces from a supermassive black hole (recall that the tidal disruption periapsis distance scales linearly with stellar radius). Also of importance in the context of TDEs is the stars' internal structure: giant stars posses a composite structure of dense core  and low-density envelope. 

\subsubsection{Disruption of Giant Stars}

The gas dynamics of the tidal disruption of giant stars was first examined extensively by \citet{MacLeod+12}. While the qualitative process of tidal disruption remains the same as for main sequence stars, the composite structure of giant stars yields differing behavior for an encounter with same impact parameter. 
Compared to a tidal disruption of a MS star, with its less-differentiated internal structure, the tidal forces of the black hole tend to disturb only the giant star's outer envelope.  Thus, the dense core generally survives the encounter, and continues on an orbit similar to that on which it first encountered the black hole. 

\begin{figure}
\begin{center}
\includegraphics[width=0.6\textwidth]{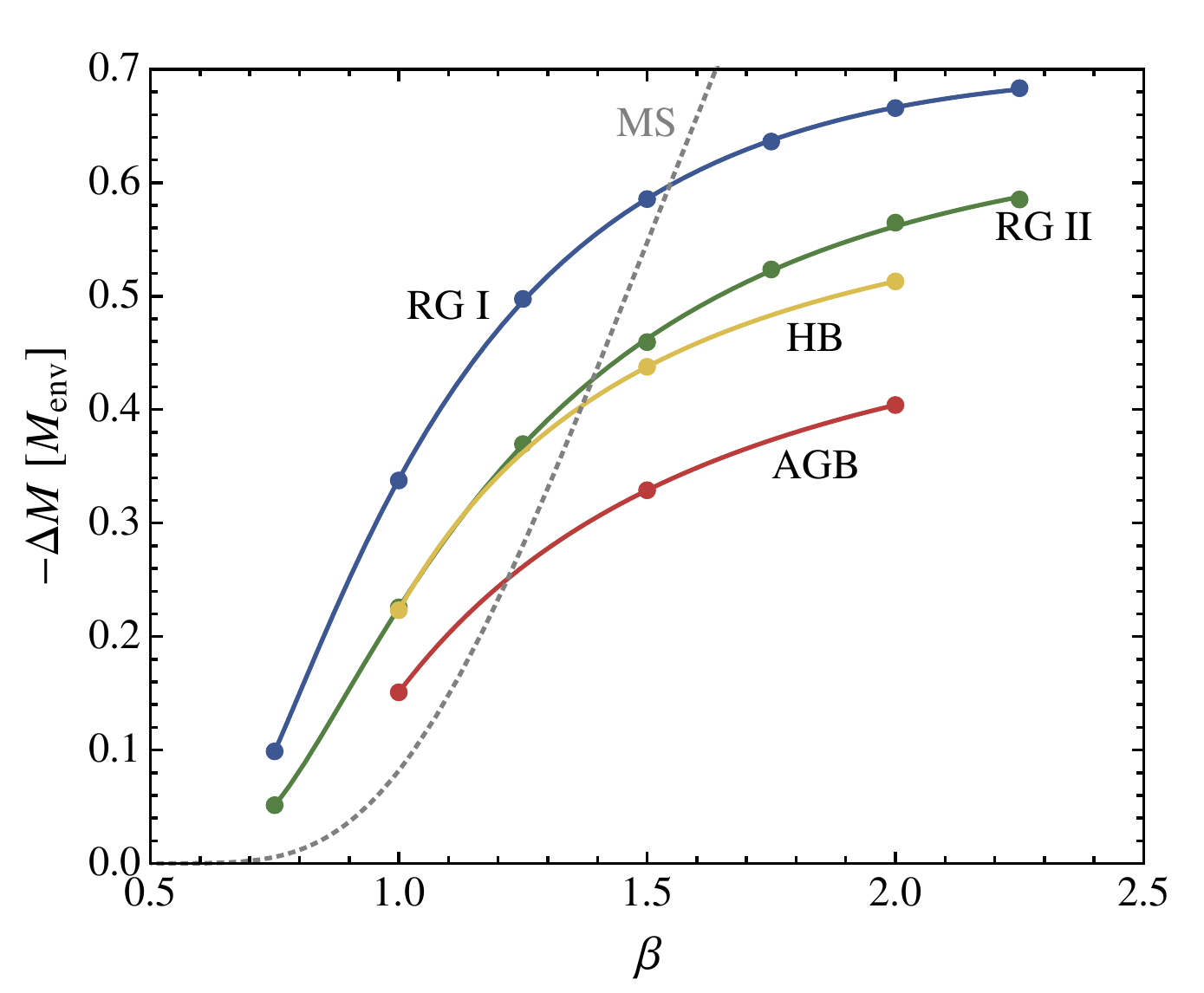}
\caption{Mass removed as a function of impact parameter, $\beta$, in four model giant stars. Here $\Delta M$ is normalized to the total envelope mass $M_{\rm env}$.  Higher impact parameter encounters remove more of the envelope material, but even relatively deeply plunging encounters with $\beta\gtrsim2$ do not remove the entire envelope. Further, models with higher core masses retain more envelope mass, because contraction of the envelope around the core upon mass loss shelters it from complete disruption.
Here MS corresponds to an $n=3/2$ polytrope \citep{Guillochon13}, while RG I (ascending the red giant branch), RG II (tip of the red giant branch), HB (horizontal branch), and AGB (tip of the asymptotic giant branch) correspond to various moments in the evolution of an originally $1.4M_\odot$ stellar model. Adapted from \citet{MacLeod+12}'s figure 6, where we refer for more information.}
\label{fig:giant_deltaM}
\end{center}
\end{figure}

This intuitive picture has been tested by hydrodynamical simulations that reveal how some surrounding envelope material is not lost to tides even in deeply-penetrating encounters, as is shown Figure~\ref{fig:giant_deltaM}. \citet{MacLeod+12} argue that this resilience to disruption can be attributed to the adiabatic change of the inner envelope on a dynamical timescale in response to the removal of the overlying layers \citep{1987ApJ...318..794H}. While the outermost layers of a giant star's stellar structure tend to expand upon mass loss, when the core becomes the dominant mass component, the remaining envelope material contracts as mass is removed, self-sheltering from further mass loss in a given encounter. Another factor that inhibits mass loss is the gravitational pull of the surviving core, that causes partial re-accretion of the surrounding envelope material, as the core sweeps through it. 

In essence, this implies that all giant-star tidal disruption events are partial tidal disruptions. Depending on the orbital dynamics, the star may return for further interactions with the black hole, as was considered in detail by \citet{MacLeod+13}. We discuss this possibility further in Section~\ref{sec:multiple_TDE}. 
Finally, \citet{2014ApJ...788...99B} consider the related scenario of a giant star disruption, where a surviving degenerate compact core
dynamically evolves because of tidal heating, as well as emission of gravitational waves. Its fate can be either a direct plunge into the massive $>10^{6} M_{\odot}$ black hole or disruption, if the tidal heating succeeds in lifting the matter degeneracy and expanding the core.

\subsubsection{Fall-back to the Black Hole}

One consequence of the extended radii and correspondingly large tidal disruption radii of giant stars is that the characteristic timescales for the periapsis passage and fall-back of material to the black hole are extended. Because the typical mass involved is of the same order as a lower main sequence star, this implies lower fall-back rates toward the black hole.  

The characteristic timescale for the encounter itself is the stellar dynamical time, because $R_{\rm t}/v_{\rm p} \approx \sqrt{R_\star^3 / 2 G M_\star} \sim T_\star/\sqrt{2}$. While a main sequence star might have a dynamical time of hours, a giant of $R_\star=100R_\odot$ and $1M_\odot$ has a dynamical time on the order of $10^6$ seconds, or a month. Thus, while the process of disruption is still rapid relative to the star, it proceeds relatively slowly from a human perspective.  Likewise, from Eq. \ref{eq:t_min}, we can see that the characteristic fall-back timescale $t_{\rm min}$ is $\sim 100$ yr for a solar-mass, $R_\star=100R_\odot$ giant star. This also carries implications for the peak fallback rate, $\dot{M} \sim \Delta M / t_{\rm fb}$. Holding fixed other properties, $\dot{M} \propto R_\star^{-3/2}$. Therefore,  TDEs of giant stars are unlikely to fuel the rapid, powerful episodes of black hole accretion that we associate with ``standard'' TDEs. Their characteristic properties are instead very extended duration, lower-level mass fallback toward the black hole. 

The key features of red giant TDE fallback are illustrated in Fig. \ref{fig:menu_mdots}, which shows results from hydrodynamic simulations by \citet{MacLeod+12} and \citet{LawSmith+17}. The figure compares main sequence and white dwarf tidal disruptions to two characteristic giant star phases. In comparison to their more compact counterparts, disruptions of giant stars yield fallback to the black hole at lower rates spread over much longer durations. The $10R_\odot$ RG I model still feeds material to a $10^6 M_\odot$ black hole above its Eddington limit (for nearly 10 years), but the $100R_\odot$ RG 2 model peaks at approximately 10\% of Eddington. 

The dilute streams of material falling back to the black hole in the red giant TDE scenario have led \citet{Bonnerot+16b} to argue that stream interaction with the gas surrounding the quiescent supermassive black hole excites the Kelvin-Helmholtz instability.  If this instability develops, the debris will fragment and dissolve into the ambient medium before they can return to the black hole, drastically decreasing the mass fall-back rate.

\subsection{White Dwarfs}

White dwarfs (WDs) have densities of $\approx 10^4$--$10^7$ g/cm$^3$ and can thus typically (for e.g., CO WDs) only be disrupted outside the event horizon for black holes of mass $\lesssim 10^5 M_\odot$, but low-mass helium WDs with hydrogen envelopes can extend this limit to $\approx 10^7 M_\odot$. The tidal disruption of WDs can thus be a unique probe of intermediate mass black holes.

The gas hydrodynamics in a WD tidal disruption is similar to the behavior of a $\gamma=5/3$ polytrope for all but the most massive WDs. WD tidal disruptions have been studied in e.g. \citet{Rosswog+09, 2013PhRvD..87j4010C, 2014ApJ...794....9M, 2016ApJ...819....3M, LawSmith+17b}. 
WD tidal disruptions are expected to produce flares with characteristic timescales of $\approx 10^2$--$10^5$ s. \citet{2013PhRvD..87j4010C} and \citet{2012ApJ...749..117H} have closely examined the effects of passages close to the black hole horizon. 
Unlike in typical MS star tidal disruptions, there is the possibility for detonation due to compression in highly-penetrating encounters. The energy release in such a detonation can approach the luminosity of a Type Ia supernova. This possibility, as well as the details of the disruption dynamics and the relative rate of WD disruptions, are discussed in detail in the \wdchap. 

WDs have an inverse mass-radius relationship, which means that it is the least massive (and thus least dense) WDs that can be tidally disrupted by the highest mass BHs. \citet{LawSmith+17b} study the disruption of helium-core hydrogen-envelope white dwarfs. These low-mass ($\lesssim 0.5 M_\odot$) WDs extend the range of BH masses that can disrupt typical WDs (see above), and offer flares with peak timescales ($\sim$1--10 days) in between those of typical WDs and MS stars. Because of their unique compositional structure, these objects can also produce flares powered by hydrogen-only fall-back material for grazing encounters, or a transition from hydrogen-rich to helium-rich fall-back material for more deeply-penetrating encounters. 

\section{Phenomenology of highly penetrating encounters}
\label{sec:high_beta}
A TDE is typically regarded as ``highly penetrating'' if $\beta = R_{\rm t}/R_{\rm p}$ is significantly greater than one.  In this case, the severe compression experienced by the star will lead to the adiabatic buildup of pressure, and the eventual reversal of the vertical collapse in a hydrodynamic rebound near pericenter.  Sometimes shocks are formed during this process.  In Newtonian gravity, the hydrodynamic pinch point is fixed in space at a true anomaly $f_{\rm c}>0$, as described in \S \ref{sec:dyn_basics}.  This indicates that the collapse and rebound will occur shortly after pericenter passage.  

Stellar TDEs may also be highly penetrating in a different sense, if the parameter $b=R_{\rm g}/R_{\rm p} \sim 1$.  In this case, the star's center of mass orbit will deviate highly from that of a closed Keplerian ellipse, as relativistic effects (such as precession and, at a higher order, gravitational radiation reaction) will come into play.  Unique phenomenology can emerge from the combination of high $\beta$ (i.e. $\gg 1$) and high $b$ (i.e. $\sim 1$).  For example, in the stronger tidal field of relativistic gravity, a star with $R_{\rm p} \sim R_{\rm g}$ may actually undergo two or more vertical compressions and bounces, the first of which is prior to pericenter passage;
\citet{Luminet&Marck85} provide a simple geometric proof that the number of bounces is roughly equal to the number of self-intersections of the center-of-mass geodesic inside the tidal radius, a prediction that is roughly borne out by combining a relativistic tidal field with the affine model \citep{Luminet&Marck85} and one-dimensional hydrodynamic simulations \citep{Brassart&Luminet10}.  

In this section, we explore three somewhat speculative predictions of high-$\beta$ and/or high-$b$ compression: high-energy shock breakout signals, gravitational wave emission, and runaway thermonuclear reactions.  At the time of writing, none of these predictions have been clearly observed, but the detection of any would be of significant value for TDE science goals. In particular, these detctions would time the disruption, providing an essential timeline for interpreting subsequent observations.

\subsection{Shock breakout and prompt X-rays}

If an outgoing shock is launched near peak compression of the star, it may acquire a large specific energy that is at most an order unity fraction of the total kinetic energy of compression, $U_{\rm c} \sim \beta^2 GM_\star^2 / R_\star$.  The timescale for the release of this energy is much longer than the time of peak compression 
($T_{\rm c} \sim T_\star \beta^{-4}$, for a $\gamma=5/3$ polytrope) because the star does not compress simultaneously; rather, the star acquires an ``hourglass'' shape as it passes through the tidal pinch point, which is fixed in space, with leading portions of the star rebounding while trailing portions are still collapsing (see also \S \ref{sec:dyn_basics}).  Assuming that each ``column'' of the star begins tidally free falling at the moment it crosses into the tidal sphere, the duration of shock breakout emission will be roughly the time it takes for the star to fully cross, $T_{\rm cr} \approx 2R_\star / \sqrt{GM_{\rm BH} / R_{\rm t}} \approx \sqrt{2}T_\star (M_{\rm BH} / M_\star)^{-1/3}$.  This estimate is equivalent to taking the time it takes the tidally distended stellar debris to cross pericenter: in the limit of high $\beta$, the star's long axis has a half-length $\approx \beta^{1/2} R_\star$ \citep{Stone+13}, so we again find $T_{\rm cr} \approx 2\beta^{1/2}R_\star / \sqrt{GM_{\rm BH} / R_{\rm p}} \approx \sqrt{2}T_\star (M_{\rm BH} / M_\star)^{-1/3}$.

Even if a large fraction of the total compressional energy budget $U_{\rm c}$ goes into shock-heating the star, most of this will not be promptly radiated.  Because the star has an optical depth at peak compression
\begin{equation}
    \tau_{\rm c} \sim \frac{M_\star \sigma_{\rm T}}{4\pi R_\star^2 m_{\rm p}},  \label{eq:opticalDepth}
\end{equation}
that is far greater than unity, only a small fraction of the shock heating can emerge as a prompt transient.  Note that in Eq. \ref{eq:opticalDepth}, we have made use of the Thompson cross-section $\sigma_{\rm T}$, the proton mass $m_{\rm p}$, and have assumed that the star's cross-sectional area at peak compression is $\approx 4\pi R_\star^2$, as is motivated by Eqs. \ref{eq:freeLong} and \ref{eq:freeShort}.  Following \citet{kobayashi+2004}, we may estimate the maximum theoretically possible bolometric luminosity by (i) assuming the shock deposits its energy uniformly through the vertical layers of the compressed star, and (ii) that during the passage of the star through the pinch point, shock-heated thermal energy diffuses out of the upper layers down to a depth 
\begin{equation}
    D \sim \sqrt{\frac{cT_{\rm cr} z_{\rm c}}{\tau_{\rm c}} },
\end{equation}
where we have made use of the height of the star at peak compression, which for a $\gamma=5/3$ polytrope is $z_{\rm c} \sim R_\star\beta^{-3}$ (Eq. \ref{eq:zc}).  This is an upper limit because in a steep stellar atmosphere, most of the shock energy will be lost in deeper layers before it arrives near the dilute surface.  Under these assumptions, the maximum peak luminosity will be 
\begin{equation}
    L_{\rm sh}^{\rm max} \sim \frac{U_{\rm c}D}{T_{\rm cr}z_{\rm c}} \approx 5\times 10^{42}~{\rm erg~s}^{-1}~\beta^{7/2} \left( \frac{M_\star}{M_\odot}\right)^{19/12} \left( \frac{R_\star}{R_\odot}\right)^{-5/4} \left( \frac{M_{\rm BH}}{10^6 M_\odot}\right)^{1/6}.
\end{equation}
This is the same upper limit as in \citet{kobayashi+2004}, except that it also considers the $\beta$-dependence of $U_{\rm c}$.  Actual estimates for the shock breakout luminosity were performed by post-processing $\beta=7$ hydrodynamical simulation results in \citet{guillochon+2009}, taking into account the realistic density profile of a stellar atmosphere (i.e. self-consistently modeling the deposition of shock energy into layers of different density).  This work found luminosities roughly one order of magnitude smaller than this upper limit.  However, it is numerically challenging to resolve the breakout layer of the tidally compressed star, and underresolution may cause an overestimate of the energy in the breakout shell.  To account for this, \citet{yalinewich+19} used an analytic model incorporating a realistic stellar atmosphere profile to derive the following, far more pessimistic estimate for the shock breakout luminosity:
\begin{equation}
    L_{\rm sh} \approx 9 \times 10^{40}~{\rm erg~s}^{-1} \beta^{1.14}\left( \frac{M_\star}{M_\odot}\right)^{0.87} \left( \frac{R_\star}{R_\odot}\right)^{-0.34} \left( \frac{M_{\rm BH}}{10^6 M_\odot}\right)^{0.33},
\end{equation}
where a $\gamma=5/3$ polytrope has been assumed.  If the shock is matter-dominated, the emitted spectrum will be quasi-thermal with a blackbody temperature $T_{\rm sh}$ in the X-rays:
\begin{equation}
    k_{\rm B}T_{\rm sh} \approx U_{\rm c}\frac{m_{\rm p}}{M_\star} \approx 1.9~{\rm keV} ~\beta^2 \left( \frac{M_\star}{M_\odot}\right) \left( \frac{R_\star}{R_\odot}\right)^{-1}.
\end{equation}
However, as was pointed out by \citet{yalinewich+19}, high-$\beta$ events will generally find themselves in either a radiation-dominated blackbody regime (which softens the growth of temperature to $T_{\rm sh} \propto \beta^{1/2}$) or a photon-starved regime in which the typical energy of the non-thermal emission is set by a balance between pair production and annihilation, i.e. $k_{\rm B}T \approx 50 ~{\rm keV}$.

The short durations and low luminosities of X-ray shock breakout signals from main sequence TDEs make their detection unlikely with current instrumentation.  However, the less frequent tidal disruption of red giant stars will produce optical/UV shock breakout flashes of much longer duration $T_{\rm cr}$, and LSST may detect these at a rate of $10^{-1}-10^1~{\rm yr}^{-1}$ \citep{yalinewich+19}.
  
\subsection{Gravitational wave emission during disruption}
\label{sec:GW}
Gravitational waves are produced when the quadrupolar moment of the mass distribution of a source is changing with time, in an accelerated fashion. During the star's closest approach, there are contributions to the variation of the mass quadrupole moment from both the changing mass quadrupole of the star-black hole system {\em and} the internal mass quadrupole of the star itself\footnote{Emission of gravitational waves during later stages of a TDE has been investigated by \cite{toscani+2019}, but in this section we will only report on gravitational wave emission linked to the disruption process.}.
In the first case, the binary components can be regarded as point masses orbiting each other, and the ultimate source of energy is their orbital energy. This signal has been investigated for both main sequence stars with SMBHs \citep{kobayashi+2004} and for white dwarfs with intermediate mass black holes \cite[e.g.][]{Sesana+08,Rosswog+09,2012ApJ...749..117H, Anninos+18}. Since the binary dynamical interaction is far from the highly relativistic merger phase, a simpler analytical description of the observed strain can be adopted: 
\begin{align}
  h \approx \frac{1}{d_{\rm L}}\frac{4G}{c^2}\frac{E_{\rm kin}}{c^2},
\label{eq:ekin}
\end{align}
\cite[e.g.][]{Thorne98}.
This signal scales with the luminosity distance $d_{\rm L}$ and it is proportional to the source's kinetic energy $E_{\rm kin}$. Since most of the emission occurs at the closest approach $R_{\rm p}=\beta^{-1} R_{\rm t}$, and the large mass ratio leaves the black hole still at the center of mass, we can substitute
$E_{\rm kin}= M_\star \left(GM_{\rm BH}/R_{\rm p}\right)$ in Eq.~\ref{eq:ekin} to estimate the strength of the signal
\begin{align}
    h&\approx \beta\times\frac{R_{\rm s}R_{\rm s\star}}{R_{\rm t}d_{\rm L}}\nonumber\\
    &\approx 2 \beta \times 10^{-22} \left(\frac{M_\star}{M_\odot}\right)^{4/3}\frac{R_{\odot}}{R_\star}\left(\frac{M_{\rm BH}}{10^6\text{M}_{\odot}} \right)^{2/3} \left(\frac{d_{\rm L}}{16\text{Mpc}} \right)^{-1},\label{eq:burst}
\end{align}
where $R_{\rm s}= 2GM_{\rm BH}/c^2$ and $R_{\rm s\star} = 2GM_\star/c^2$. This "point particle" description of the signal was verified by numerical simulations, even in $\beta \ll 1$ encounters \citep{kobayashi+2004}. The signal duration is roughly the orbital period $T(R_{\rm p})$ at pericenter and the associated frequency is $f_{\rm GW} \sim 1/T$,
\begin{align}
    f_{\rm GW}&\approx\frac{\beta^{3/2}}{2{\rm \pi}}\left(\frac{GM_{\rm BH}}{R^3_{\rm t}}\right)^{1/2}\nonumber\\
    &\approx 10^{-4}\,\text{Hz} \,\beta^{3/2} \left(\frac{M_\star}{M_\odot}\right)^{1/2} \left(\frac{R_\star}{R_\odot}\right)^{-3/2}.\label{eq:freqkepl}
\end{align}
TDEs involving WDs are necessarily associated with smaller mass black holes ($ M_{\rm BH} \le 10^{5} \text{M}_{\odot}$), as otherwise disruption does not take place (see Eq. \ref{eq:Mmax}). The involvement of a lower mass black hole would cause by itself a suppression of the signal (all other conditions the same), however WDs are also more compact than MS stars by a factor of $\sim 100$: overall the signal amplitude from a WD disruption can span from about one order of magnitude {\em above} to one order of magnitude {\em below} the value reported in Eq. \ref{eq:burst}, when considering black hole masses between $10^{3} M_\odot-10^{5} M_\odot$, and a WD with mass $0.5 M_\odot$ and radius $0.01 R_\odot$. The signal duration decreases from $\sim 2.7 \beta^{-3/2}$ hr for solar type stars down to $\sim 14 \beta^{-3/2}$ sec or less for our fiducial WD, implying that WD signals should be expected at higher frequencies, in the range $\sim 7 \times 10^{-2} - 10$ Hz for the parameters considered here. These frequency and strain estimations suggest that disruption of WDs by intermediate mass black holes are within detection reach of future space-based interferometers like the DECI-hertz inteferometer Gravitational wave Observatory \cite[DECIGO][]{Sato+17} and ALIA \citep{baker+19}, but would remain subthreshold events for the Laser Interferometer Space Antenna \cite[LISA][]{Amaro-Seoane+17} and TianQin \citep{Luo+16}. The much lower frequency range for TDEs involving MS stars, on the other hand, precludes detection by any planned mission.

We now turn our attention to the second source of gravitational wave emission: the time-verying quadrupole moment of the {\em star's} mass distribution, during the vertical compression/in-plane stretching of the star at pericenter \cite[see Section 2, and][]{guillochon+2009,stone+2013}.  Under the assumption of simultaneous vertical collapse of all stellar layers to a ``pancake" shape, the duration of the gravitational wave burst is roughly the duration of maximal compression $T_{\rm c}\approx 8.5 T_\star \beta^{-4}$ (for a $\gamma=5/3$ polytrope), with a characteristic frequency $f_{\rm GW} \sim 1/T_{\rm c}$,
\begin{equation}
    f_{\rm GW} \approx 4 \times 10^{-5}~{\rm Hz}~\beta^{4} \left(\frac{M_\star}{M_\odot}\right)^{1/2} \left(\frac{R_\star}{R_\odot}\right)^{-3/2}.
    \label{eq:fgw_starquadrupole}
\end{equation}
For main sequence stars and grazing events, this frequency is too low for any planned space-based interferometers and too high for the Pulsar Timing Array\footnote{http://ipta4gw.org}. However the strong dependence on $\beta$ implies that nearly plunging events would emit {\em just} within the LIGO/VIRGO ground-based interferometer frequency band $\sim 10$ Hz-$10$ kHz: e.g. $f_{\rm GW}\approx 15.6$ Hz for $\beta=25$. Likewise, WDs with their smaller radius would produce events at higher frequencies spanning from the LISA band ($10^{-4}-10^{-2}$ Hz) for $\beta \sim 1$ to the LIGO/VIRGO band for $\beta > 1$ (e.g. for\footnote{Note that $\beta>6$ for WDs is possible only for black hole masses smaller than $10^{4} M_\odot$.} $\beta=10$, $f_{\rm GW} \approx 280$ Hz). 
The rapid vertical collapse ($v_{\rm z}/T_{\rm c} \propto \beta^{5} \sqrt{G M_\star/R_\star}/T_\star$) causes an accelerated quadrupolar variation that produces a gravitational wave strain highly dependent on $\beta$, with $h \propto \beta^{2}$ \citep{stone+2013}. If one instead considers the less dramatic stretching of the star in the orbital plane that occurs on a slower time scale --the orbital time $T(r_{\rm p})$, rather than the compression timescale $T_{\rm c}$ -- then the dependence of the strain on $\beta$ is even steeper: $h \sim \beta^{3}$ \citep{guillochon+2009}. Let's now elaborate further on the former case, that produces emission at a frequency given by Eq. \ref{eq:fgw_starquadrupole}. When the strain is computed under the assumption of synchronous vertical collapse to a stellar pancake, it reads \citep{stone+2013}
\begin{equation}
h \simeq 2 \times 10^{-27} \left(\frac{M_\star}{M_\odot}\right)^{2} \left(\frac{R_\star}{R_\odot}\right)^{-1} \beta^{2} \left(\frac{d_{\rm L}}{16\text{Mpc}} \right)^{-1} .
\end{equation}
This is a very weak signal for a MS star event, hardly detectable by LIGO/VIRGO even when considering highly penetrating events (e.g. $\beta=25$, $h \sim 10^{-24}$ at $f_{\rm GW} \approx 15.6$ Hz). On the other hand, the same nearly plunging ($\beta \gg 1$) events involving WDs would be boosted to higher frequencies and strains (e.g. $\beta=10$, $h \sim 5 \times 10^{-24}$ at $f_{\rm GW} \approx 280$ Hz), placing them above the detection threshold for future ground-based facilities, such as the Einstein Telescope\footnote{http://www.et-gw.eu/}.

\subsection{Nuclear reactions}

During high-$\beta$ encounters, stars suffer large temperature and density increases from adiabatic compression.  The degree of temperature and density increase can be even larger if shocks form during either the compression or early on in the rebound.  In this sense, a deeply penetrating tidal disruption event is analogous to an inertial confinement fusion reactor, where the inertia of an imploding plasma can -- in principle -- pin the plasma in place long enough for dynamical thermonuclear fusion reactions to occur.  On Earth, inertial confinement fusion is achieved by using powerful lasers to compress fuel pellets, but in galactic nuclei, the ``tidal piston'' of the SMBH may serve the same purpose.

This possibility was first investigated in the context of the affine model by \citet{Luminet&Pichon89}.  This early study found that only a small fraction of a MS star's mass could undergo fusion, even for extreme parameter choices, and that the energetics of the resulting radioisotopes would be subdominant to the impulsive disruption energy spread, $\Delta E$.  These conclusions have been qualitatively confirmed by a handful of subsequent hydrodynamical simulations that investigated thermonuclear reactions in high-$\beta$ TDEs in greater detail \citep{guillochon+2009}.  Thermonuclear burning does not achieve dynamical importance in MS TDEs because the primary reactions at peak compression are those on the hot CNO cycle, which is $\beta$-decay limited.  The short duration of peak compression is therefore an insurmountable bottleneck to a thermonuclear runaway.

Runaway fusion appears more promising, however, in the compression of WDs, where the burning can, depending on composition, proceed via either the triple-$\alpha$ reaction or carbon-oxygen fusion. Studies using both the affine model \citep{Luminet&Pichon89b} and hydrodynamical simulations \citep{Rosswog+09} have found evidence for thermonuclear detonation waves that burn an order unity fraction of the star, substantially alter the spread in debris energy, and synthesize sufficient quantities of radioisotopes to power a radioactive transient in the unbound ejecta.  However, more recent high-resolution studies have shown that the formation of shocks that generate detonation waves during WD compression is highly sensitive to numerical resolution \citep{Tanikawa+17}, and may be artificially triggered by spurious heating produced by underresolution of the tidal compression phase.  While the most recent sets of WD disruption simulations indicate a broad parameter space for ignition and thermonuclear burning \citep{Tanikawa18, Anninos+18, Kawana+18}, the exact parameters required to tidally detonate a WD remain contested.  These issues are addressed in greater detail in the \wdchap .

\section{Unbound debris}
\label{sec:unbound}
As mentioned earlier, part of the stellar debris acquires a positive orbital energy and leaves the vicinity of the black hole on unbound orbits. The exact unbound fraction of stellar mass depends 
on whether the disruption was total or partial (Section \ref{sec:partial_TDE}), and on the initial stellar orbit -- or equivalently on the initial star's {\em orbital} energy $E_\star$. For a star approaching on a parabolic orbit ($E_\star=0$) that is completely disrupted, half of the debris is expected to gain energy and be lost from the system. For the same fully disrupted star approaching instead on either an elliptical or a hyperbolic orbit, the unbound fraction would depend on the ratio $|\Delta E/E_\star|$ between the tidal energy spread $\Delta E$ (eq.~\ref{eq:energy_spread}) and the orbital energy. For instance, for $|\Delta E/E_\star| \ll 1$, the whole debris remains respectively bound for $E_\star<0$
and unbound for $E_\star >0$. These considerations have been explored both in the context of tidal separation of stellar binaries and tidal disruption of a single star \citep{kobayashi+2012,Hayasaki16}.

Because effectively parabolic encounters are expected to dominate overall TDE rates (see e.g. the \ratechap), we discuss the canonical case of a star with mass $M_\star$, on a parabolic orbit with pericenter equal to the tidal radius $R_{\rm p}=R_{\rm t}$, that undergoes a complete disruption. We will also first assume a constant specific energy distribution $dE/dM = \Delta E/(M_\star/2)$ between $-\Delta E$ and $+\Delta E$ (Eq. \ref{eq:distr}). A simple estimate of the maximum terminal velocity is then
\begin{equation}
        v_{\rm max} = \sqrt{2 \, \Delta E} \approx 8,000 ~{\rm km~s^{-1}}~\left(\frac{M_\star}{M_\odot}\right)^{1/3} \left(\frac{R_\star}{R_\odot}\right)^{-1/2} \left(\frac{M_{\rm BH}}{4 \times 10^{6} M_{\odot}}\right)^{1/6},
        \label{eq:vmax_unbound}
\end{equation}
where we consider a SMBH with mass similar to that of Sagittarius A* (a.k.a Sgr A*), the SMBH in the Centre of our Galaxy. If not specified otherwise, these will be our fiducial parameters for this section. For such a system the tidal radius is $R_{\rm t} \approx 10^{13}$ cm. Some fraction of the unbound debris then escapes at a few percent of the speed of light, comparable to the speed of a supernova blastwave. Likewise, the total available kinetic energy and momentum are  $\lesssim (1/2) v_{\rm max}^2 (M_\star/2) \approx 3 \times 10^{50}$ erg and $\lesssim v_{\rm max} (M_\star/2) \approx 8 \times 10^{41}$ g cm s$^{-1}$. This significant reservoir of energy and momentum stimulated the studies that we are summarizing here, as it is intriguing to explore whether they can produce observable signatures when deposited into the ambient medium. Special attention has been given to observable signatures in the Galactic Centre \citep{khoklovMelia1996,guillochon16}, because of the unique possibility to directly image a ``remnant'' of this event and because of the presence of Sagittarius A East, a radio source with unclear origin that engulfs Sgr A*'s gravitational sphere of influence.

Deceleration of the debris as it expands into the black hole surroundings and sweeps up the intervening gas is a possible way to tap its kinetic energy and convert it into radiation, in analogy with supernova remnants. The effective deceleration or ``stopping" length $R_{\rm st}$ can be defined as the distance from the black hole at which the debris velocity is half of its initial value.  
This is equivalent to the distance within which an amount of gas equal to the debris's mass has been swept up along the way, i.e. $M_\star/2 = \rho_{\rm ism} (4 \pi/3) R_{\rm st}^{3} (\Omega/4 \pi)$, where $\rho_{\rm ism}$ is the interstellar medium mass density, which has been assumed constant. Re-arranging the mass equality to evaluate the deceleration length yields
\begin{equation}
\frac{R_{\rm st}}{R_{\rm t}} \approx q^{-1/3} \left(\frac{\langle\rho \rangle}{\rho_{\rm ism}}\right)^{1/3}  \left(\frac{\pi}{\Omega} \right)^{1/3},
\end{equation}
where the mean stellar density is $\langle \rho \rangle = 3 \rho_\star/ 4 \pi$ and the mass ratio is defined as $q=M_{\rm BH}/M_*$. Assuming radial expansion into a constant solid angle equal to that under which the star is seen by the black hole $\Omega \approx \pi (R_\star/R_{\rm t})^2 \equiv \pi \theta_\star^2 \approx \pi q^{-2/3}$, we get a stopping length of $\approx 70$ pc for the mean density of the Sun ($\langle \rho \rangle=1.4$ g cm$^{-3}$) and the canonical ISM mean density (1 particle cm$^{-3}$)\footnote{This is also the distance at which the mean density within the debris equal that of the surrounding medium in this free expanding scenario}.

In fact, debris with positive energy does not move radially outward but on hyperbolic orbits with a range of energies. This causes the streams to spread in the orbital plane and trace a ``fan" shape. The sweeping area is therefore larger than in the radial case and the stopping length consequently smaller. We can simply estimate it by assuming that the black hole's gravity is the only force in place and ignoring relativistic effects. The true anomaly $\theta_{\infty}$ of the stream's orbits at $r \gg R_{\rm t}$ obeys $\cos(\theta_{\infty}) = -1/e$, where $e$ is the orbital eccentricity. The maximum eccentricity of the unbound streams, corresponding to a specific energy $+\Delta E$, is $e_{\rm max} = 1+2 q^{-1/3}$, therefore $\theta_{\infty}= \arccos{(-1/(1+2 q^{-1/3}))} \approx \pi- \sqrt{2} q^{-1/6}$. On the other hand, the funnel of debris should continuously connect to material moving on a parabolic orbit with $e=1$ (and $\theta_\infty= \pi$) and the overall opening angle in the orbital plane is roughly $\theta_{\phi} \approx \sqrt{2} q^{-1/6}$. The cross-sectional area of the ``fan" is therefore an ellips and its solid angle can be estimated as $\Omega \approx \pi \theta_{\phi} \theta_*$. The increase in the sweeping area by $\theta_{\phi}/\theta_* = \sqrt{2} q^{1/6} \approx 20$ gives a shorter stopping length of a few parsecs for our fiducial parameters.
So the expectation is that unbound debris streams deposit $\approx 10^{50}$ erg at and beyond the black hole's sphere of influence after $R_{\rm st}/v_{\rm max} \gtrsim 10^{3}$ yr.  

This approximate picture should, however, be verified in more realistic conditions. Indeed, 
the above estimates neglect several physical ingredients. First of all, the shape of the unbound debris ``fan" should be reconsidered to include the effect of self-gravity within the debris, which may be important in both the shallow-penetration and deep-penetration regimes \citep{kochanek94,coughlin&nixon15,coughlin+16a,steinberg+19}. The self-confinement of most of the mass in the stream may lead to a lower sweeping area and therefore to a longer deceleration scale with respect to the ballistic case. The situation may be most favourable in highly penetrating events, when 
an order unity fraction of the unbound debris stream remains unconstrained by self-gravity \citep{steinberg+19,yalinewich+19}. However, self-gravity may not play an important role in setting the deceleration scale, if suppressed at an earlier stage by the the energy released in the streams from hydrogen recombination weeks to months after disruption \citep{kasen+RR10,guillochon16}. In addition, the shape of the energy distribution within the debris is not truly flat, but instead it depends on both the stellar internal structure and the penetration factor (see Sections ~\ref{sec:impulse} and \ref{subsec:hydro}). Finally, partial disruptions would see less massive unbound ejecta, which would result in both a shorter free-expansion phase (when $r< R_{\rm st}$) and, on the other hand, less energy deposited upon deceleration.

Other physical ingredients to reconsider pertain to the environment. The estimated stopping radius is of the same order of magnitude of the black hole sphere of influence (e.g. $\sim 1-3$ pc for Sgr A*) and therefore the stellar gravitational potential should be also be taken into account along with the debris evolution.  The density profile in the sphere of influence of a SMBH is far from constant, given the deep potential well, and may also be aspherical if the SMBH harbours an accretion disc.

When analytically accounting for all of the above effects -- modelling both the free expansion and the deceleration phases in a Galactic Centre type of environment -- \citet{guillochon16} found that the typical energy and momentum deposited in the ISM are $ 5 \times 10^{49}$ ergs and $2 \times 10^{41}$ cm g s$^{-1}$ for a MS star and an order of magnitude less for a giant star disruption. The typical stopping length is around $R_{\rm st} \sim 20$ pc for a MS star and nearly 10 times smaller for giants, while the transverse size $R_{\rm st} \theta_{\phi}$ is around a few parsecs. The free-expansion, mostly adiabatic phase typically lasts $\sim 10^{4}$ yr. 

The energy and momentum deposited by the unbound debris streams can give rise to observational signatures (``unbound debris remnants''), such as X-ray and radio emission by the shocked interstellar medium, analogously to supernova remnants \citep{guillochon16,yalinewich+19,krolik+16}. The long energy deposition timescale of several thousand years implies a peak bolometric luminosity of only $\lesssim 10^{40}$ erg s$^{-1}$, limiting the prospects for an extragalactic detection. With radio observations of our Galactic Centre, however one may constrain the TDE rates by detecting unbound debris remnants. \cite{guillochon16} argue that a few such remnants should be present in our Galactic Centre within the innermost $\sim 100 \rm pc$ and that Sgr A East is in fact one of them. Beside the aforementioned ``supernova-like" emission, other observational signatures include the optical flare from hydrogen recombination \citep{kasen+RR10} within the unbound debris streams, and optical emission lines from the reprocessing of light coming from the accreted bound material \citep{StrubbeQuataert09}. A detailed description of all these emission processes can be found in the \emischap~in this book.    

\section{Variations on the classical tune}
\label{sec:variations}
In this section, we review works investigating variations of the classical study case explored in the preceding sections, where a single star on a nearly parabolic orbit is fully or partially tidally disrupted during a single pericenter passage. In particular, we will focus on the physical implications of elliptic and circular orbits (Sections \ref{sec:eccentric orbits} and \ref{sec:roche-lobe_overflow}), of recursive partial disruption episodes for the same star (Section \ref{sec:multiple_TDE}) and of stellar binarity (Section \ref{sec:bianries}). The conditions for these different scenarios are also discussed.

\subsection{Disruption of stars on eccentric orbits}
\label{sec:eccentric orbits}
In the standard loss cone theory, where only scatterings between stars (in vacuum) around a SMBH are considered, a star typically enters the loss cone with marginal binding energy and therefore is disrupted on a nearly parabolic orbit, with an orbital eccentricity $1- e\sim 10^{-6}$. However, there are other channels which can supply stars to be disrupted from more tightly bound orbits. When does a finite value of $1-e$ begin to affect the dynamics of disruption?  We can determine the answer to this question by comparing two specific energy scales: the frozen-in spread of debris specific energy, $\Delta E$ (see Eq. \ref{eq:energy_spread}), and the pre-disruption specific orbital energy of the star, $|E_\star| \approx \frac{1}{2}GM_{\rm BH}/a_0$, where $a_0$ is the semimajor axis of the star's final orbit.  Normally, $|E_\star| \ll \Delta E$ and the orbit can be treated as effectively parabolic, but below a critical orbital eccentricity,
\begin{equation}
e_{\rm crit}~\approx~1-2/\beta\times(m_\star/M_{\rm BH})^{1/3},
\end{equation}
$|E_\star| \gtrsim \Delta E$ \citep{Hayasaki13}, and we are in the qualitatively different regime of an ``eccentric'' tidal disruption\footnote{For a similar discussion, impacting the amount of unbound debris, see beginning of Section \ref{sec:unbound}.}. For example, $e_{\rm crit}=0.98$ for a solar-type star disrupted by a $10^6~M_\odot$ SMBH along an orbit with $\beta=1$.

In an eccentric TDE, 
the debris fallback timescale decreases from years ($e = 1$) to weeks ($e = 0.99$), days ($e = 0.9$), or even hours or minutes ($e < 0.7$), leading to a much higher peak fallback rate and a time evolution dramatically deviating from the canonical $t^{-5/3}$ pattern \citep{Hayasaki13, Dai13}. Therefore, it is expected that flares produced in eccentric TDEs should evolve on timescales faster than years. The accretion level should largely exceed that of standard parabolic TDEs, though the luminosity may still be regulated by the Eddington limit. Also, when the eccentricity of the initial stellar orbit is sufficiently small (less than $e_{\rm crit}$), all the debris will be bound so all of the stellar mass can, in principle,  accrete onto the SMBH.

N-body simulations directly modeling stellar clusters around SMBHs generally confirm the semi-analytic expectation that most stars are disrupted from effectively parabolic orbits, if the main source of loss cone refueling is stellar two-body relaxation \citep{Zhong14, Hayasaki18}. However, if there exists a massive perturber deep inside the SMBH influence radius, such as an intermediate mass black hole, or if the TDE happens in a tight binary SMBH system, then one encounter between the perturber and a very tightly-bound star (one with specific orbital energy $|E_\star| \gtrsim \Delta E$) can remove enough angular momentum from the star to create an eccentric TDE. Such encounters occur naturally during the final stages of a SMBH binary inspiral \citep{Chen+11}, and the rate may be enhanced by trapping of stars in mean-motion resonances \citep{Seto&Muto10, Seto&Muto11}.  A final burst of such TDEs may occur following the merger of the SMBH with a secondary massive black hole, when anisotropic gravitational wave emission provides a recoil kick that tilts the phase space loss cone to overlap with the orbits of surviving, tightly bound stars \citep{Stone&Loeb11}. Eccentric TDEs may also be produced following the tidal separation of a binary stellar system by a SMBH. This process places one star on a tightly -- although initially highly eccentric -- bound orbit \citep{Sari+10}, while ejecting the other star on a hyperbolic (and hypervelocity) orbit.  The bound star detached from its binary companion will have a much larger $|E_\star|$ than is typical in the stellar cusp, and a cluster of such stars will be more favorable for generating eccentric TDEs \citep{Amaro12}. 

Disruption of stars on eccentric orbits have also been simulated for understanding the disk formation process in TDEs \citep[e.g.,][]{Bonnerot16,Hayasaki16}.  
Due to the reduced dynamical spread of the debris orbital energy, the normally prohibitive computational expense of simulating the long-term aftermath of tidal disruption by SMBHs is greatly reduced (in comparison to the more astrophysically common case of a parabolic orbit TDE).  In the eccentric TDE limit, the hydrodynamical dissipation or nozzle shocks associated with compression and shearing at the pericenter will be less energetically important compared to standard parabolic TDEs. On the other hand, for eccentric TDEs the self-intersection of the debris stream induced by GR apsidal precession will always happen close to the SMBH, which promotes prompt disk formation, although the delays induced by the Lense-Thirring effect have not yet been thoroughly studied (\citealt{Dai13, Hayasaki16} -- but see \citealt{Liptai19}). We refer the readers to the \flowchap~in this book for more details.

\subsection{Roche-lobe overflow of stars on nearly circular orbits}
\label{sec:roche-lobe_overflow}
Under certain circumstances a star can approach a massive black hole on a close orbit and form an extreme-mass-ratio-inspiral (EMRI) system close to the SMBH \citep{Miller05, Amaro12}.
The orbital radius and eccentricity of the stellar orbit can both quickly decrease 
if gravitational radiation is sufficiently efficient,
until the star eventually fills its Roche lobe on a nearly circular orbit. 
The star will then keep orbiting the SMBH for a long time while its envelope is steadily stripped by the tidal force from the black hole. The stripped stellar material will then accrete onto the black hole producing quasi-periodic X-ray signals. 

Similar stable mass transfer processes between a star and a black hole via Roche lobe overflow have been extensively studied in the context of normal X-ray binaries \citep[e.g.,][]{Rappaport82, Webbink83}. The effective radius $R_L$ of the Roche lobe of the star has been calculated in \citet{Kopal59, Paczynski71, Eggleton83}, and is the same as the parabolic-encounter tidal radius $R_{\rm t}$ up to a factor of $\approx 2$.
Although much of the theoretical formalism for X-ray binaries can be adopted to study the mass-transfer between a star and a SMBH, the latter scenario is distinct in having a much more relativistic $R_{\rm L}$ (i.e. $R_{\rm L}$ is a few or a few tens of SMBH gravitational radii) due to the extreme mass ratio. Therefore, the dynamics and the evolution of the star--SMBH binary is more dramatically affected by general relativity as compared to the case of normal X-ray binaries. 

Stable Roche-lobe overflow in a star-SMBH binary was first explored by \citet{King93} and \citet{Hameury94}. Later, \citet{Dai13b} did a more rigorous calculation including general relativistic effects. Since the mass transfer happens on a timescale that is faster than the thermal (Kelvin--Helmholtz) timescale of the star but slower than the dynamical
timescale, the interior structure of the star evolves adiabatically as its envelope is stripped \citep{Dai13a}. The star fills its Roche lobe throughout this process, so from Kepler's law we see that its orbital period $T\propto~R^{3/2}_t~M^{-1/2}_{\rm BH}\propto\langle \rho \rangle^{-1/2}$, where $\langle \rho \rangle$ is the mean density of the star being tidally stripped. Therefore, 
the evolution of the stellar orbit 
is controlled by how the stellar mean density changes: low-mass stars will recede from the SMBH during Roche-lobe overflow, while high-mass stars will first continue to spiral in for some time after reaching the Roche limit, and later will spiral out. 

If mass transfer is conservative (i.e. the angular momentum of the binary only changes through the torque of gravitational radiation), one can calculate the mass transfer rate to be $\sim 10^{21-23}~{\rm g~s}^{-1}$, depending on the exact black hole and stellar masses \citep{Dai13b}. The stripped stellar material will form an accretion disk around the SMBH, and the stream of the stripped stellar material flowing through the inner Lagrange point L1 can continuously hit this small accretion disk and produce a hot spot. The hot spot will then orbit the black hole at the orbital frequency of the star and produce quasi-periodic signals, likely in X-rays. \citet{Linial17} further consider the mass leakage through the outer Lagrange point L2, and analytically estimate the conditions in which it is comparable to the canonical mass transfer rate through the first Lagrange point L1, discussed so far.

Since a Roche-lobe filling star spends most of the time (at least hundreds of thousands of years) receding from the SMBH, collisions can happen between one star in the slow outward migration and another star that enters the inspiral phase at a later time. \citet{Metzger17} considered such colliding EMRIs and found that the timescale and the luminosity of the flare produced during the collision are similar to those of TDEs. Therefore, such EMRI collisions can serve as TDE imposters if the rate is high enough.

Gravitational waves emitted by star--SMBH Roche-lobe overflow systems would have a strain given by Eq.~\ref{eq:ekin} and frequency twice the orbital one, $f_{\rm GW} = 2/T$ \citep{Linial17}. Contrary to classical single TDEs, these events will not be bursts of gravitational radiation but rather persistent sources, and this fact might help detection by allowing to accumulate the signal over many orbits observed with a spaced based interferometer. This leads to a signal-to-noise enhancement by a factor of  $\approx \sqrt{T_{\rm obs}f_{\rm GW}}$, which can reach $\sim 100$ for a $T_{\rm obs} =4$ yr mission duration (e.g. the nominal mission duration of LISA). The emission frequency for MS stars remains is rather low, $\sim 10^{-4}$ Hz.  Therefore -- conditional on the actual performances of LISA in the lower-frequency portion of its spectrum -- the gravitational radiation produced by these systems may be detectable from the Galactic Center or from nearby galaxies, for example Andromeda.

\subsection{Multiple Encounters}
\label{sec:multiple_TDE}
\begin{figure*}[t]
\centering\includegraphics[width=\linewidth,clip=true]{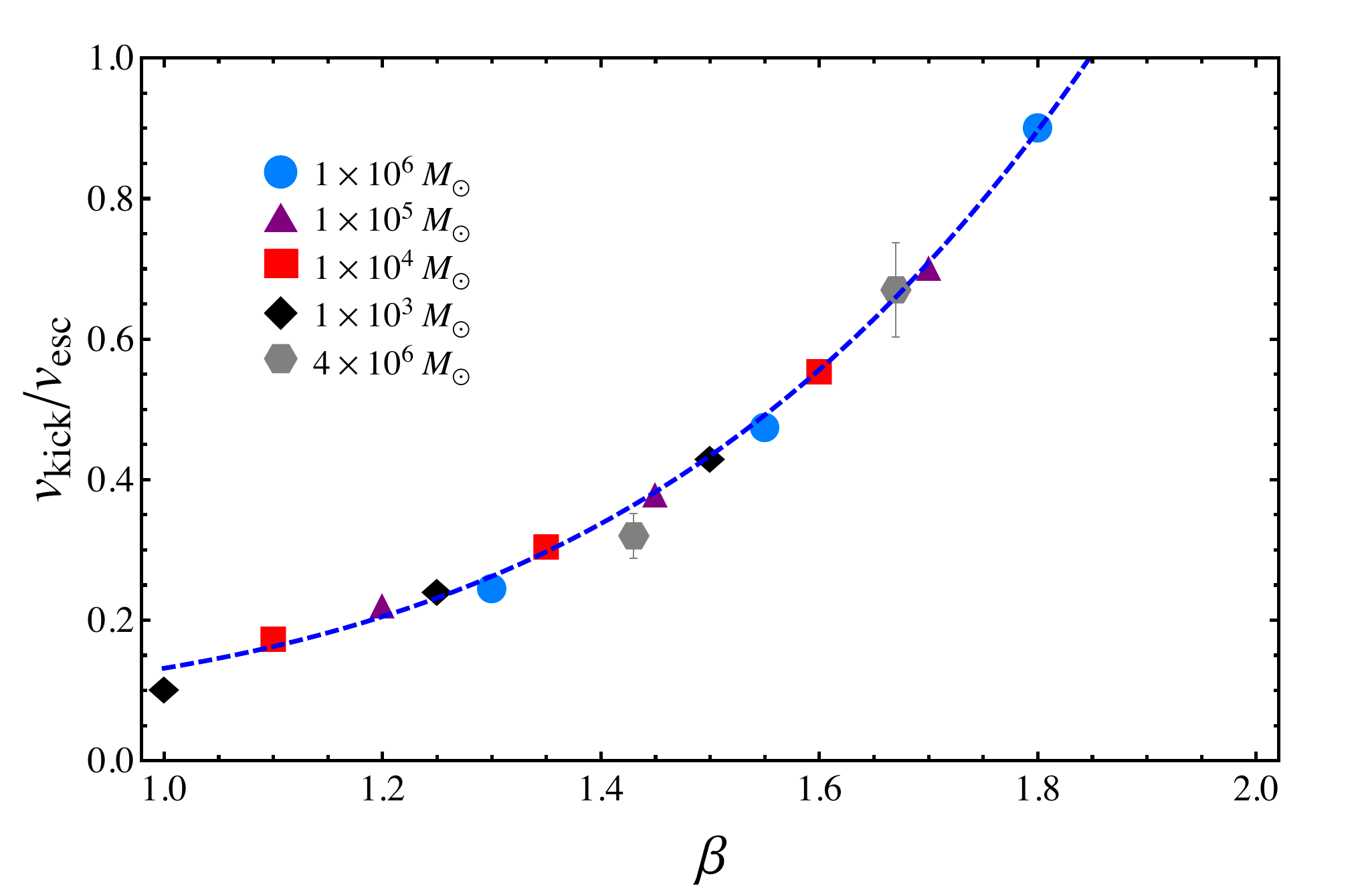}
\caption{The kick velocity $v_{\rm kick}$ delivered to the surviving core after partial disruption over the escape speed from the core $v_{\rm esc}$ as a function of the penetration factor $\beta$. The blue dashed line is an analytical fit to the simulation results: the kick velocity is found to be dependent on $\beta$ but largely independent of the black hole to stellar mass ratio, probed here in the range $10^{3}-10^{6}$ (see legend). This Figure is from \citep{2013ApJ...771L..28M} -- their figure 3 -- where a complete description of the hydrodynamical simulations that produced these results can be found.}
\label{fig:core_kick}
\end{figure*}

In certain scenarios a star may have multiple encounters with a black hole and thus experience multiple (partial) disruptions. For this to take place in practice, several conditions must be met. On the first passage, the star must be only partially disrupted, and remain in a bound orbit. This implies that any kick imparted to the core by mass-loss asymmetry \citep{2013ApJ...771L..28M} must be small enough that the surviving core stays bound to the black hole (see Figure \ref{fig:core_kick}).  The star cannot be scattered away from its disruptive orbit by dynamical processes like two-body gravitational interactions. Finally, the mass--radius relationship of the star must be such that the star becomes increasingly tidally vulnerable (less dense) following mass loss, rather than more dense. 

An argument long presented in favor of multiple encounters is the possible process of capture into eccentric orbits by tidal dissipation \citep{1975MNRAS.172P..15F,1977ApJ...213..183P,1986ApJ...310..176L}. The specific case of tidal capture by SMBHs has been examined by \citet{1992MNRAS.255..276N,1992MNRAS.258..715K,1995MNRAS.275..498D} in the context of the affine model. \citet{2006MNRAS.372..467B} analyzed the related problem of tidal captures by intermediate-mass black holes in N-body dynamical models. A key point in these analyses -- the spatial distribution of tidal dissipation within the star -- was raised by \citet{1996MNRAS.279.1104P}: the supplied energy can be quickly radiated away if deposited in the stellar atmosphere, while it does work and causes stellar expansion if deposited in deeper layers. In this latter scenario, the star is increasingly susceptible to tidal interaction. These details of tidal dissipation remain highly uncertain areas of active discussion, particularly in the context of possible tidal captures of Hot Jupiters by their host stars \citep[e.g.][]{2018AJ....155..118W}. 

In the limit when $R_{\rm p}$ remains too large for even partial mass loss to occur, \citet{2003ApJ...590L..25A} predict the existence of ``squeezars'': stars on highly eccentric orbits around a SMBH, with luminosities approaching the star's Eddington limit, that are powered by tidal interactions with the black hole. These stars undergo orbital decay from tidal heating, and the inspiral terminates in a tidal disruption once the star's orbital energy exceeds its own binding energy.  Squeezar formation is estimated to occur at $\sim 5\%$ of the TDE rate, but with lifetimes of $\sim10^5$ yr, a mean number of 0.1--1 squeezars are likely orbiting our Galactic Center SMBH.

While numerous uncertainties remain, there are several situations in which conditions for repeated tidal-stripping encounters might be met. MS stars, especially low-mass stars with isentropic structures, expand upon losing mass. If these are in a sufficiently bound orbit that they are unlikely to be scattered by gravitational interactions, the star will expand following each tidal mass loss episode and will undergo runaway disruption over several orbits. \citet{2011ApJ...732...74G} discuss and simulate a very similar scenario with particular focus on the slightly different science case of giant planets scattered close to their host stars. 

 Because giant stars are never completely disrupted in a single passage (see e.g. \S \ref{sec:giant}), they have the opportunity to return for subsequent encounters, making them strong candidates for multiple disruptive passages. Furthermore, because the loss cone angular momenta of giant stars are larger than those of their more-compact companions (for a given semi-major axis), these objects experience less scatter in $\beta$ due to two-body relaxation over the course of an orbit. This implies that they are more likely to return for a subsequent passage. The extreme limit of this occurs when the orbit of a giant star is evolving so slowly that their radial growth (from stellar evolution) controls the time evolution of $R_{\rm p}/R_{\rm t}(t)$ \citep{SyerUlmer98,MacLeod+13,Merritt13}.  Through hydrodynamic simulations and stellar evolution modeling, \citet{MacLeod+13} find that giant stars on such orbits around a SMBH produce low-level flares that repeat on the orbital timescale. This gradual, piece-by-piece feeding of the SMBH is comparable to the feeding rate provided by stellar winds from all stars in the surrounding nuclear star cluster.
We further note that the dynamics of these encounters are the high-mass ratio limit of the more-thoroughly considered scenario of Roche lobe overflow in eccentric stellar binaries \citep[e.g.][]{MateseWhitmire83,HamersDosopoulou19}.

Finally, WDs may likewise undergo multiple passages while in an eccentric orbit \citep{2010MNRAS.409L..25Z,2014ApJ...794....9M}. These systems can be captured or decay through a combination of gravitational radiation and tidal interaction \citep{2017MNRAS.468.2296V}. A more complete discussion of the possible phenomenology and implications for multi-messenger detection is given in the \wdchap. 

\begin{figure}
  \includegraphics[width=\linewidth]{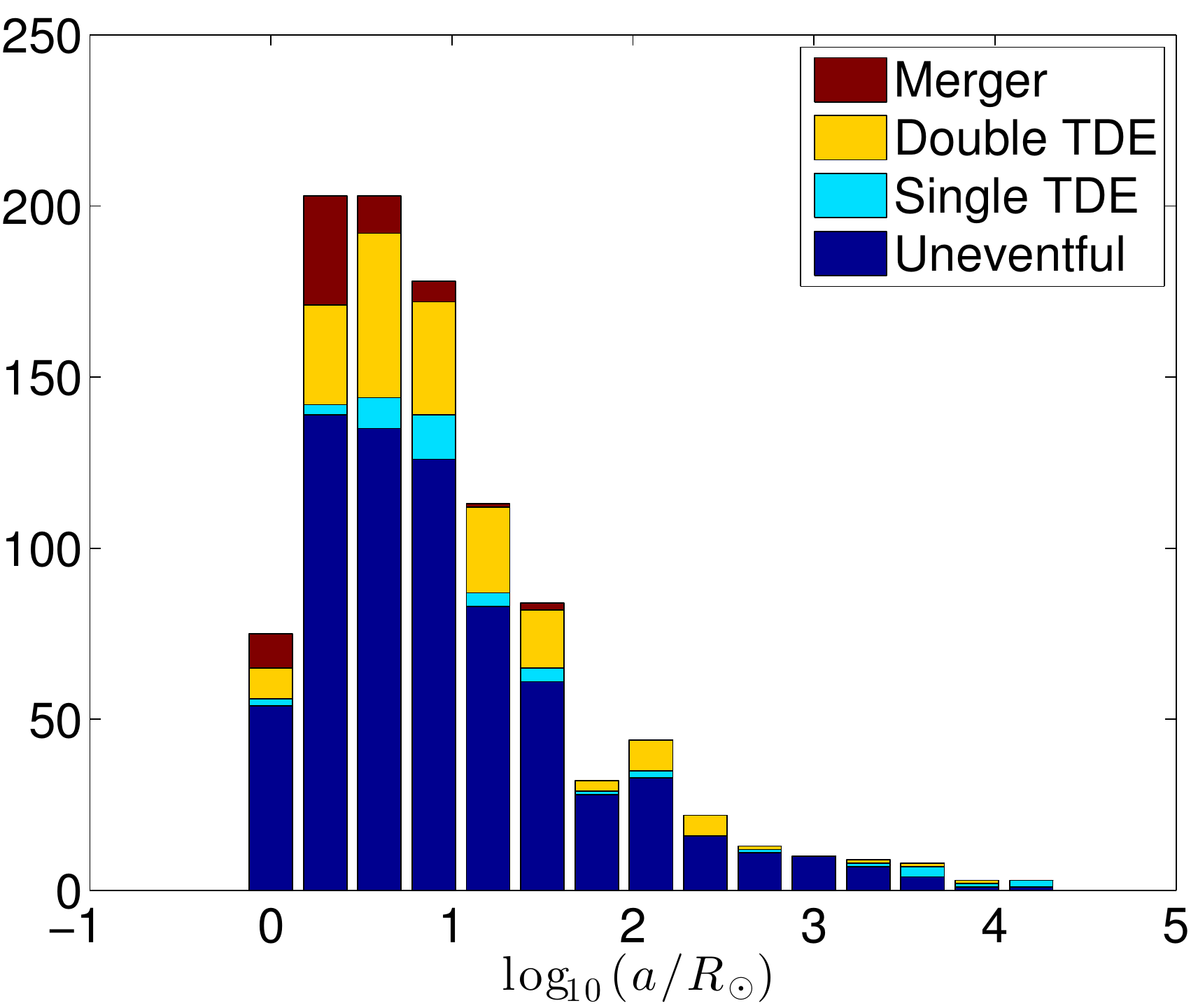}
  \caption{\label{fig:niaryfate} The distribution -- as a function of the binary semi-major axis $a$ -- of the outcomes of 1000 numerical simulations of a stellar binary encounter with an SMBH performed by \cite{mandel&levin15} in the full loss cone regime.  From the top down, the outcomes are: a merger of the binary components, a double tidal disruption, a single tidal disruption, and an uneventful fly-by. Tidal disruptions of both stars become prominent for initial binary separations smaller than $\sim 100 R_{\odot}$, while mergers are common only for binary separations smaller than $\sim 10 R_\odot$. Overall among these 1000 trials, 18\% produced sequential tidal disruptions of both stars, 5\% resulted in single TDEs with typically the more massive star being disrupted, and binary components merged in 6\% of simulations before either was disrupted. This is figure 2, lower panel in \cite{mandel&levin15}.} 
\end{figure}

\subsection{Stellar binary TDEs}
\label{sec:bianries}

When two stars in a binary of semimajor axis $a$ and total mass $M_{\rm b}$ approach a massive black hole to within the tidal separation radius\footnote{Note that this is identical to equation (\ref{eq:rt}), but with $R_\star$ and $M_*$ replaced by $a$ and $M_{\rm b}$ respectively, so that binary tidal separation occurs when the binary is at least a few stellar tidal disruption radii away from the SMBH.} 
\begin{equation}
    R_{\rm ts} \approx a \left( \frac{M_{\rm BH}}{M_{\rm b}} \right)^{1/3}\ , \label{eq:rts}
\end{equation}
 several outcomes are possible: (i) both stars may be sequentially disrupted by the black hole; (ii) the stars may merge, with the merger product possibly being tidally disrupted by the black hole; (iii) one star may be gravitationally captured by the black hole (possibly to be subsequently tidally disrupted) with the other ejected as a hypervelocity star \citep{mandel&levin15,bradnick+17,bonnerot&rossi19}.  All three outcomes are potentially informative, with unique signatures and opportunities to extract information about nuclear cluster dynamics, tidal disruption events, and stellar evolution.

The loss cone for tidal separation is characterised by a specific angular momentum of
\begin{equation}
  l_{\rm ts} \approx \sqrt{2G M_{\rm BH} a} \left(\frac{M_{\rm BH}}{M_{\rm b}} \right)^{1/6}\, .
  \label{eq:llc}
\end{equation}
The stellar binary will change its orbital angular momentum over time due to relaxational interactions far from the black hole. If the binary's typical per-orbit change in angular momentum $\Delta l$ is large, i.e. $\Delta l \gtrsim l_{\rm lc}$, the binary is in the full loss cone regime, meaning that the entire space of possible angular momenta is sampled.  In this case, it is possible that the binary plunges toward the black hole on a trajectory with a small impact parameter, coming within the companion stars' individual tidal disruption radii.  This leads to a sequential tidal disruption of two stars, outcome (i), which could produce an unusual double-peaked light-curve or at least enhance interactions between tidal streams \citep{mandel&levin15}.  In fact, \citet{bonnerot&rossi19} analytically estimate that tidal streams from the two stellar disruptions following the tidal separation of a binary will collide in nearly half of such encounters.  The burst of radiation associated with shock heating from these stream collisions may yield a detectable precursor to the TDE light curve \citep{bonnerot&rossi19}.

Full tidal separation loss cone encounters occasionally lead to outcome (ii), stellar mergers  \citep{mandel&levin15}.  Figure \ref{fig:niaryfate} shows quantitatively that smaller initial binary separations are more conducive to mergers and double TDEs.  Mergers are even more likely as outcomes of empty loss cone encounters, in which the angular momentum is only gradually perturbed during each orbit, $\Delta l \ll l_{\rm lc}$.  In this case, the binary passes by the massive black hole on a very eccentric orbit, and is tidally perturbed over many periapsis passages with $R_{\rm p} \sim {\rm few} \times R_{\rm ts}$.  Such encounters barely affect the semi-major axis (energy) of the stellar binary, but the exchange of angular momentum between the stellar binary and its orbit around the black hole can significantly drive up the binary's eccentricity \citep{bradnick+17}.  Stellar tides can then decrease the internal binary separation while the binary is far from the black hole.  A sequence of eccentricity excitations and tidal damping further increases the likelihood of merger as a possible outcome, particularly for initially close binaries \citep{bradnick+17}.   The likely outcome of an interaction is sensitive to the initial properties of the binary (masses and separation) and the efficiency of binary tides in circularising the orbit. Meanwhile, a key environmental factor is the efficiency of relaxation of the orbit of the binary around the massive black hole.  

Some of the merger products may subsequently approach the massive black hole closely enough to be tidally disrupted.  A recent merger remnant could have a much larger magnetic field than would be typical for a single star; the subsequent disruption of such a magnetised object may make it easier to produce jets associated with some TDEs such as Swift J164449.3+573451 \citep{mandel&levin15,bradnick+17}.

Meanwhile, the tidal separation of a binary, outcome (iii), generally leads to the rapid ejection of one of the companions.  By analogy with Eq.~(\ref{eq:vej}), the velocity of the ejected star with mass $M_\star$ is of order
\begin{equation}
v_{\rm ej} \sim \left(\frac{M_{\rm BH}}{M_\star}\right)^{1/6} v_{\rm bin}\ ,
\label{eq:vejbin}
\end{equation}
where $v_{\rm bin}$ is the binary's orbital velocity, of order $100$ km s$^{-1}$ for a compact binary (two solar-type stars at ten solar radii) and the two stars are assumed to have a similar mass.  For a $10^6 M_\odot$ black hole, the ejection velocity is then of order $v_{\rm ej,\ hvs} \sim 1000$  km s$^{-1}$: a hypervelocity star.  \citet{Rossi:2014,Rossi:2017} found that the observed population of hypervelocity stars is missing a high-velocity tail expected from models.  \citet{bradnick+17} conjectured that mergers of the the most compact, high orbital velocity binaries could be responsible for the observed dearth of particularly rapid ejections.  

\section{Conclusions}
\label{sec:conclusions}
The process of stellar tidal disruption has been studied since the pioneering work of \citet{Hills75}.  In the almost half-century of theoretical work since then, our understanding of the central features of the disruption process has largely converged.  The position-dependent tidal field felt by an extended object can be written in closed form, both in Newtonian gravity and also in the general relativistic gravity of the Schwarzschild and Kerr metrics (Eqs. \ref{eq:tidalTensorN} and \ref{eq:tidalTensorGR}).  These tidal fields can then be used to build analytic, semi-analytic, or fully numerical models for the dynamics of tidal disruption. Because the full process of disruption is a nonlinear hydrodynamics problem, a numerical simulation -- if/when adequate resolution can be achieved -- provides the most precise answers, but a variety of simpler models provide useful physical intuition and some degree of accuracy.

We have reviewed these models in \S \ref{sec:dyn_basics}.  The simplest ones can be constructed by assuming impulsive disruption of the star at the moment it enters the tidal radius, with the star's fluid elements ``freezing in'' to ballistic, or geodesic, trajectories in the aftermath of this disruption; this approximation can be used to make predictions about the mass fallback rate in a fully analytic way. Increasing accuracy comes from applying the free solutions to the parabolic Hill equations (Eq. \ref{eq:freeEnergy}), which also allows limited study of the process of tidal compression; or from adding an integral equation to account for the internal structure of the star (Eq. \ref{eq:salami}). The semi-analytic ``affine model'' provides significantly greater physical realism by using the tensor virial theorem (and simplifying geometrical assumptions) to model the hydrodynamics of the disruption process, rather than neglecting hydrodynamics entirely as in the impulse approximation. The affine model has been used extensively to study the tidal compression suffered by stars undergoing high-$\beta$ TDEs, and makes surprisingly accurate predictions for mass lost in partial tidal disruptions.

We have also surveyed the extensive literature of numerical (magneto)-hydrodynamics simulations of TDEs, the results of which are presented in \S \ref{subsec:hydro}, \S \ref{sec:partial_TDE}, and \S \ref{sec:stellar_types}.  At this point, the tidal disruption of non-rotating polytropic stars in Newtonian gravity is a largely solved problem: at least two large parameter studies employing different numerical techniques \citep{Guillochon13, Mainetti+17} are converged on the outcomes of both full and partial disruptions under these assumptions.  However, there are many active frontiers of numerical disruption simulations that have not yet fully explored the parameter space of real TDEs.  While tidal disruption in general relativistic gravity has been simulated since the work of \citet{Diener+97}, broad parameter surveys have only recently appeared \citep{Ryu+20d}, and comparatively few simulations have been done in the Kerr (rather than Schwarzschild) metric. A full understanding of the energetics and dynamics of a star disrupted on an inclined trajectory with respect to the SMBH's spin is for instance missing. Likewise, it is only very recently that a number of groups  \citep{Golightly+2019b, Law-Smith+2019, Goicovic19, Ryu+20a} have begun simulating the disruption of stars generated with fully realistic internal structure (usually employing \texttt{MESA} models). Pursuing these studies further  would improve our understanding of how detectable spectral lines -- their strengths and temporal behaviour -- can be used to explore the disrupted star and gas thermodynamics during the time between disruption and spectral line observations. Likewise, recent -- and few -- investigations of the disruption of rotating \citep{Golightly19, Sacchi19} or magnetised \citep{Guillochon17,Bonnerot17} stars offer leads for more in depth analysis. The rapid progress of the last two years seems promising for our future understanding of astrophysically realistic TDEs.

One of the original motivations for studying TDEs relates to the exotic phenomena that may occur during extreme tidal compression in high-$\beta$ events, as we discuss in \S \ref{sec:high_beta}.  The onset of runaway nuclear fusion (particularly in WD TDEs), the emission of gravitational waves, or the production of a bright shock breakout signal remain somewhat speculative possibilities, largely due to the computational challenge of resolving severe compression in high-$\beta$ TDEs.  If signals like these exist, however, they would pin down the precise moment of disruption, which might help to resolve open questions related to accretion flow formation. Looking into the future, observation of GW signals are particularly promising, but little work has been done so far in exploring the scientific gain from the synergy between gravitational-wave and electromagnetic observations.

We further discuss in \S \ref{sec:unbound} the dynamics and possible signatures of the dynamically unbound half of the star; the current assessment is that its observability remains elusive. Even if more exotic scenarios discussed in \S \ref{sec:variations}, such as repeated disruptions, binary disruptions, $e<1$ TDEs, or quasi-circular Roche-lobe overflow are intrinsically rare or hard to observe, it seems plausible that the flood of data from upcoming time domain surveys may include some of these variations on more typical TDE flares. 

\begin{acknowledgements}
IM is a recipient of the Australian Research Council Future Fellowship FT190100574.
\end{acknowledgements}

\bibliography{biblio}

\begin{thebibliography}{146}
\ifx \bisbn   \undefined \def \bisbn  #1{ISBN #1}\fi
\ifx \binits  \undefined \def \binits#1{#1} \fi
\ifx \bauthor  \undefined \def \bauthor#1{#1} \fi
\ifx \bjtitle  \undefined \def \bjtitle#1{\textrm{#1}}\fi
\ifx \batitle  \undefined \def \batitle#1{#1} \fi
\ifx \bctitle  \undefined \def \bctitle#1{#1} \fi
\ifx \bvolume  \undefined \def \bvolume#1{\textbf{#1}}\fi
\ifx \byear  \undefined \def \byear#1{#1} \fi
\ifx \bissue  \undefined \def \bissue#1{#1} \fi
\ifx \bfpage  \undefined \def \bfpage#1{#1} \fi
\ifx \blpage  \undefined \def \blpage #1{#1} \fi
\ifx \burl  \undefined \def \burl#1{#1} \fi
\ifx \doiurl  \undefined \def \doiurl#1{#1} \fi
\ifx \betal  \undefined \def \betal{et al.} \fi
\ifx \binstitute  \undefined \def \binstitute#1{#1} \fi
\ifx \beditor  \undefined \def \beditor#1{#1} \fi
\ifx \bpublisher  \undefined \def \bpublisher#1{#1} \fi
\ifx \bbtitle  \undefined \def \bbtitle#1{\textit{#1}} \fi
\ifx \bedition  \undefined \def \bedition#1{#1} \fi
\ifx \bseriesno  \undefined \def \bseriesno#1{#1} \fi
\ifx \blocation  \undefined \def \blocation#1{#1} \fi
\ifx \bsertitle  \undefined \def \bsertitle#1{#1} \fi
\ifx \bsnm \undefined \def \bsnm#1{#1} \fi
\ifx \bsuffix \undefined \def \bsuffix#1{#1} \fi
\ifx \bparticle \undefined \def \bparticle#1{#1} \fi
\ifx \barticle \undefined \def \barticle#1{#1} \fi
\ifx \botherref \undefined \def \botherref #1{#1} \fi
\ifx \url \undefined \def \url#1{#1} \fi
\ifx \bchapter \undefined \def \bchapter#1{#1} \fi
\ifx \bbook \undefined \def \bbook#1{#1} \fi
\ifx \bcomment \undefined \def \bcomment#1{#1} \fi
\ifx \oauthor \undefined \def \oauthor#1{#1} \fi
\ifx \citeauthoryear \undefined \def \citeauthoryear#1{#1} \fi
\ifx \texttildelow  \undefined \def \texttildelow{\symbol{126}} \fi
\def \endbibitem {}
\ifx \bconflocation  \undefined \def \bconflocation#1{#1} \fi

\bibitem[\protect\citeauthoryear{{Alexander} and
  {Morris}}{2003}]{2003ApJ...590L..25A}
\begin{barticle}
\bauthor{\binits{T.} \bsnm{{Alexander}}},
\bauthor{\binits{M.} \bsnm{{Morris}}},
\batitle{{Squeezars: Tidally Powered Stars Orbiting a Massive Black Hole}}.
\bjtitle{\apj}
\bvolume{590}(\bissue{1}),
\bfpage{25}--\blpage{28}
(\byear{2003}).
doi:\doiurl{10.1086/376671}
\end{barticle}
\endbibitem

\bibitem[\protect\citeauthoryear{{Amaro-Seoane} et~al.}{2017}]{Amaro-Seoane+17}
\begin{botherref}
\oauthor{\binits{P.} \bsnm{{Amaro-Seoane}}},
\oauthor{\binits{H.} \bsnm{{Audley}}},
\oauthor{\binits{S.} \bsnm{{Babak}}},
\oauthor{\binits{J.} \bsnm{{Baker}}},
\oauthor{\binits{E.} \bsnm{{Barausse}}},
\oauthor{\binits{P.} \bsnm{{Bender}}},
\oauthor{\binits{E.} \bsnm{{Berti}}},
\oauthor{\binits{P.} \bsnm{{Binetruy}}},
\oauthor{\binits{M.} \bsnm{{Born}}},
\oauthor{\binits{D.} \bsnm{{Bortoluzzi}}},
\oauthor{\binits{J.} \bsnm{{Camp}}},
\oauthor{\binits{C.} \bsnm{{Caprini}}},
\oauthor{\binits{V.} \bsnm{{Cardoso}}},
\oauthor{\binits{M.} \bsnm{{Colpi}}},
\oauthor{\binits{J.} \bsnm{{Conklin}}},
\oauthor{\binits{N.} \bsnm{{Cornish}}},
\oauthor{\binits{C.} \bsnm{{Cutler}}},
\oauthor{\binits{K.} \bsnm{{Danzmann}}},
\oauthor{\binits{R.} \bsnm{{Dolesi}}},
\oauthor{\binits{L.} \bsnm{{Ferraioli}}},
\oauthor{\binits{V.} \bsnm{{Ferroni}}},
\oauthor{\binits{E.} \bsnm{{Fitzsimons}}},
\oauthor{\binits{J.} \bsnm{{Gair}}},
\oauthor{\binits{L.} \bsnm{{Gesa Bote}}},
\oauthor{\binits{D.} \bsnm{{Giardini}}},
\oauthor{\binits{F.} \bsnm{{Gibert}}},
\oauthor{\binits{C.} \bsnm{{Grimani}}},
\oauthor{\binits{H.} \bsnm{{Halloin}}},
\oauthor{\binits{G.} \bsnm{{Heinzel}}},
\oauthor{\binits{T.} \bsnm{{Hertog}}},
\oauthor{\binits{M.} \bsnm{{Hewitson}}},
\oauthor{\binits{K.} \bsnm{{Holley-Bockelmann}}},
\oauthor{\binits{D.} \bsnm{{Hollington}}},
\oauthor{\binits{M.} \bsnm{{Hueller}}},
\oauthor{\binits{H.} \bsnm{{Inchauspe}}},
\oauthor{\binits{P.} \bsnm{{Jetzer}}},
\oauthor{\binits{N.} \bsnm{{Karnesis}}},
\oauthor{\binits{C.} \bsnm{{Killow}}},
\oauthor{\binits{A.} \bsnm{{Klein}}},
\oauthor{\binits{B.} \bsnm{{Klipstein}}},
\oauthor{\binits{N.} \bsnm{{Korsakova}}},
\oauthor{\binits{S.L.} \bsnm{{Larson}}},
\oauthor{\binits{J.} \bsnm{{Livas}}},
\oauthor{\binits{I.} \bsnm{{Lloro}}},
\oauthor{\binits{N.} \bsnm{{Man}}},
\oauthor{\binits{D.} \bsnm{{Mance}}},
\oauthor{\binits{J.} \bsnm{{Martino}}},
\oauthor{\binits{I.} \bsnm{{Mateos}}},
\oauthor{\binits{K.} \bsnm{{McKenzie}}},
\oauthor{\binits{S.T.} \bsnm{{McWilliams}}},
\oauthor{\binits{C.} \bsnm{{Miller}}},
\oauthor{\binits{G.} \bsnm{{Mueller}}},
\oauthor{\binits{G.} \bsnm{{Nardini}}},
\oauthor{\binits{G.} \bsnm{{Nelemans}}},
\oauthor{\binits{M.} \bsnm{{Nofrarias}}},
\oauthor{\binits{A.} \bsnm{{Petiteau}}},
\oauthor{\binits{P.} \bsnm{{Pivato}}},
\oauthor{\binits{E.} \bsnm{{Plagnol}}},
\oauthor{\binits{E.} \bsnm{{Porter}}},
\oauthor{\binits{J.} \bsnm{{Reiche}}},
\oauthor{\binits{D.} \bsnm{{Robertson}}},
\oauthor{\binits{N.} \bsnm{{Robertson}}},
\oauthor{\binits{E.} \bsnm{{Rossi}}},
\oauthor{\binits{G.} \bsnm{{Russano}}},
\oauthor{\binits{B.} \bsnm{{Schutz}}},
\oauthor{\binits{A.} \bsnm{{Sesana}}},
\oauthor{\binits{D.} \bsnm{{Shoemaker}}},
\oauthor{\binits{J.} \bsnm{{Slutsky}}},
\oauthor{\binits{C.F.} \bsnm{{Sopuerta}}},
\oauthor{\binits{T.} \bsnm{{Sumner}}},
\oauthor{\binits{N.} \bsnm{{Tamanini}}},
\oauthor{\binits{I.} \bsnm{{Thorpe}}},
\oauthor{\binits{M.} \bsnm{{Troebs}}},
\oauthor{\binits{M.} \bsnm{{Vallisneri}}},
\oauthor{\binits{A.} \bsnm{{Vecchio}}},
\oauthor{\binits{D.} \bsnm{{Vetrugno}}},
\oauthor{\binits{S.} \bsnm{{Vitale}}},
\oauthor{\binits{M.} \bsnm{{Volonteri}}},
\oauthor{\binits{G.} \bsnm{{Wanner}}},
\oauthor{\binits{H.} \bsnm{{Ward}}},
\oauthor{\binits{P.} \bsnm{{Wass}}},
\oauthor{\binits{W.} \bsnm{{Weber}}},
\oauthor{\binits{J.} \bsnm{{Ziemer}}},
\oauthor{\binits{P.} \bsnm{{Zweifel}}},
{Laser Interferometer Space Antenna}.
arXiv e-prints
(2017)
\end{botherref}
\endbibitem

\bibitem[\protect\citeauthoryear{{Amaro-Seoane} et~al.}{2012}]{Amaro12}
\begin{barticle}
\bauthor{\binits{P.} \bsnm{{Amaro-Seoane}}},
\bauthor{\binits{M.C.} \bsnm{{Miller}}},
\bauthor{\binits{G.F.} \bsnm{{Kennedy}}},
\batitle{{Tidal disruptions of separated binaries in galactic nuclei}}.
\bjtitle{\mnras}
\bvolume{425}(\bissue{4}),
\bfpage{2401}--\blpage{2406}
(\byear{2012}).
doi:\doiurl{10.1111/j.1365-2966.2012.21162.x}
\end{barticle}
\endbibitem

\bibitem[\protect\citeauthoryear{{Anninos} et~al.}{2018}]{Anninos+18}
\begin{barticle}
\bauthor{\binits{P.} \bsnm{{Anninos}}},
\bauthor{\binits{P.C.} \bsnm{{Fragile}}},
\bauthor{\binits{S.S.} \bsnm{{Olivier}}},
\bauthor{\binits{R.} \bsnm{{Hoffman}}},
\bauthor{\binits{B.} \bsnm{{Mishra}}},
\bauthor{\binits{K.} \bsnm{{Camarda}}},
\batitle{{Relativistic Tidal Disruption and Nuclear Ignition of White Dwarf
  Stars by Intermediate-mass Black Holes}}.
\bjtitle{\apj}
\bvolume{865}(\bissue{1}),
\bfpage{3}
(\byear{2018}).
doi:\doiurl{10.3847/1538-4357/aadad9}
\end{barticle}
\endbibitem

\bibitem[\protect\citeauthoryear{{Baker} et~al.}{2019}]{baker+19}
\begin{botherref}
\oauthor{\binits{J.} \bsnm{{Baker}}},
\oauthor{\binits{S.F.} \bsnm{{Barke}}},
\oauthor{\binits{P.L.} \bsnm{{Bender}}},
\oauthor{\binits{E.} \bsnm{{Berti}}},
\oauthor{\binits{R.} \bsnm{{Caldwell}}},
\oauthor{\binits{J.W.} \bsnm{{Conklin}}},
\oauthor{\binits{N.} \bsnm{{Cornish}}},
\oauthor{\binits{E.C.} \bsnm{{Ferrara}}},
\oauthor{\binits{K.} \bsnm{{Holley-Bockelmann}}},
\oauthor{\binits{B.} \bsnm{{Kamai}}},
\oauthor{\binits{S.L.} \bsnm{{Larson}}},
\oauthor{\binits{J.} \bsnm{{Livas}}},
\oauthor{\binits{S.T.} \bsnm{{McWilliams}}},
\oauthor{\binits{G.} \bsnm{{Mueller}}},
\oauthor{\binits{P.} \bsnm{{Natarajan}}},
\oauthor{\binits{N.} \bsnm{{Rioux}}},
\oauthor{\binits{S.R.} \bsnm{{Sankar}}},
\oauthor{\binits{J.} \bsnm{{Schnittman}}},
\oauthor{\binits{D.} \bsnm{{Shoemaker}}},
\oauthor{\binits{J.} \bsnm{{Slutsky}}},
\oauthor{\binits{R.} \bsnm{{Stebbins}}},
\oauthor{\binits{I.} \bsnm{{Thorpe}}},
\oauthor{\binits{J.} \bsnm{{Ziemer}}},
Space based gravitational wave astronomy beyond lisa.
ArXiv e-prints
(2019)
\end{botherref}
\endbibitem

\bibitem[\protect\citeauthoryear{{Baumgardt}
  et~al.}{2006}]{2006MNRAS.372..467B}
\begin{barticle}
\bauthor{\binits{H.} \bsnm{{Baumgardt}}},
\bauthor{\binits{C.} \bsnm{{Hopman}}},
\bauthor{\binits{S.} \bsnm{{Portegies Zwart}}},
\bauthor{\binits{J.} \bsnm{{Makino}}},
\batitle{{Tidal capture of stars by intermediate-mass black holes}}.
\bjtitle{\mnras}
\bvolume{372}(\bissue{1}),
\bfpage{467}--\blpage{478}
(\byear{2006}).
doi:\doiurl{10.1111/j.1365-2966.2006.10885.x}
\end{barticle}
\endbibitem

\bibitem[\protect\citeauthoryear{{Beloborodov} et~al.}{1992}]{Beloborodov+92}
\begin{barticle}
\bauthor{\binits{A.M.} \bsnm{{Beloborodov}}},
\bauthor{\binits{A.F.} \bsnm{{Illarionov}}},
\bauthor{\binits{P.B.} \bsnm{{Ivanov}}},
\bauthor{\binits{A.G.} \bsnm{{Polnarev}}},
\batitle{{Angular momentum of a supermassive black hole in a dense star
  cluster}}.
\bjtitle{\mnras}
\bvolume{259},
\bfpage{209}--\blpage{217}
(\byear{1992}).
doi:\doiurl{10.1093/mnras/259.2.209}
\end{barticle}
\endbibitem

\bibitem[\protect\citeauthoryear{{Bogdanovi{\'c}}
  et~al.}{2014}]{2014ApJ...788...99B}
\begin{barticle}
\bauthor{\binits{T.} \bsnm{{Bogdanovi{\'c}}}},
\bauthor{\binits{R.M.} \bsnm{{Cheng}}},
\bauthor{\binits{P.} \bsnm{{Amaro-Seoane}}},
\batitle{{Disruption of a Red Giant Star by a Supermassive Black Hole and the
  Case of PS1-10jh}}.
\bjtitle{\apj}
\bvolume{788}(\bissue{2}),
\bfpage{99}
(\byear{2014}).
doi:\doiurl{10.1088/0004-637X/788/2/99}
\end{barticle}
\endbibitem

\bibitem[\protect\citeauthoryear{{Bonnerot} and
  {Rossi}}{2019}]{bonnerot&rossi19}
\begin{barticle}
\bauthor{\binits{C.} \bsnm{{Bonnerot}}},
\bauthor{\binits{E.M.} \bsnm{{Rossi}}},
\batitle{{Streams collision as possible precursor of double tidal disruption
  events}}.
\bjtitle{\mnras}
\bvolume{484},
\bfpage{1301}--\blpage{1316}
(\byear{2019}).
doi:\doiurl{10.1093/mnras/stz062}
\end{barticle}
\endbibitem

\bibitem[\protect\citeauthoryear{{Bonnerot} et~al.}{2016a}]{Bonnerot+16b}
\begin{barticle}
\bauthor{\binits{C.} \bsnm{{Bonnerot}}},
\bauthor{\binits{E.M.} \bsnm{{Rossi}}},
\bauthor{\binits{G.} \bsnm{{Lodato}}},
\batitle{{Bad prospects for the detection of giant stars' tidal disruption:
  effect of the ambient medium on bound debris}}.
\bjtitle{\mnras}
\bvolume{458}(\bissue{3}),
\bfpage{3324}--\blpage{3330}
(\byear{2016}a).
doi:\doiurl{10.1093/mnras/stw486}
\end{barticle}
\endbibitem

\bibitem[\protect\citeauthoryear{{Bonnerot} et~al.}{2016b}]{Bonnerot16}
\begin{barticle}
\bauthor{\binits{C.} \bsnm{{Bonnerot}}},
\bauthor{\binits{E.M.} \bsnm{{Rossi}}},
\bauthor{\binits{G.} \bsnm{{Lodato}}},
\bauthor{\binits{D.J.} \bsnm{{Price}}},
\batitle{{Disc formation from tidal disruptions of stars on eccentric orbits by
  Schwarzschild black holes}}.
\bjtitle{\mnras}
\bvolume{455}(\bissue{2}),
\bfpage{2253}--\blpage{2266}
(\byear{2016}b).
doi:\doiurl{10.1093/mnras/stv2411}
\end{barticle}
\endbibitem

\bibitem[\protect\citeauthoryear{{Bonnerot} et~al.}{2017}]{Bonnerot17}
\begin{barticle}
\bauthor{\binits{C.} \bsnm{{Bonnerot}}},
\bauthor{\binits{D.J.} \bsnm{{Price}}},
\bauthor{\binits{G.} \bsnm{{Lodato}}},
\bauthor{\binits{E.M.} \bsnm{{Rossi}}},
\batitle{{Magnetic field evolution in tidal disruption events}}.
\bjtitle{\mnras}
\bvolume{469}(\bissue{4}),
\bfpage{4879}--\blpage{4888}
(\byear{2017}).
doi:\doiurl{10.1093/mnras/stx1210}
\end{barticle}
\endbibitem

\bibitem[\protect\citeauthoryear{{Bradnick} et~al.}{2017}]{bradnick+17}
\begin{barticle}
\bauthor{\binits{B.} \bsnm{{Bradnick}}},
\bauthor{\binits{I.} \bsnm{{Mandel}}},
\bauthor{\binits{Y.} \bsnm{{Levin}}},
\batitle{{Stellar binaries in galactic nuclei: tidally stimulated mergers
  followed by tidal disruptions}}.
\bjtitle{\mnras}
\bvolume{469},
\bfpage{2042}--\blpage{2048}
(\byear{2017}).
doi:\doiurl{10.1093/mnras/stx1007}
\end{barticle}
\endbibitem

\bibitem[\protect\citeauthoryear{{Brassart} and
  {Luminet}}{2008}]{Brassart&Luminet08}
\begin{barticle}
\bauthor{\binits{M.} \bsnm{{Brassart}}},
\bauthor{\binits{J.-P.} \bsnm{{Luminet}}},
\batitle{{Shock waves in tidally compressed stars by massive black holes}}.
\bjtitle{\aap}
\bvolume{481}(\bissue{2}),
\bfpage{259}--\blpage{277}
(\byear{2008}).
doi:\doiurl{10.1051/0004-6361:20078264}
\end{barticle}
\endbibitem

\bibitem[\protect\citeauthoryear{{Brassart} and
  {Luminet}}{2010}]{Brassart&Luminet10}
\begin{barticle}
\bauthor{\binits{M.} \bsnm{{Brassart}}},
\bauthor{\binits{J.-P.} \bsnm{{Luminet}}},
\batitle{{Relativistic tidal compressions of a star by a massive black hole}}.
\bjtitle{\aap}
\bvolume{511},
\bfpage{80}
(\byear{2010}).
doi:\doiurl{10.1051/0004-6361/200913442}
\end{barticle}
\endbibitem

\bibitem[\protect\citeauthoryear{{Carter} and
  {Luminet}}{1982}]{Carter&Luminet82}
\begin{barticle}
\bauthor{\binits{B.} \bsnm{{Carter}}},
\bauthor{\binits{J.P.} \bsnm{{Luminet}}},
\batitle{{Pancake detonation of stars by black holes in galactic nuclei}}.
\bjtitle{\nat}
\bvolume{296}(\bissue{5854}),
\bfpage{211}--\blpage{214}
(\byear{1982}).
doi:\doiurl{10.1038/296211a0}
\end{barticle}
\endbibitem

\bibitem[\protect\citeauthoryear{{Carter} and
  {Luminet}}{1983}]{Carter&Luminet83}
\begin{barticle}
\bauthor{\binits{B.} \bsnm{{Carter}}},
\bauthor{\binits{J.-P.} \bsnm{{Luminet}}},
\batitle{{Tidal compression of a star by a large black hole. I Mechanical
  evolution and nuclear energy release by proton capture}}.
\bjtitle{\aap}
\bvolume{121},
\bfpage{97}--\blpage{113}
(\byear{1983})
\end{barticle}
\endbibitem

\bibitem[\protect\citeauthoryear{{Carter} and
  {Luminet}}{1985}]{Carter&Luminet85}
\begin{barticle}
\bauthor{\binits{B.} \bsnm{{Carter}}},
\bauthor{\binits{J.P.} \bsnm{{Luminet}}},
\batitle{{Mechanics of the affine star model}}.
\bjtitle{\mnras}
\bvolume{212},
\bfpage{23}--\blpage{55}
(\byear{1985}).
doi:\doiurl{10.1093/mnras/212.1.23}
\end{barticle}
\endbibitem

\bibitem[\protect\citeauthoryear{{Chen} et~al.}{2011}]{Chen+11}
\begin{barticle}
\bauthor{\binits{X.} \bsnm{{Chen}}},
\bauthor{\binits{A.} \bsnm{{Sesana}}},
\bauthor{\binits{P.} \bsnm{{Madau}}},
\bauthor{\binits{F.K.} \bsnm{{Liu}}},
\batitle{{Tidal Stellar Disruptions by Massive Black Hole Pairs. II. Decaying
  Binaries}}.
\bjtitle{\apj}
\bvolume{729},
\bfpage{13}
(\byear{2011}).
doi:\doiurl{10.1088/0004-637X/729/1/13}
\end{barticle}
\endbibitem

\bibitem[\protect\citeauthoryear{{Cheng} and
  {Evans}}{2013}]{2013PhRvD..87j4010C}
\begin{barticle}
\bauthor{\binits{R.M.} \bsnm{{Cheng}}},
\bauthor{\binits{C.R.} \bsnm{{Evans}}},
\batitle{{Relativistic effects in the tidal interaction between a white dwarf
  and a massive black hole in Fermi normal coordinates}}.
\bjtitle{\prd}
\bvolume{87}(\bissue{10}),
\bfpage{104010}
(\byear{2013}).
doi:\doiurl{10.1103/PhysRevD.87.104010}
\end{barticle}
\endbibitem

\bibitem[\protect\citeauthoryear{{Coughlin} and {Armitage}}{2018}]{Coughlin18}
\begin{barticle}
\bauthor{\binits{E.R.} \bsnm{{Coughlin}}},
\bauthor{\binits{P.J.} \bsnm{{Armitage}}},
\batitle{{Tidal disruption by extreme mass ratio binaries and application to
  ASASSN-15lh}}.
\bjtitle{\mnras}
\bvolume{474}(\bissue{3}),
\bfpage{3857}--\blpage{3865}
(\byear{2018}).
doi:\doiurl{10.1093/mnras/stx3039}
\end{barticle}
\endbibitem

\bibitem[\protect\citeauthoryear{{Coughlin} and {Nixon}}{2019}]{Coughlin19}
\begin{barticle}
\bauthor{\binits{E.R.} \bsnm{{Coughlin}}},
\bauthor{\binits{C.J.} \bsnm{{Nixon}}},
\batitle{{Partial Stellar Disruption by a Supermassive Black Hole: Is the Light
  Curve Really Proportional to t $^{-9/4}$?}}
\bjtitle{\apjl}
\bvolume{883}(\bissue{1}),
\bfpage{17}
(\byear{2019}).
doi:\doiurl{10.3847/2041-8213/ab412d}
\end{barticle}
\endbibitem

\bibitem[\protect\citeauthoryear{{Coughlin} and
  {Nixon}}{2015}]{coughlin&nixon15}
\begin{barticle}
\bauthor{\binits{E.R.} \bsnm{{Coughlin}}},
\bauthor{\binits{C.} \bsnm{{Nixon}}},
\batitle{{Variability in Tidal Disruption Events: Gravitationally Unstable
  Streams}}.
\bjtitle{\apj}
\bvolume{808}(\bissue{1}),
\bfpage{11}
(\byear{2015}).
doi:\doiurl{10.1088/2041-8205/808/1/L11}
\end{barticle}
\endbibitem

\bibitem[\protect\citeauthoryear{{Coughlin} et~al.}{2016}]{coughlin+16a}
\begin{barticle}
\bauthor{\binits{E.R.} \bsnm{{Coughlin}}},
\bauthor{\binits{C.} \bsnm{{Nixon}}},
\bauthor{\binits{M.C.} \bsnm{{Begelman}}},
\bauthor{\binits{P.J.} \bsnm{{Armitage}}},
\bauthor{\binits{D.J.} \bsnm{{Price}}},
\batitle{{Post-periapsis pancakes: sustenance for self-gravity in tidal
  disruption events}}.
\bjtitle{\mnras}
\bvolume{455}(\bissue{4}),
\bfpage{3612}--\blpage{3627}
(\byear{2016}).
doi:\doiurl{10.1093/mnras/stv2511}
\end{barticle}
\endbibitem

\bibitem[\protect\citeauthoryear{{Coughlin} et~al.}{2017}]{Coughlin17}
\begin{barticle}
\bauthor{\binits{E.R.} \bsnm{{Coughlin}}},
\bauthor{\binits{P.J.} \bsnm{{Armitage}}},
\bauthor{\binits{C.} \bsnm{{Nixon}}},
\bauthor{\binits{M.C.} \bsnm{{Begelman}}},
\batitle{{Tidal disruption events from supermassive black hole binaries}}.
\bjtitle{\mnras}
\bvolume{465}(\bissue{4}),
\bfpage{3840}--\blpage{3864}
(\byear{2017}).
doi:\doiurl{10.1093/mnras/stw2913}
\end{barticle}
\endbibitem

\bibitem[\protect\citeauthoryear{{Dai} and {Blandford}}{2013}]{Dai13b}
\begin{barticle}
\bauthor{\binits{L.} \bsnm{{Dai}}},
\bauthor{\binits{R.} \bsnm{{Blandford}}},
\batitle{{Roche accretion of stars close to massive black holes}}.
\bjtitle{\mnras}
\bvolume{434}(\bissue{4}),
\bfpage{2948}--\blpage{2960}
(\byear{2013}).
doi:\doiurl{10.1093/mnras/stt1209}
\end{barticle}
\endbibitem

\bibitem[\protect\citeauthoryear{{Dai} et~al.}{2013a}]{Dai13a}
\begin{barticle}
\bauthor{\binits{L.} \bsnm{{Dai}}},
\bauthor{\binits{R.D.} \bsnm{{Blandford}}},
\bauthor{\binits{P.P.} \bsnm{{Eggleton}}},
\batitle{{Adiabatic evolution of mass-losing stars}}.
\bjtitle{\mnras}
\bvolume{434}(\bissue{4}),
\bfpage{2940}--\blpage{2947}
(\byear{2013}a).
doi:\doiurl{10.1093/mnras/stt1208}
\end{barticle}
\endbibitem

\bibitem[\protect\citeauthoryear{{Dai} et~al.}{2013b}]{Dai13}
\begin{barticle}
\bauthor{\binits{L.} \bsnm{{Dai}}},
\bauthor{\binits{A.} \bsnm{{Escala}}},
\bauthor{\binits{P.} \bsnm{{Coppi}}},
\batitle{{The Impact of Bound Stellar Orbits and General Relativity on the
  Temporal Behavior of Tidal Disruption Flares}}.
\bjtitle{\apj}
\bvolume{775}(\bissue{1}),
\bfpage{9}
(\byear{2013}b).
doi:\doiurl{10.1088/2041-8205/775/1/L9}
\end{barticle}
\endbibitem

\bibitem[\protect\citeauthoryear{{Diener} et~al.}{1995}]{1995MNRAS.275..498D}
\begin{barticle}
\bauthor{\binits{P.} \bsnm{{Diener}}},
\bauthor{\binits{A.G.} \bsnm{{Kosovichev}}},
\bauthor{\binits{E.V.} \bsnm{{Kotok}}},
\bauthor{\binits{I.D.} \bsnm{{Novikov}}},
\bauthor{\binits{C.J.} \bsnm{{Pethick}}},
\batitle{{Non-linear effects at tidal capture of stars by a massive black hole
  - II. Compressible affine models and tidal interaction after capture}}.
\bjtitle{\mnras}
\bvolume{275}(\bissue{2}),
\bfpage{498}--\blpage{506}
(\byear{1995}).
doi:\doiurl{10.1093/mnras/275.2.498}
\end{barticle}
\endbibitem

\bibitem[\protect\citeauthoryear{{Diener} et~al.}{1997}]{Diener+97}
\begin{barticle}
\bauthor{\binits{P.} \bsnm{{Diener}}},
\bauthor{\binits{V.P.} \bsnm{{Frolov}}},
\bauthor{\binits{A.M.} \bsnm{{Khokhlov}}},
\bauthor{\binits{I.D.} \bsnm{{Novikov}}},
\bauthor{\binits{C.J.} \bsnm{{Pethick}}},
\batitle{{Relativistic Tidal Interaction of Stars with a Rotating Black Hole}}.
\bjtitle{\apj}
\bvolume{479}(\bissue{1}),
\bfpage{164}--\blpage{178}
(\byear{1997}).
doi:\doiurl{10.1086/303875}
\end{barticle}
\endbibitem

\bibitem[\protect\citeauthoryear{{Eggleton}}{1983}]{Eggleton83}
\begin{barticle}
\bauthor{\binits{P.P.} \bsnm{{Eggleton}}},
\batitle{{Aproximations to the radii of Roche lobes.}}
\bjtitle{\apj}
\bvolume{268},
\bfpage{368}--\blpage{369}
(\byear{1983}).
doi:\doiurl{10.1086/160960}
\end{barticle}
\endbibitem

\bibitem[\protect\citeauthoryear{{Evans} and {Kochanek}}{1989}]{Evans89}
\begin{barticle}
\bauthor{\binits{C.R.} \bsnm{{Evans}}},
\bauthor{\binits{C.S.} \bsnm{{Kochanek}}},
\batitle{{The Tidal Disruption of a Star by a Massive Black Hole}}.
\bjtitle{\apj}
\bvolume{346},
\bfpage{13}
(\byear{1989}).
doi:\doiurl{10.1086/185567}
\end{barticle}
\endbibitem

\bibitem[\protect\citeauthoryear{{Evans} et~al.}{2015}]{Evans15}
\begin{barticle}
\bauthor{\binits{C.} \bsnm{{Evans}}},
\bauthor{\binits{P.} \bsnm{{Laguna}}},
\bauthor{\binits{M.} \bsnm{{Eracleous}}},
\batitle{{Ultra-close Encounters of Stars with Massive Black Holes: Tidal
  Disruption Events with Prompt Hyperaccretion}}.
\bjtitle{\apj}
\bvolume{805}(\bissue{2}),
\bfpage{19}
(\byear{2015}).
doi:\doiurl{10.1088/2041-8205/805/2/L19}
\end{barticle}
\endbibitem

\bibitem[\protect\citeauthoryear{{Fabian} et~al.}{1975}]{1975MNRAS.172P..15F}
\begin{barticle}
\bauthor{\binits{A.C.} \bsnm{{Fabian}}},
\bauthor{\binits{J.E.} \bsnm{{Pringle}}},
\bauthor{\binits{M.J.} \bsnm{{Rees}}},
\batitle{{Tidal capture formation of binary systems and X-ray sources in
  globular clusters.}}
\bjtitle{\mnras}
\bvolume{172},
\bfpage{15}
(\byear{1975}).
doi:\doiurl{10.1093/mnras/172.1.15P}
\end{barticle}
\endbibitem

\bibitem[\protect\citeauthoryear{{Ferrarese} and
  {Merritt}}{2000}]{FerrareseMerritt2000}
\begin{barticle}
\bauthor{\binits{L.} \bsnm{{Ferrarese}}},
\bauthor{\binits{D.} \bsnm{{Merritt}}},
\batitle{{A Fundamental Relation between Supermassive Black Holes and Their
  Host Galaxies}}.
\bjtitle{\apjl}
\bvolume{539}(\bissue{1}),
\bfpage{9}--\blpage{12}
(\byear{2000}).
doi:\doiurl{10.1086/312838}
\end{barticle}
\endbibitem

\bibitem[\protect\citeauthoryear{{Frank} and {Rees}}{1976}]{Frank&Rees76}
\begin{barticle}
\bauthor{\binits{J.} \bsnm{{Frank}}},
\bauthor{\binits{M.J.} \bsnm{{Rees}}},
\batitle{{Effects of massive central black holes on dense stellar systems}}.
\bjtitle{\mnras}
\bvolume{176},
\bfpage{633}--\blpage{647}
(\byear{1976}).
doi:\doiurl{10.1093/mnras/176.3.633}
\end{barticle}
\endbibitem

\bibitem[\protect\citeauthoryear{{Fryxell} et~al.}{2000}]{2000ApJS..131..273F}
\begin{barticle}
\bauthor{\binits{B.} \bsnm{{Fryxell}}},
\bauthor{\binits{K.} \bsnm{{Olson}}},
\bauthor{\binits{P.} \bsnm{{Ricker}}},
\bauthor{\binits{F.X.} \bsnm{{Timmes}}},
\bauthor{\binits{M.} \bsnm{{Zingale}}},
\bauthor{\binits{D.Q.} \bsnm{{Lamb}}},
\bauthor{\binits{P.} \bsnm{{MacNeice}}},
\bauthor{\binits{R.} \bsnm{{Rosner}}},
\bauthor{\binits{J.W.} \bsnm{{Truran}}},
\bauthor{\binits{H.} \bsnm{{Tufo}}},
\batitle{{FLASH: An Adaptive Mesh Hydrodynamics Code for Modeling Astrophysical
  Thermonuclear Flashes}}.
\bjtitle{\apjs}
\bvolume{131}(\bissue{1}),
\bfpage{273}--\blpage{334}
(\byear{2000}).
doi:\doiurl{10.1086/317361}
\end{barticle}
\endbibitem

\bibitem[\protect\citeauthoryear{{Gafton} and {Rosswog}}{2019}]{Gafton19}
\begin{botherref}
\oauthor{\binits{E.} \bsnm{{Gafton}}},
\oauthor{\binits{S.} \bsnm{{Rosswog}}},
{Tidal disruptions by rotating black holes: effects of spin and impact
  parameter}.
\mnras,
1458
(2019).
doi:\doiurl{10.1093/mnras/stz1530}
\end{botherref}
\endbibitem

\bibitem[\protect\citeauthoryear{{Gallegos-Garcia}
  et~al.}{2018}]{2018ApJ...857..109G}
\begin{barticle}
\bauthor{\binits{M.} \bsnm{{Gallegos-Garcia}}},
\bauthor{\binits{J.} \bsnm{{Law-Smith}}},
\bauthor{\binits{E.} \bsnm{{Ramirez-Ruiz}}},
\batitle{{Tidal Disruptions of Main-sequence Stars of Varying Mass and Age:
  Inferences from the Composition of the Fallback Material}}.
\bjtitle{\apj}
\bvolume{857}(\bissue{2}),
\bfpage{109}
(\byear{2018}).
doi:\doiurl{10.3847/1538-4357/aab5b8}
\end{barticle}
\endbibitem

\bibitem[\protect\citeauthoryear{{Georgiev} and
  {B{\"o}ker}}{2014}]{Georgiev&Boker14}
\begin{barticle}
\bauthor{\binits{I.Y.} \bsnm{{Georgiev}}},
\bauthor{\binits{T.} \bsnm{{B{\"o}ker}}},
\batitle{{Nuclear star clusters in 228 spiral galaxies in the HST/WFPC2
  archive: catalogue and comparison to other stellar systems}}.
\bjtitle{\mnras}
\bvolume{441}(\bissue{4}),
\bfpage{3570}--\blpage{3590}
(\byear{2014}).
doi:\doiurl{10.1093/mnras/stu797}
\end{barticle}
\endbibitem

\bibitem[\protect\citeauthoryear{{Goicovic} et~al.}{2019}]{Goicovic19}
\begin{barticle}
\bauthor{\binits{F.G.} \bsnm{{Goicovic}}},
\bauthor{\binits{V.} \bsnm{{Springel}}},
\bauthor{\binits{S.T.} \bsnm{{Ohlmann}}},
\bauthor{\binits{R.} \bsnm{{Pakmor}}},
\batitle{{Hydrodynamical moving-mesh simulations of the tidal disruption of
  stars by supermassive black holes}}.
\bjtitle{\mnras}
\bvolume{487}(\bissue{1}),
\bfpage{981}--\blpage{992}
(\byear{2019}).
doi:\doiurl{10.1093/mnras/stz1368}
\end{barticle}
\endbibitem

\bibitem[\protect\citeauthoryear{{Golightly} et~al.}{2019a}]{Golightly+2019b}
\begin{barticle}
\bauthor{\binits{E.C.A.} \bsnm{{Golightly}}},
\bauthor{\binits{C.J.} \bsnm{{Nixon}}},
\bauthor{\binits{E.R.} \bsnm{{Coughlin}}},
\batitle{{On the Diversity of Fallback Rates from Tidal Disruption Events with
  Accurate Stellar Structure}}.
\bjtitle{\apjl}
\bvolume{882}(\bissue{2}),
\bfpage{26}
(\byear{2019}a).
doi:\doiurl{10.3847/2041-8213/ab380d}
\end{barticle}
\endbibitem

\bibitem[\protect\citeauthoryear{{Golightly} et~al.}{2019b}]{Golightly19}
\begin{barticle}
\bauthor{\binits{E.C.A.} \bsnm{{Golightly}}},
\bauthor{\binits{E.R.} \bsnm{{Coughlin}}},
\bauthor{\binits{C.J.} \bsnm{{Nixon}}},
\batitle{{Tidal Disruption Events: The Role of Stellar Spin}}.
\bjtitle{\apj}
\bvolume{872}(\bissue{2}),
\bfpage{163}
(\byear{2019}b).
doi:\doiurl{10.3847/1538-4357/aafd2f}
\end{barticle}
\endbibitem

\bibitem[\protect\citeauthoryear{{Guillochon} and
  {Ramirez-Ruiz}}{2013}]{Guillochon&RamirezRuiz13}
\begin{barticle}
\bauthor{\binits{J.} \bsnm{{Guillochon}}},
\bauthor{\binits{E.} \bsnm{{Ramirez-Ruiz}}},
\batitle{{Hydrodynamical Simulations to Determine the Feeding Rate of Black
  Holes by the Tidal Disruption of Stars: The Importance of the Impact
  Parameter and Stellar Structure}}.
\bjtitle{\apj}
\bvolume{767},
\bfpage{25}
(\byear{2013}).
doi:\doiurl{10.1088/0004-637X/767/1/25}
\end{barticle}
\endbibitem

\bibitem[\protect\citeauthoryear{{Guillochon} et~al.}{2016}]{guillochon16}
\begin{barticle}
\bauthor{\binits{J.} \bsnm{{Guillochon}}},
\bauthor{\binits{M.} \bsnm{{McCourt}}},
\bauthor{\binits{X.} \bsnm{{Chen}}},
\bauthor{\binits{M.D.} \bsnm{{Johnson}}},
\bauthor{\binits{E.} \bsnm{{Berger}}},
\batitle{{Unbound Debris Streams and Remnants Resulting from the Tidal
  Disruptions of Stars by Supermassive Black Holes}}.
\bjtitle{\apj}
\bvolume{822},
\bfpage{48}
(\byear{2016}).
doi:\doiurl{10.3847/0004-637X/822/1/48}
\end{barticle}
\endbibitem

\bibitem[\protect\citeauthoryear{{Guillochon} and
  {McCourt}}{2017}]{Guillochon17}
\begin{barticle}
\bauthor{\binits{J.} \bsnm{{Guillochon}}},
\bauthor{\binits{M.} \bsnm{{McCourt}}},
\batitle{{Simulations of Magnetic Fields in Tidally Disrupted Stars}}.
\bjtitle{\apj}
\bvolume{834}(\bissue{2}),
\bfpage{19}
(\byear{2017}).
doi:\doiurl{10.3847/2041-8213/834/2/L19}
\end{barticle}
\endbibitem

\bibitem[\protect\citeauthoryear{{Guillochon} and
  {Ramirez-Ruiz}}{2013}]{Guillochon13}
\begin{barticle}
\bauthor{\binits{J.} \bsnm{{Guillochon}}},
\bauthor{\binits{E.} \bsnm{{Ramirez-Ruiz}}},
\batitle{{Hydrodynamical Simulations to Determine the Feeding Rate of Black
  Holes by the Tidal Disruption of Stars: The Importance of the Impact
  Parameter and Stellar Structure}}.
\bjtitle{\apj}
\bvolume{767}(\bissue{1}),
\bfpage{25}
(\byear{2013}).
doi:\doiurl{10.1088/0004-637X/767/1/25}
\end{barticle}
\endbibitem

\bibitem[\protect\citeauthoryear{{Guillochon}
  et~al.}{2011}]{2011ApJ...732...74G}
\begin{barticle}
\bauthor{\binits{J.} \bsnm{{Guillochon}}},
\bauthor{\binits{E.} \bsnm{{Ramirez-Ruiz}}},
\bauthor{\binits{D.} \bsnm{{Lin}}},
\batitle{{Consequences of the Ejection and Disruption of Giant Planets}}.
\bjtitle{\apj}
\bvolume{732}(\bissue{2}),
\bfpage{74}
(\byear{2011}).
doi:\doiurl{10.1088/0004-637X/732/2/74}
\end{barticle}
\endbibitem

\bibitem[\protect\citeauthoryear{{Guillochon} et~al.}{2009}]{guillochon+2009}
\begin{barticle}
\bauthor{\binits{J.} \bsnm{{Guillochon}}},
\bauthor{\binits{E.} \bsnm{{Ramirez-Ruiz}}},
\bauthor{\binits{S.} \bsnm{{Rosswog}}},
\bauthor{\binits{D.} \bsnm{{Kasen}}},
\batitle{{Three-dimensional Simulations of Tidally Disrupted Solar-type Stars
  and the Observational Signatures of Shock Breakout}}.
\bjtitle{\apj}
\bvolume{705}(\bissue{1}),
\bfpage{844}--\blpage{853}
(\byear{2009}).
doi:\doiurl{10.1088/0004-637X/705/1/844}
\end{barticle}
\endbibitem

\bibitem[\protect\citeauthoryear{{Haas} et~al.}{2012a}]{Haas12}
\begin{barticle}
\bauthor{\binits{R.} \bsnm{{Haas}}},
\bauthor{\binits{R.V.} \bsnm{{Shcherbakov}}},
\bauthor{\binits{T.} \bsnm{{Bode}}},
\bauthor{\binits{P.} \bsnm{{Laguna}}},
\batitle{{Tidal Disruptions of White Dwarfs from Ultra-close Encounters with
  Intermediate-mass Spinning Black Holes}}.
\bjtitle{\apj}
\bvolume{749}(\bissue{2}),
\bfpage{117}
(\byear{2012}a).
doi:\doiurl{10.1088/0004-637X/749/2/117}
\end{barticle}
\endbibitem

\bibitem[\protect\citeauthoryear{{Haas} et~al.}{2012b}]{2012ApJ...749..117H}
\begin{barticle}
\bauthor{\binits{R.} \bsnm{{Haas}}},
\bauthor{\binits{R.V.} \bsnm{{Shcherbakov}}},
\bauthor{\binits{T.} \bsnm{{Bode}}},
\bauthor{\binits{P.} \bsnm{{Laguna}}},
\batitle{{Tidal Disruptions of White Dwarfs from Ultra-close Encounters with
  Intermediate-mass Spinning Black Holes}}.
\bjtitle{\apj}
\bvolume{749}(\bissue{2}),
\bfpage{117}
(\byear{2012}b).
doi:\doiurl{10.1088/0004-637X/749/2/117}
\end{barticle}
\endbibitem

\bibitem[\protect\citeauthoryear{{Hamers} and
  {Dosopoulou}}{2019}]{HamersDosopoulou19}
\begin{barticle}
\bauthor{\binits{A.S.} \bsnm{{Hamers}}},
\bauthor{\binits{F.} \bsnm{{Dosopoulou}}},
\batitle{{An Analytic Model for Mass Transfer in Binaries with Arbitrary
  Eccentricity, with Applications to Triple-star Systems}}.
\bjtitle{\apj}
\bvolume{872}(\bissue{2}),
\bfpage{119}
(\byear{2019}).
doi:\doiurl{10.3847/1538-4357/ab001d}
\end{barticle}
\endbibitem

\bibitem[\protect\citeauthoryear{{Hameury} et~al.}{1994}]{Hameury94}
\begin{barticle}
\bauthor{\binits{J.-M.} \bsnm{{Hameury}}},
\bauthor{\binits{A.R.} \bsnm{{King}}},
\bauthor{\binits{J.-P.} \bsnm{{Lasota}}},
\bauthor{\binits{M.} \bsnm{{Auvergne}}},
\batitle{{Black-hole stellar accretion in active galactic nuclei.}}
\bjtitle{\aap}
\bvolume{292},
\bfpage{404}--\blpage{408}
(\byear{1994})
\end{barticle}
\endbibitem

\bibitem[\protect\citeauthoryear{{Hayasaki} et~al.}{2013}]{Hayasaki13}
\begin{barticle}
\bauthor{\binits{K.} \bsnm{{Hayasaki}}},
\bauthor{\binits{N.} \bsnm{{Stone}}},
\bauthor{\binits{A.} \bsnm{{Loeb}}},
\batitle{{Finite, intense accretion bursts from tidal disruption of stars on
  bound orbits}}.
\bjtitle{\mnras}
\bvolume{434}(\bissue{2}),
\bfpage{909}--\blpage{924}
(\byear{2013}).
doi:\doiurl{10.1093/mnras/stt871}
\end{barticle}
\endbibitem

\bibitem[\protect\citeauthoryear{{Hayasaki} et~al.}{2016}]{Hayasaki16}
\begin{barticle}
\bauthor{\binits{K.} \bsnm{{Hayasaki}}},
\bauthor{\binits{N.} \bsnm{{Stone}}},
\bauthor{\binits{A.} \bsnm{{Loeb}}},
\batitle{{Circularization of tidally disrupted stars around spinning
  supermassive black holes}}.
\bjtitle{\mnras}
\bvolume{461}(\bissue{4}),
\bfpage{3760}--\blpage{3780}
(\byear{2016}).
doi:\doiurl{10.1093/mnras/stw1387}
\end{barticle}
\endbibitem

\bibitem[\protect\citeauthoryear{{Hayasaki} et~al.}{2018}]{Hayasaki18}
\begin{barticle}
\bauthor{\binits{K.} \bsnm{{Hayasaki}}},
\bauthor{\binits{S.} \bsnm{{Zhong}}},
\bauthor{\binits{S.} \bsnm{{Li}}},
\bauthor{\binits{P.} \bsnm{{Berczik}}},
\bauthor{\binits{R.} \bsnm{{Spurzem}}},
\batitle{{Classification of Tidal Disruption Events Based on Stellar Orbital
  Properties}}.
\bjtitle{\apj}
\bvolume{855}(\bissue{2}),
\bfpage{129}
(\byear{2018}).
doi:\doiurl{10.3847/1538-4357/aab0a5}
\end{barticle}
\endbibitem

\bibitem[\protect\citeauthoryear{{Hills}}{1975}]{Hills75}
\begin{barticle}
\bauthor{\binits{J.G.} \bsnm{{Hills}}},
\batitle{{Possible power source of Seyfert galaxies and QSOs}}.
\bjtitle{\nat}
\bvolume{254},
\bfpage{295}--\blpage{298}
(\byear{1975}).
doi:\doiurl{10.1038/254295a0}
\end{barticle}
\endbibitem

\bibitem[\protect\citeauthoryear{{Hjellming} and
  {Webbink}}{1987}]{1987ApJ...318..794H}
\begin{barticle}
\bauthor{\binits{M.S.} \bsnm{{Hjellming}}},
\bauthor{\binits{R.F.} \bsnm{{Webbink}}},
\batitle{{Thresholds for Rapid Mass Transfer in Binary System. I. Polytropic
  Models}}.
\bjtitle{\apj}
\bvolume{318},
\bfpage{794}
(\byear{1987}).
doi:\doiurl{10.1086/165412}
\end{barticle}
\endbibitem

\bibitem[\protect\citeauthoryear{{Ivanov} and
  {Chernyakova}}{2006}]{Ivanov&Chernyakova06}
\begin{barticle}
\bauthor{\binits{P.B.} \bsnm{{Ivanov}}},
\bauthor{\binits{M.A.} \bsnm{{Chernyakova}}},
\batitle{{Relativistic cross sections of mass stripping and tidal disruption of
  a star by a super-massive rotating black hole}}.
\bjtitle{\aap}
\bvolume{448}(\bissue{3}),
\bfpage{843}--\blpage{852}
(\byear{2006}).
doi:\doiurl{10.1051/0004-6361:20053409}
\end{barticle}
\endbibitem

\bibitem[\protect\citeauthoryear{{Ivanov} et~al.}{2003}]{Ivanov+03}
\begin{barticle}
\bauthor{\binits{P.B.} \bsnm{{Ivanov}}},
\bauthor{\binits{M.A.} \bsnm{{Chernyakova}}},
\bauthor{\binits{I.D.} \bsnm{{Novikov}}},
\batitle{{The new model of a tidally disrupted star: further development and
  relativistic calculations}}.
\bjtitle{\mnras}
\bvolume{338}(\bissue{1}),
\bfpage{147}--\blpage{155}
(\byear{2003}).
doi:\doiurl{10.1046/j.1365-8711.2003.06028.x}
\end{barticle}
\endbibitem

\bibitem[\protect\citeauthoryear{{Ivanov} and
  {Novikov}}{2001}]{Ivanov&Novikov01}
\begin{barticle}
\bauthor{\binits{P.B.} \bsnm{{Ivanov}}},
\bauthor{\binits{I.D.} \bsnm{{Novikov}}},
\batitle{{A New Model of a Tidally Disrupted Star}}.
\bjtitle{\apj}
\bvolume{549}(\bissue{1}),
\bfpage{467}--\blpage{482}
(\byear{2001}).
doi:\doiurl{10.1086/319050}
\end{barticle}
\endbibitem

\bibitem[\protect\citeauthoryear{{Kasen} and {Ramirez-Ruiz}}{2010}]{kasen+RR10}
\begin{barticle}
\bauthor{\binits{D.} \bsnm{{Kasen}}},
\bauthor{\binits{E.} \bsnm{{Ramirez-Ruiz}}},
\batitle{{Optical Transients from the Unbound Debris of Tidal Disruption}}.
\bjtitle{\apj}
\bvolume{714}(\bissue{1}),
\bfpage{155}--\blpage{162}
(\byear{2010}).
doi:\doiurl{10.1088/0004-637X/714/1/155}
\end{barticle}
\endbibitem

\bibitem[\protect\citeauthoryear{{Kawana} et~al.}{2018}]{Kawana+18}
\begin{barticle}
\bauthor{\binits{K.} \bsnm{{Kawana}}},
\bauthor{\binits{A.} \bsnm{{Tanikawa}}},
\bauthor{\binits{N.} \bsnm{{Yoshida}}},
\batitle{{Tidal disruption of a white dwarf by a black hole: the diversity of
  nucleosynthesis, explosion energy, and the fate of debris streams}}.
\bjtitle{\mnras}
\bvolume{477}(\bissue{3}),
\bfpage{3449}--\blpage{3460}
(\byear{2018}).
doi:\doiurl{10.1093/mnras/sty842}
\end{barticle}
\endbibitem

\bibitem[\protect\citeauthoryear{{Kesden}}{2012a}]{Kesden12}
\begin{barticle}
\bauthor{\binits{M.} \bsnm{{Kesden}}},
\batitle{{Tidal-disruption rate of stars by spinning supermassive black
  holes}}.
\bjtitle{\prd}
\bvolume{85}(\bissue{2}),
\bfpage{024037}
(\byear{2012}a).
doi:\doiurl{10.1103/PhysRevD.85.024037}
\end{barticle}
\endbibitem

\bibitem[\protect\citeauthoryear{{Kesden}}{2012b}]{Kesden12b}
\begin{barticle}
\bauthor{\binits{M.} \bsnm{{Kesden}}},
\batitle{{Black-hole spin dependence in the light curves of tidal disruption
  events}}.
\bjtitle{\prd}
\bvolume{86}(\bissue{6}),
\bfpage{064026}
(\byear{2012}b).
doi:\doiurl{10.1103/PhysRevD.86.064026}
\end{barticle}
\endbibitem

\bibitem[\protect\citeauthoryear{{Khokhlov} and
  {Melia}}{1996}]{khoklovMelia1996}
\begin{barticle}
\bauthor{\binits{A.} \bsnm{{Khokhlov}}},
\bauthor{\binits{F.} \bsnm{{Melia}}},
\batitle{{Powerful Ejection of Matter from Tidally Disrupted Stars near Massive
  Black Holes and a Possible Application to Sagittarius A East}}.
\bjtitle{\apjl}
\bvolume{457},
\bfpage{61}
(\byear{1996}).
doi:\doiurl{10.1086/309895}
\end{barticle}
\endbibitem

\bibitem[\protect\citeauthoryear{{King} and {Done}}{1993}]{King93}
\begin{barticle}
\bauthor{\binits{A.R.} \bsnm{{King}}},
\bauthor{\binits{C.} \bsnm{{Done}}},
\batitle{{Stellar accretion in active galactic nuclei.}}
\bjtitle{\mnras}
\bvolume{264},
\bfpage{388}--\blpage{394}
(\byear{1993}).
doi:\doiurl{10.1093/mnras/264.2.388}
\end{barticle}
\endbibitem

\bibitem[\protect\citeauthoryear{{Kobayashi} et~al.}{2004}]{kobayashi+2004}
\begin{barticle}
\bauthor{\binits{S.} \bsnm{{Kobayashi}}},
\bauthor{\binits{P.} \bsnm{{Laguna}}},
\bauthor{\binits{E.S.} \bsnm{{Phinney}}},
\bauthor{\binits{P.} \bsnm{{M{\'e}sz{\'a}ros}}},
\batitle{{Gravitational Waves and X-Ray Signals from Stellar Disruption by a
  Massive Black Hole}}.
\bjtitle{\apj}
\bvolume{615}(\bissue{2}),
\bfpage{855}--\blpage{865}
(\byear{2004}).
doi:\doiurl{10.1086/424684}
\end{barticle}
\endbibitem

\bibitem[\protect\citeauthoryear{{Kobayashi} et~al.}{2012}]{kobayashi+2012}
\begin{barticle}
\bauthor{\binits{S.} \bsnm{{Kobayashi}}},
\bauthor{\binits{Y.} \bsnm{{Hainick}}},
\bauthor{\binits{R.} \bsnm{{Sari}}},
\bauthor{\binits{E.M.} \bsnm{{Rossi}}},
\batitle{{Ejection and Capture Dynamics in Restricted Three-body Encounters}}.
\bjtitle{\apj}
\bvolume{748}(\bissue{2}),
\bfpage{105}
(\byear{2012}).
doi:\doiurl{10.1088/0004-637X/748/2/105}
\end{barticle}
\endbibitem

\bibitem[\protect\citeauthoryear{{Kochanek}}{1994}]{kochanek94}
\begin{barticle}
\bauthor{\binits{C.S.} \bsnm{{Kochanek}}},
\batitle{{The aftermath of tidal disruption: The dynamics of thin gas
  streams}}.
\bjtitle{\apj}
\bvolume{422},
\bfpage{508}--\blpage{520}
(\byear{1994}).
doi:\doiurl{10.1086/173745}
\end{barticle}
\endbibitem

\bibitem[\protect\citeauthoryear{{Kochanek}}{2016}]{2016MNRAS.458..127K}
\begin{barticle}
\bauthor{\binits{C.S.} \bsnm{{Kochanek}}},
\batitle{{Abundance anomalies in tidal disruption events}}.
\bjtitle{\mnras}
\bvolume{458}(\bissue{1}),
\bfpage{127}--\blpage{134}
(\byear{2016}).
doi:\doiurl{10.1093/mnras/stw267}
\end{barticle}
\endbibitem

\bibitem[\protect\citeauthoryear{{Kopal}}{1959}]{Kopal59}
\begin{bbook}
\bauthor{\binits{Z.} \bsnm{{Kopal}}},
\bbtitle{{Close binary systems}}
\byear{1959}
\end{bbook}
\endbibitem

\bibitem[\protect\citeauthoryear{{Kosovichev} and
  {Novikov}}{1992}]{1992MNRAS.258..715K}
\begin{barticle}
\bauthor{\binits{A.G.} \bsnm{{Kosovichev}}},
\bauthor{\binits{I.D.} \bsnm{{Novikov}}},
\batitle{{Non-linear effects at tidal capture of stars by a massive black hole.
  I - Incompressible affine model}}.
\bjtitle{\mnras}
\bvolume{258}(\bissue{4}),
\bfpage{715}--\blpage{724}
(\byear{1992}).
doi:\doiurl{10.1093/mnras/258.4.715}
\end{barticle}
\endbibitem

\bibitem[\protect\citeauthoryear{{Krolik} et~al.}{2016}]{krolik+16}
\begin{barticle}
\bauthor{\binits{J.} \bsnm{{Krolik}}},
\bauthor{\binits{T.} \bsnm{{Piran}}},
\bauthor{\binits{G.} \bsnm{{Svirski}}},
\bauthor{\binits{R.M.} \bsnm{{Cheng}}},
\batitle{{ASASSN-14li: A Model Tidal Disruption Event}}.
\bjtitle{\apj}
\bvolume{827}(\bissue{2}),
\bfpage{127}
(\byear{2016}).
doi:\doiurl{10.3847/0004-637X/827/2/127}
\end{barticle}
\endbibitem

\bibitem[\protect\citeauthoryear{{Law-Smith} et~al.}{2017a}]{LawSmith+17b}
\begin{barticle}
\bauthor{\binits{J.} \bsnm{{Law-Smith}}},
\bauthor{\binits{M.} \bsnm{{MacLeod}}},
\bauthor{\binits{J.} \bsnm{{Guillochon}}},
\bauthor{\binits{P.} \bsnm{{Macias}}},
\bauthor{\binits{E.} \bsnm{{Ramirez-Ruiz}}},
\batitle{{Low-mass White Dwarfs with Hydrogen Envelopes as a Missing Link in
  the Tidal Disruption Menu}}.
\bjtitle{\apj}
\bvolume{841},
\bfpage{132}
(\byear{2017}a).
doi:\doiurl{10.3847/1538-4357/aa6ffb}
\end{barticle}
\endbibitem

\bibitem[\protect\citeauthoryear{{Law-Smith} et~al.}{2017b}]{LawSmith+17}
\begin{barticle}
\bauthor{\binits{J.} \bsnm{{Law-Smith}}},
\bauthor{\binits{E.} \bsnm{{Ramirez-Ruiz}}},
\bauthor{\binits{S.L.} \bsnm{{Ellison}}},
\bauthor{\binits{R.J.} \bsnm{{Foley}}},
\batitle{{Tidal Disruption Event Host Galaxies in the Context of the Local
  Galaxy Population}}.
\bjtitle{\apj}
\bvolume{850},
\bfpage{22}
(\byear{2017}b).
doi:\doiurl{10.3847/1538-4357/aa94c7}
\end{barticle}
\endbibitem

\bibitem[\protect\citeauthoryear{{Law-Smith} et~al.}{2019}]{Law-Smith+2019}
\begin{barticle}
\bauthor{\binits{J.} \bsnm{{Law-Smith}}},
\bauthor{\binits{J.} \bsnm{{Guillochon}}},
\bauthor{\binits{E.} \bsnm{{Ramirez-Ruiz}}},
\batitle{{The Tidal Disruption of Sun-like Stars by Massive Black Holes}}.
\bjtitle{\apjl}
\bvolume{882}(\bissue{2}),
\bfpage{25}
(\byear{2019}).
doi:\doiurl{10.3847/2041-8213/ab379a}
\end{barticle}
\endbibitem

\bibitem[\protect\citeauthoryear{{Lee} and
  {Ostriker}}{1986}]{1986ApJ...310..176L}
\begin{barticle}
\bauthor{\binits{H.M.} \bsnm{{Lee}}},
\bauthor{\binits{J.P.} \bsnm{{Ostriker}}},
\batitle{{Cross Sections for Tidal Capture Binary Formation and Stellar
  Merger}}.
\bjtitle{\apj}
\bvolume{310},
\bfpage{176}
(\byear{1986}).
doi:\doiurl{10.1086/164674}
\end{barticle}
\endbibitem

\bibitem[\protect\citeauthoryear{{Leloudas} et~al.}{2016}]{Leloudas+16}
\begin{barticle}
\bauthor{\binits{G.} \bsnm{{Leloudas}}},
\bauthor{\binits{M.} \bsnm{{Fraser}}},
\bauthor{\binits{N.C.} \bsnm{{Stone}}},
\bauthor{\binits{S.} \bsnm{{van Velzen}}},
\bauthor{\binits{P.G.} \bsnm{{Jonker}}},
\bauthor{\binits{I.} \bsnm{{Arcavi}}},
\bauthor{\binits{C.} \bsnm{{Fremling}}},
\bauthor{\binits{J.R.} \bsnm{{Maund}}},
\bauthor{\binits{S.J.} \bsnm{{Smartt}}},
\bauthor{\binits{T.} \bsnm{{Kr{\`\i}hler}}},
\bauthor{\binits{J.C.A.} \bsnm{{Miller-Jones}}},
\bauthor{\binits{P.M.} \bsnm{{Vreeswijk}}},
\bauthor{\binits{A.} \bsnm{{Gal-Yam}}},
\bauthor{\binits{P.A.} \bsnm{{Mazzali}}},
\bauthor{\binits{A.} \bsnm{{De Cia}}},
\bauthor{\binits{D.A.} \bsnm{{Howell}}},
\bauthor{\binits{C.} \bsnm{{Inserra}}},
\bauthor{\binits{F.} \bsnm{{Patat}}},
\bauthor{\binits{A.} \bsnm{{de Ugarte Postigo}}},
\bauthor{\binits{O.} \bsnm{{Yaron}}},
\bauthor{\binits{C.} \bsnm{{Ashall}}},
\bauthor{\binits{I.} \bsnm{{Bar}}},
\bauthor{\binits{H.} \bsnm{{Campbell}}},
\bauthor{\binits{T.-W.} \bsnm{{Chen}}},
\bauthor{\binits{M.} \bsnm{{Childress}}},
\bauthor{\binits{N.} \bsnm{{Elias-Rosa}}},
\bauthor{\binits{J.} \bsnm{{Harmanen}}},
\bauthor{\binits{G.} \bsnm{{Hosseinzadeh}}},
\bauthor{\binits{J.} \bsnm{{Johansson}}},
\bauthor{\binits{T.} \bsnm{{Kangas}}},
\bauthor{\binits{E.} \bsnm{{Kankare}}},
\bauthor{\binits{S.} \bsnm{{Kim}}},
\bauthor{\binits{H.} \bsnm{{Kuncarayakti}}},
\bauthor{\binits{J.} \bsnm{{Lyman}}},
\bauthor{\binits{M.R.} \bsnm{{Magee}}},
\bauthor{\binits{K.} \bsnm{{Maguire}}},
\bauthor{\binits{D.} \bsnm{{Malesani}}},
\bauthor{\binits{S.} \bsnm{{Mattila}}},
\bauthor{\binits{C.V.} \bsnm{{McCully}}},
\bauthor{\binits{M.} \bsnm{{Nicholl}}},
\bauthor{\binits{S.} \bsnm{{Prentice}}},
\bauthor{\binits{C.} \bsnm{{Romero-Ca{\~n}izales}}},
\bauthor{\binits{S.} \bsnm{{Schulze}}},
\bauthor{\binits{K.W.} \bsnm{{Smith}}},
\bauthor{\binits{J.} \bsnm{{Sollerman}}},
\bauthor{\binits{M.} \bsnm{{Sullivan}}},
\bauthor{\binits{B.E.} \bsnm{{Tucker}}},
\bauthor{\binits{S.} \bsnm{{Valenti}}},
\bauthor{\binits{J.C.} \bsnm{{Wheeler}}},
\bauthor{\binits{D.R.} \bsnm{{Young}}},
\batitle{{The superluminous transient ASASSN-15lh as a tidal disruption event
  from a Kerr black hole}}.
\bjtitle{Nature Astronomy}
\bvolume{1},
\bfpage{0002}
(\byear{2016}).
doi:\doiurl{10.1038/s41550-016-0002}
\end{barticle}
\endbibitem

\bibitem[\protect\citeauthoryear{{Linial} and {Sari}}{2017}]{Linial17}
\begin{barticle}
\bauthor{\binits{I.} \bsnm{{Linial}}},
\bauthor{\binits{R.} \bsnm{{Sari}}},
\batitle{{Mass-loss through the L2 Lagrange point - application to
  main-sequence EMRI}}.
\bjtitle{\mnras}
\bvolume{469}(\bissue{2}),
\bfpage{2441}--\blpage{2454}
(\byear{2017}).
doi:\doiurl{10.1093/mnras/stx1041}
\end{barticle}
\endbibitem

\bibitem[\protect\citeauthoryear{{Liptai} et~al.}{2019}]{Liptai19}
\begin{botherref}
\oauthor{\binits{D.} \bsnm{{Liptai}}},
\oauthor{\binits{D.J.} \bsnm{{Price}}},
\oauthor{\binits{I.} \bsnm{{Mandel}}},
\oauthor{\binits{G.} \bsnm{{Lodato}}},
{Disc formation from tidal disruption of stars on eccentric orbits by Kerr
  black holes using GRSPH}.
arXiv e-prints,
1910--10154
(2019)
\end{botherref}
\endbibitem

\bibitem[\protect\citeauthoryear{{Liu} et~al.}{2009}]{Liu09}
\begin{barticle}
\bauthor{\binits{F.K.} \bsnm{{Liu}}},
\bauthor{\binits{S.} \bsnm{{Li}}},
\bauthor{\binits{X.} \bsnm{{Chen}}},
\batitle{{Interruption of Tidal-Disruption Flares by Supermassive Black Hole
  Binaries}}.
\bjtitle{\apj}
\bvolume{706}(\bissue{1}),
\bfpage{133}--\blpage{137}
(\byear{2009}).
doi:\doiurl{10.1088/0004-637X/706/1/L133}
\end{barticle}
\endbibitem

\bibitem[\protect\citeauthoryear{{Liu} et~al.}{2014}]{Liu14}
\begin{barticle}
\bauthor{\binits{F.K.} \bsnm{{Liu}}},
\bauthor{\binits{S.} \bsnm{{Li}}},
\bauthor{\binits{S.} \bsnm{{Komossa}}},
\batitle{{A Milliparsec Supermassive Black Hole Binary Candidate in the Galaxy
  SDSS J120136.02+300305.5}}.
\bjtitle{\apj}
\bvolume{786}(\bissue{2}),
\bfpage{103}
(\byear{2014}).
doi:\doiurl{10.1088/0004-637X/786/2/103}
\end{barticle}
\endbibitem

\bibitem[\protect\citeauthoryear{{Lodato} et~al.}{2009}]{Lodato09}
\begin{barticle}
\bauthor{\binits{G.} \bsnm{{Lodato}}},
\bauthor{\binits{A.R.} \bsnm{{King}}},
\bauthor{\binits{J.E.} \bsnm{{Pringle}}},
\batitle{{Stellar disruption by a supermassive black hole: is the light curve
  really proportional to t$^{-5/3}$?}}
\bjtitle{\mnras}
\bvolume{392}(\bissue{1}),
\bfpage{332}--\blpage{340}
(\byear{2009}).
doi:\doiurl{10.1111/j.1365-2966.2008.14049.x}
\end{barticle}
\endbibitem

\bibitem[\protect\citeauthoryear{{Luminet} and
  {Carter}}{1986}]{Luminet&Carter86}
\begin{barticle}
\bauthor{\binits{J.-P.} \bsnm{{Luminet}}},
\bauthor{\binits{B.} \bsnm{{Carter}}},
\batitle{{Dynamics of an Affine Star Model in a Black Hole Tidal Field}}.
\bjtitle{\apjs}
\bvolume{61},
\bfpage{219}
(\byear{1986}).
doi:\doiurl{10.1086/191113}
\end{barticle}
\endbibitem

\bibitem[\protect\citeauthoryear{{Luminet} and {Marck}}{1985}]{Luminet&Marck85}
\begin{barticle}
\bauthor{\binits{J.-P.} \bsnm{{Luminet}}},
\bauthor{\binits{J.-A.} \bsnm{{Marck}}},
\batitle{{Tidal squeezing of stars by Schwarzschild black holes}}.
\bjtitle{\mnras}
\bvolume{212},
\bfpage{57}--\blpage{75}
(\byear{1985}).
doi:\doiurl{10.1093/mnras/212.1.57}
\end{barticle}
\endbibitem

\bibitem[\protect\citeauthoryear{{Luminet} and
  {Pichon}}{1989a}]{Luminet&Pichon89b}
\begin{barticle}
\bauthor{\binits{J.-P.} \bsnm{{Luminet}}},
\bauthor{\binits{B.} \bsnm{{Pichon}}},
\batitle{{Tidal pinching of white dwarfs}}.
\bjtitle{\aap}
\bvolume{209}(\bissue{1-2}),
\bfpage{103}--\blpage{110}
(\byear{1989}a)
\end{barticle}
\endbibitem

\bibitem[\protect\citeauthoryear{{Luminet} and
  {Pichon}}{1989b}]{Luminet&Pichon89}
\begin{barticle}
\bauthor{\binits{J.-P.} \bsnm{{Luminet}}},
\bauthor{\binits{B.} \bsnm{{Pichon}}},
\batitle{{Tidally-detonated nuclear reactions in main sequence stars passing
  near a large black hole}}.
\bjtitle{\aap}
\bvolume{209}(\bissue{1-2}),
\bfpage{85}--\blpage{102}
(\byear{1989}b)
\end{barticle}
\endbibitem

\bibitem[\protect\citeauthoryear{{Luo} et~al.}{2016}]{Luo+16}
\begin{barticle}
\bauthor{\binits{J.} \bsnm{{Luo}}},
\bauthor{\binits{L.-S.} \bsnm{{Chen}}},
\bauthor{\binits{H.-Z.} \bsnm{{Duan}}},
\bauthor{\binits{Y.-G.} \bsnm{{Gong}}},
\bauthor{\binits{S.} \bsnm{{Hu}}},
\bauthor{\binits{J.} \bsnm{{Ji}}},
\bauthor{\binits{Q.} \bsnm{{Liu}}},
\bauthor{\binits{J.} \bsnm{{Mei}}},
\bauthor{\binits{V.} \bsnm{{Milyukov}}},
\bauthor{\binits{M.} \bsnm{{Sazhin}}},
\bauthor{\binits{C.-G.} \bsnm{{Shao}}},
\bauthor{\binits{V.T.} \bsnm{{Toth}}},
\bauthor{\binits{H.-B.} \bsnm{{Tu}}},
\bauthor{\binits{Y.} \bsnm{{Wang}}},
\bauthor{\binits{Y.} \bsnm{{Wang}}},
\bauthor{\binits{H.-C.} \bsnm{{Yeh}}},
\bauthor{\binits{M.-S.} \bsnm{{Zhan}}},
\bauthor{\binits{Y.} \bsnm{{Zhang}}},
\bauthor{\binits{V.} \bsnm{{Zharov}}},
\bauthor{\binits{Z.-B.} \bsnm{{Zhou}}},
\batitle{Tianqin: a space-borne gravitational wave detector}.
\bjtitle{Classical and Quantum Gravity}
\bvolume{33}(\bissue{3}),
\bfpage{035010}
(\byear{2016})
\end{barticle}
\endbibitem

\bibitem[\protect\citeauthoryear{{Lynden-Bell}}{1969}]{lynden-bell69}
\begin{barticle}
\bauthor{\binits{D.} \bsnm{{Lynden-Bell}}},
\batitle{{Galactic Nuclei as Collapsed Old Quasars}}.
\bjtitle{\nat}
\bvolume{223}(\bissue{5207}),
\bfpage{690}--\blpage{694}
(\byear{1969}).
doi:\doiurl{10.1038/223690a0}
\end{barticle}
\endbibitem

\bibitem[\protect\citeauthoryear{{MacLeod} et~al.}{2012}]{MacLeod+12}
\begin{barticle}
\bauthor{\binits{M.} \bsnm{{MacLeod}}},
\bauthor{\binits{J.} \bsnm{{Guillochon}}},
\bauthor{\binits{E.} \bsnm{{Ramirez-Ruiz}}},
\batitle{{The Tidal Disruption of Giant Stars and their Contribution to the
  Flaring Supermassive Black Hole Population}}.
\bjtitle{\apj}
\bvolume{757},
\bfpage{134}
(\byear{2012}).
doi:\doiurl{10.1088/0004-637X/757/2/134}
\end{barticle}
\endbibitem

\bibitem[\protect\citeauthoryear{{MacLeod} et~al.}{2013}]{MacLeod+13}
\begin{barticle}
\bauthor{\binits{M.} \bsnm{{MacLeod}}},
\bauthor{\binits{E.} \bsnm{{Ramirez-Ruiz}}},
\bauthor{\binits{S.} \bsnm{{Grady}}},
\bauthor{\binits{J.} \bsnm{{Guillochon}}},
\batitle{{Spoon-feeding Giant Stars to Supermassive Black Holes: Episodic Mass
  Transfer from Evolving Stars and their Contribution to the Quiescent Activity
  of Galactic Nuclei}}.
\bjtitle{\apj}
\bvolume{777},
\bfpage{133}
(\byear{2013}).
doi:\doiurl{10.1088/0004-637X/777/2/133}
\end{barticle}
\endbibitem

\bibitem[\protect\citeauthoryear{{MacLeod} et~al.}{2014}]{2014ApJ...794....9M}
\begin{barticle}
\bauthor{\binits{M.} \bsnm{{MacLeod}}},
\bauthor{\binits{J.} \bsnm{{Goldstein}}},
\bauthor{\binits{E.} \bsnm{{Ramirez-Ruiz}}},
\bauthor{\binits{J.} \bsnm{{Guillochon}}},
\bauthor{\binits{J.} \bsnm{{Samsing}}},
\batitle{{Illuminating Massive Black Holes with White Dwarfs: Orbital Dynamics
  and High-energy Transients from Tidal Interactions}}.
\bjtitle{\apj}
\bvolume{794}(\bissue{1}),
\bfpage{9}
(\byear{2014}).
doi:\doiurl{10.1088/0004-637X/794/1/9}
\end{barticle}
\endbibitem

\bibitem[\protect\citeauthoryear{{MacLeod} et~al.}{2016}]{2016ApJ...819....3M}
\begin{barticle}
\bauthor{\binits{M.} \bsnm{{MacLeod}}},
\bauthor{\binits{J.} \bsnm{{Guillochon}}},
\bauthor{\binits{E.} \bsnm{{Ramirez-Ruiz}}},
\bauthor{\binits{D.} \bsnm{{Kasen}}},
\bauthor{\binits{S.} \bsnm{{Rosswog}}},
\batitle{{Optical Thermonuclear Transients from Tidal Compression of White
  Dwarfs as Tracers of the Low End of the Massive Black Hole Mass Function}}.
\bjtitle{\apj}
\bvolume{819}(\bissue{1}),
\bfpage{3}
(\byear{2016}).
doi:\doiurl{10.3847/0004-637X/819/1/3}
\end{barticle}
\endbibitem

\bibitem[\protect\citeauthoryear{{Mainetti} et~al.}{2017}]{Mainetti+17}
\begin{barticle}
\bauthor{\binits{D.} \bsnm{{Mainetti}}},
\bauthor{\binits{A.} \bsnm{{Lupi}}},
\bauthor{\binits{S.} \bsnm{{Campana}}},
\bauthor{\binits{M.} \bsnm{{Colpi}}},
\bauthor{\binits{E.R.} \bsnm{{Coughlin}}},
\bauthor{\binits{J.} \bsnm{{Guillochon}}},
\bauthor{\binits{E.} \bsnm{{Ramirez-Ruiz}}},
\batitle{{The fine line between total and partial tidal disruption events}}.
\bjtitle{\aap}
\bvolume{600},
\bfpage{124}
(\byear{2017}).
doi:\doiurl{10.1051/0004-6361/201630092}
\end{barticle}
\endbibitem

\bibitem[\protect\citeauthoryear{{Mandel} and {Levin}}{2015}]{mandel&levin15}
\begin{barticle}
\bauthor{\binits{I.} \bsnm{{Mandel}}},
\bauthor{\binits{Y.} \bsnm{{Levin}}},
\batitle{{Double Tidal Disruptions in Galactic Nuclei}}.
\bjtitle{\apjl}
\bvolume{805},
\bfpage{4}
(\byear{2015}).
doi:\doiurl{10.1088/2041-8205/805/1/L4}
\end{barticle}
\endbibitem

\bibitem[\protect\citeauthoryear{{Manukian} et~al.}{2013}]{2013ApJ...771L..28M}
\begin{barticle}
\bauthor{\binits{H.} \bsnm{{Manukian}}},
\bauthor{\binits{J.} \bsnm{{Guillochon}}},
\bauthor{\binits{E.} \bsnm{{Ramirez-Ruiz}}},
\bauthor{\binits{R.M.} \bsnm{{O'Leary}}},
\batitle{{Turbovelocity Stars: Kicks Resulting from the Tidal Disruption of
  Solitary Stars}}.
\bjtitle{\apjl}
\bvolume{771}(\bissue{2}),
\bfpage{28}
(\byear{2013}).
doi:\doiurl{10.1088/2041-8205/771/2/L28}
\end{barticle}
\endbibitem

\bibitem[\protect\citeauthoryear{{Marck}}{1983}]{Marck+83}
\begin{barticle}
\bauthor{\binits{J.} \bsnm{{Marck}}},
\batitle{{Solution to the Equations of Parallel Transport in Kerr Geometry;
  Tidal Tensor}}.
\bjtitle{Royal Society of London Proceedings Series A}
\bvolume{385},
\bfpage{431}--\blpage{438}
(\byear{1983}).
doi:\doiurl{10.1098/rspa.1983.0021}
\end{barticle}
\endbibitem

\bibitem[\protect\citeauthoryear{{Matese} and
  {Whitmire}}{1983}]{MateseWhitmire83}
\begin{barticle}
\bauthor{\binits{J.J.} \bsnm{{Matese}}},
\bauthor{\binits{D.P.} \bsnm{{Whitmire}}},
\batitle{{Conservative mass transfer in close binary systems. I. Equations of
  motion for spin and orbital angular momenta.}}
\bjtitle{\apj}
\bvolume{266},
\bfpage{776}--\blpage{786}
(\byear{1983}).
doi:\doiurl{10.1086/160825}
\end{barticle}
\endbibitem

\bibitem[\protect\citeauthoryear{{Merritt}}{2013}]{Merritt13}
\begin{bbook}
\bauthor{\binits{D.} \bsnm{{Merritt}}},
\bbtitle{{Dynamics and Evolution of Galactic Nuclei}}
(\bpublisher{Princeton University Press},
\blocation{Princeton, NJ, USA}, \byear{2013})
\end{bbook}
\endbibitem

\bibitem[\protect\citeauthoryear{{Metzger} and {Stone}}{2017}]{Metzger17}
\begin{barticle}
\bauthor{\binits{B.D.} \bsnm{{Metzger}}},
\bauthor{\binits{N.C.} \bsnm{{Stone}}},
\batitle{{Periodic Accretion-powered Flares from Colliding EMRIs as TDE
  Imposters}}.
\bjtitle{\apj}
\bvolume{844}(\bissue{1}),
\bfpage{75}
(\byear{2017}).
doi:\doiurl{10.3847/1538-4357/aa7a16}
\end{barticle}
\endbibitem

\bibitem[\protect\citeauthoryear{{Michel}}{1988}]{Curtis88}
\begin{barticle}
\bauthor{\binits{F.C.} \bsnm{{Michel}}},
\batitle{{Neutron star disk formation from supernova fall-back and possible
  observational consequences}}.
\bjtitle{\nat}
\bvolume{333}(\bissue{6174}),
\bfpage{644}--\blpage{645}
(\byear{1988}).
doi:\doiurl{10.1038/333644a0}
\end{barticle}
\endbibitem

\bibitem[\protect\citeauthoryear{{Miller} et~al.}{2005}]{Miller05}
\begin{barticle}
\bauthor{\binits{M.C.} \bsnm{{Miller}}},
\bauthor{\binits{M.} \bsnm{{Freitag}}},
\bauthor{\binits{D.P.} \bsnm{{Hamilton}}},
\bauthor{\binits{V.M.} \bsnm{{Lauburg}}},
\batitle{{Binary Encounters with Supermassive Black Holes: Zero-Eccentricity
  LISA Events}}.
\bjtitle{\apjl}
\bvolume{631}(\bissue{2}),
\bfpage{117}--\blpage{120}
(\byear{2005}).
doi:\doiurl{10.1086/497335}
\end{barticle}
\endbibitem

\bibitem[\protect\citeauthoryear{{Novikov} et~al.}{1992}]{1992MNRAS.255..276N}
\begin{barticle}
\bauthor{\binits{I.D.} \bsnm{{Novikov}}},
\bauthor{\binits{C.J.} \bsnm{{Pethick}}},
\bauthor{\binits{A.G.} \bsnm{{Polnarev}}},
\batitle{{Tidal capture of stars by a massive black hole.}}
\bjtitle{\mnras}
\bvolume{255},
\bfpage{276}--\blpage{284}
(\byear{1992}).
doi:\doiurl{10.1093/mnras/255.2.276}
\end{barticle}
\endbibitem

\bibitem[\protect\citeauthoryear{{Paczy{\'n}ski}}{1971}]{Paczynski71}
\begin{barticle}
\bauthor{\binits{B.} \bsnm{{Paczy{\'n}ski}}},
\batitle{{Evolutionary Processes in Close Binary Systems}}.
\bjtitle{\araa}
\bvolume{9},
\bfpage{183}
(\byear{1971}).
doi:\doiurl{10.1146/annurev.aa.09.090171.001151}
\end{barticle}
\endbibitem

\bibitem[\protect\citeauthoryear{{Paxton} et~al.}{2011}]{2011ApJS..192....3P}
\begin{barticle}
\bauthor{\binits{B.} \bsnm{{Paxton}}},
\bauthor{\binits{L.} \bsnm{{Bildsten}}},
\bauthor{\binits{A.} \bsnm{{Dotter}}},
\bauthor{\binits{F.} \bsnm{{Herwig}}},
\bauthor{\binits{P.} \bsnm{{Lesaffre}}},
\bauthor{\binits{F.} \bsnm{{Timmes}}},
\batitle{{Modules for Experiments in Stellar Astrophysics (MESA)}}.
\bjtitle{\apjs}
\bvolume{192}(\bissue{1}),
\bfpage{3}
(\byear{2011}).
doi:\doiurl{10.1088/0067-0049/192/1/3}
\end{barticle}
\endbibitem

\bibitem[\protect\citeauthoryear{{Phinney}}{1989}]{Phinney89}
\begin{bchapter}
\bauthor{\binits{E.S.} \bsnm{{Phinney}}},
\bctitle{{Manifestations of a Massive Black Hole in the Galactic Center}},
in \bbtitle{The Center of the Galaxy},
ed. by \beditor{\binits{M.} \bsnm{{Morris}}}
\bsertitle{IAU Symposium},
vol. \bseriesno{136},
\byear{1989},
p. \bfpage{543}
\end{bchapter}
\endbibitem

\bibitem[\protect\citeauthoryear{{Podsiadlowski}}{1996}]{1996MNRAS.279.1104P}
\begin{barticle}
\bauthor{\binits{P.} \bsnm{{Podsiadlowski}}},
\batitle{{The response of tidally heated stars}}.
\bjtitle{\mnras}
\bvolume{279},
\bfpage{1104}
(\byear{1996}).
doi:\doiurl{10.1093/mnras/279.4.1104}
\end{barticle}
\endbibitem

\bibitem[\protect\citeauthoryear{{Press} and
  {Teukolsky}}{1977a}]{Press&Teukolsky77}
\begin{barticle}
\bauthor{\binits{W.H.} \bsnm{{Press}}},
\bauthor{\binits{S.A.} \bsnm{{Teukolsky}}},
\batitle{{On formation of close binaries by two-body tidal capture.}}
\bjtitle{\apj}
\bvolume{213},
\bfpage{183}--\blpage{192}
(\byear{1977}a).
doi:\doiurl{10.1086/155143}
\end{barticle}
\endbibitem

\bibitem[\protect\citeauthoryear{{Press} and
  {Teukolsky}}{1977b}]{1977ApJ...213..183P}
\begin{barticle}
\bauthor{\binits{W.H.} \bsnm{{Press}}},
\bauthor{\binits{S.A.} \bsnm{{Teukolsky}}},
\batitle{{On formation of close binaries by two-body tidal capture.}}
\bjtitle{\apj}
\bvolume{213},
\bfpage{183}--\blpage{192}
(\byear{1977}b).
doi:\doiurl{10.1086/155143}
\end{barticle}
\endbibitem

\bibitem[\protect\citeauthoryear{{Rappaport} et~al.}{1982}]{Rappaport82}
\begin{barticle}
\bauthor{\binits{S.} \bsnm{{Rappaport}}},
\bauthor{\binits{P.C.} \bsnm{{Joss}}},
\bauthor{\binits{R.F.} \bsnm{{Webbink}}},
\batitle{{The evolution of highly compact binary stellar systems.}}
\bjtitle{\apj}
\bvolume{254},
\bfpage{616}--\blpage{640}
(\byear{1982}).
doi:\doiurl{10.1086/159772}
\end{barticle}
\endbibitem

\bibitem[\protect\citeauthoryear{{Rees}}{1988}]{Rees88}
\begin{barticle}
\bauthor{\binits{M.J.} \bsnm{{Rees}}},
\batitle{{Tidal disruption of stars by black holes of 10 to the 6th-10 to the
  8th solar masses in nearby galaxies}}.
\bjtitle{\nat}
\bvolume{333},
\bfpage{523}--\blpage{528}
(\byear{1988}).
doi:\doiurl{10.1038/333523a0}
\end{barticle}
\endbibitem

\bibitem[\protect\citeauthoryear{{Rossi} et~al.}{2014}]{Rossi:2014}
\begin{barticle}
\bauthor{\binits{E.M.} \bsnm{{Rossi}}},
\bauthor{\binits{S.} \bsnm{{Kobayashi}}},
\bauthor{\binits{R.} \bsnm{{Sari}}},
\batitle{{The Velocity Distribution of Hypervelocity Stars}}.
\bjtitle{\apj}
\bvolume{795},
\bfpage{125}
(\byear{2014}).
doi:\doiurl{10.1088/0004-637X/795/2/125}
\end{barticle}
\endbibitem

\bibitem[\protect\citeauthoryear{{Rossi} et~al.}{2017}]{Rossi:2017}
\begin{barticle}
\bauthor{\binits{E.M.} \bsnm{{Rossi}}},
\bauthor{\binits{T.} \bsnm{{Marchetti}}},
\bauthor{\binits{M.} \bsnm{{Cacciato}}},
\bauthor{\binits{M.} \bsnm{{Kuiack}}},
\bauthor{\binits{R.} \bsnm{{Sari}}},
\batitle{{Joint constraints on the Galactic dark matter halo and Galactic
  Centre from hypervelocity stars}}.
\bjtitle{\mnras}
\bvolume{467},
\bfpage{1844}--\blpage{1856}
(\byear{2017}).
doi:\doiurl{10.1093/mnras/stx098}
\end{barticle}
\endbibitem

\bibitem[\protect\citeauthoryear{{Rosswog} et~al.}{2009}]{Rosswog+09}
\begin{barticle}
\bauthor{\binits{S.} \bsnm{{Rosswog}}},
\bauthor{\binits{E.} \bsnm{{Ramirez-Ruiz}}},
\bauthor{\binits{W.R.} \bsnm{{Hix}}},
\batitle{{Tidal Disruption and Ignition of White Dwarfs by Moderately Massive
  Black Holes}}.
\bjtitle{\apj}
\bvolume{695}(\bissue{1}),
\bfpage{404}--\blpage{419}
(\byear{2009}).
doi:\doiurl{10.1088/0004-637X/695/1/404}
\end{barticle}
\endbibitem

\bibitem[\protect\citeauthoryear{{Ryu} et~al.}{2020a}]{Ryu+20a}
\begin{botherref}
\oauthor{\binits{T.} \bsnm{{Ryu}}},
\oauthor{\binits{J.} \bsnm{{Krolik}}},
\oauthor{\binits{T.} \bsnm{{Piran}}},
\oauthor{\binits{S.C.} \bsnm{{Noble}}},
{Tidal Disruptions of Main Sequence Stars -- I. Observable Quantities and their
  Dependence on Stellar and Black Hole Mass}.
arXiv e-prints,
2001--03501
(2020a)
\end{botherref}
\endbibitem

\bibitem[\protect\citeauthoryear{{Ryu} et~al.}{2020b}]{Ryu+20c}
\begin{botherref}
\oauthor{\binits{T.} \bsnm{{Ryu}}},
\oauthor{\binits{J.} \bsnm{{Krolik}}},
\oauthor{\binits{T.} \bsnm{{Piran}}},
\oauthor{\binits{S.C.} \bsnm{{Noble}}},
{Tidal disruptions of main sequence stars -- III. Stellar mass dependence of
  the character of partial disruptions}.
arXiv e-prints,
2001--03503
(2020b)
\end{botherref}
\endbibitem

\bibitem[\protect\citeauthoryear{{Ryu} et~al.}{2020c}]{Ryu+20d}
\begin{botherref}
\oauthor{\binits{T.} \bsnm{{Ryu}}},
\oauthor{\binits{J.} \bsnm{{Krolik}}},
\oauthor{\binits{T.} \bsnm{{Piran}}},
\oauthor{\binits{S.C.} \bsnm{{Noble}}},
{Tidal disruptions of main sequence stars -- IV. Relativistic effects and
  dependence on black hole mass}.
arXiv e-prints,
2001--03504
(2020c)
\end{botherref}
\endbibitem

\bibitem[\protect\citeauthoryear{{Sacchi} and {Lodato}}{2019}]{Sacchi19}
\begin{barticle}
\bauthor{\binits{A.} \bsnm{{Sacchi}}},
\bauthor{\binits{G.} \bsnm{{Lodato}}},
\batitle{{`Failed' tidal disruption events and X-ray flares from the Galactic
  Centre}}.
\bjtitle{\mnras}
\bvolume{486}(\bissue{2}),
\bfpage{1833}--\blpage{1839}
(\byear{2019}).
doi:\doiurl{10.1093/mnras/stz981}
\end{barticle}
\endbibitem

\bibitem[\protect\citeauthoryear{{Sari} et~al.}{2010}]{Sari+10}
\begin{barticle}
\bauthor{\binits{R.} \bsnm{{Sari}}},
\bauthor{\binits{S.} \bsnm{{Kobayashi}}},
\bauthor{\binits{E.M.} \bsnm{{Rossi}}},
\batitle{{Hypervelocity Stars and the Restricted Parabolic Three-Body
  Problem}}.
\bjtitle{\apj}
\bvolume{708}(\bissue{1}),
\bfpage{605}--\blpage{614}
(\byear{2010}).
doi:\doiurl{10.1088/0004-637X/708/1/605}
\end{barticle}
\endbibitem

\bibitem[\protect\citeauthoryear{Sato et~al.}{2017}]{Sato+17}
\begin{barticle}
\bauthor{\binits{S.} \bsnm{Sato}},
\bauthor{\binits{S.} \bsnm{Kawamura}},
\bauthor{\binits{M.} \bsnm{Ando}},
\bauthor{\binits{T.} \bsnm{Nakamura}},
\bauthor{\binits{K.} \bsnm{Tsubono}},
\bauthor{\binits{A.} \bsnm{Araya}},
\bauthor{\binits{I.} \bsnm{Funaki}},
\bauthor{\binits{K.} \bsnm{Ioka}},
\bauthor{\binits{N.} \bsnm{Kanda}},
\bauthor{\binits{S.} \bsnm{Moriwaki}},
\bauthor{\binits{M.} \bsnm{Musha}},
\bauthor{\binits{K.} \bsnm{Nakazawa}},
\bauthor{\binits{K.} \bsnm{Numata}},
\bauthor{\binits{S.-i.} \bsnm{Sakai}},
\bauthor{\binits{N.} \bsnm{Seto}},
\bauthor{\binits{T.} \bsnm{Takashima}},
\bauthor{\binits{T.} \bsnm{Tanaka}},
\bauthor{\binits{K.} \bsnm{Agatsuma}},
\bauthor{\binits{K.-s.} \bsnm{Aoyanagi}},
\bauthor{\binits{K.} \bsnm{Arai}},
\bauthor{\binits{H.} \bsnm{Asada}},
\bauthor{\binits{Y.} \bsnm{Aso}},
\bauthor{\binits{T.} \bsnm{Chiba}},
\bauthor{\binits{T.} \bsnm{Ebisuzaki}},
\bauthor{\binits{Y.} \bsnm{Ejiri}},
\bauthor{\binits{M.} \bsnm{Enoki}},
\bauthor{\binits{Y.} \bsnm{Eriguchi}},
\bauthor{\binits{M.-K.} \bsnm{Fujimoto}},
\bauthor{\binits{R.} \bsnm{Fujita}},
\bauthor{\binits{M.} \bsnm{Fukushima}},
\bauthor{\binits{T.} \bsnm{Futamase}},
\bauthor{\binits{K.} \bsnm{Ganzu}},
\bauthor{\binits{T.} \bsnm{Harada}},
\bauthor{\binits{T.} \bsnm{Hashimoto}},
\bauthor{\binits{K.} \bsnm{Hayama}},
\bauthor{\binits{W.} \bsnm{Hikida}},
\bauthor{\binits{Y.} \bsnm{Himemoto}},
\bauthor{\binits{H.} \bsnm{Hirabayashi}},
\bauthor{\binits{T.} \bsnm{Hiramatsu}},
\bauthor{\binits{F.-L.} \bsnm{Hong}},
\bauthor{\binits{H.} \bsnm{Horisawa}},
\bauthor{\binits{M.} \bsnm{Hosokawa}},
\bauthor{\binits{K.} \bsnm{Ichiki}},
\bauthor{\binits{T.} \bsnm{Ikegami}},
\bauthor{\binits{K.T.} \bsnm{Inoue}},
\bauthor{\binits{K.} \bsnm{Ishidoshiro}},
\bauthor{\binits{H.} \bsnm{Ishihara}},
\bauthor{\binits{T.} \bsnm{Ishikawa}},
\bauthor{\binits{H.} \bsnm{Ishizaki}},
\bauthor{\binits{H.} \bsnm{Ito}},
\bauthor{\binits{Y.} \bsnm{Itoh}},
\bauthor{\binits{N.} \bsnm{Kawashima}},
\bauthor{\binits{F.} \bsnm{Kawazoe}},
\bauthor{\binits{N.} \bsnm{Kishimoto}},
\bauthor{\binits{K.} \bsnm{Kiuchi}},
\bauthor{\binits{S.} \bsnm{Kobayashi}},
\bauthor{\binits{K.} \bsnm{Kohri}},
\bauthor{\binits{H.} \bsnm{Koizumi}},
\bauthor{\binits{Y.} \bsnm{Kojima}},
\bauthor{\binits{K.} \bsnm{Kokeyama}},
\bauthor{\binits{W.} \bsnm{Kokuyama}},
\bauthor{\binits{K.} \bsnm{Kotake}},
\bauthor{\binits{Y.} \bsnm{Kozai}},
\bauthor{\binits{H.} \bsnm{Kudoh}},
\bauthor{\binits{H.} \bsnm{Kunimori}},
\bauthor{\binits{H.} \bsnm{Kuninaka}},
\bauthor{\binits{K.} \bsnm{Kuroda}},
\bauthor{\binits{K.-i.} \bsnm{Maeda}},
\bauthor{\binits{H.} \bsnm{Matsuhara}},
\bauthor{\binits{Y.} \bsnm{Mino}},
\bauthor{\binits{O.} \bsnm{Miyakawa}},
\bauthor{\binits{S.} \bsnm{Miyoki}},
\bauthor{\binits{M.Y.} \bsnm{Morimoto}},
\bauthor{\binits{T.} \bsnm{Morioka}},
\bauthor{\binits{T.} \bsnm{Morisawa}},
\bauthor{\binits{S.} \bsnm{Mukohyama}},
\bauthor{\binits{S.} \bsnm{Nagano}},
\bauthor{\binits{I.} \bsnm{Naito}},
\bauthor{\binits{K.} \bsnm{Nakamura}},
\bauthor{\binits{H.} \bsnm{Nakano}},
\bauthor{\binits{K.} \bsnm{Nakao}},
\bauthor{\binits{S.} \bsnm{Nakasuka}},
\bauthor{\binits{Y.} \bsnm{Nakayama}},
\bauthor{\binits{E.} \bsnm{Nishida}},
\bauthor{\binits{K.} \bsnm{Nishiyama}},
\bauthor{\binits{A.} \bsnm{Nishizawa}},
\bauthor{\binits{Y.} \bsnm{Niwa}},
\bauthor{\binits{T.} \bsnm{Noumi}},
\bauthor{\binits{Y.} \bsnm{Obuchi}},
\bauthor{\binits{M.} \bsnm{Ohashi}},
\bauthor{\binits{N.} \bsnm{Ohishi}},
\bauthor{\binits{M.} \bsnm{Ohkawa}},
\bauthor{\binits{N.} \bsnm{Okada}},
\bauthor{\binits{K.} \bsnm{Onozato}},
\bauthor{\binits{K.} \bsnm{Oohara}},
\bauthor{\binits{N.} \bsnm{Sago}},
\bauthor{\binits{M.} \bsnm{Saijo}},
\bauthor{\binits{M.} \bsnm{Sakagami}},
\bauthor{\binits{S.} \bsnm{Sakata}},
\bauthor{\binits{M.} \bsnm{Sasaki}},
\bauthor{\binits{T.} \bsnm{Sato}},
\bauthor{\binits{M.} \bsnm{Shibata}},
\bauthor{\binits{H.} \bsnm{Shinkai}},
\bauthor{\binits{K.} \bsnm{Somiya}},
\bauthor{\binits{H.} \bsnm{Sotani}},
\bauthor{\binits{N.} \bsnm{Sugiyama}},
\bauthor{\binits{Y.} \bsnm{Suwa}},
\bauthor{\binits{R.} \bsnm{Suzuki}},
\bauthor{\binits{H.} \bsnm{Tagoshi}},
\bauthor{\binits{F.} \bsnm{Takahashi}},
\bauthor{\binits{K.} \bsnm{Takahashi}},
\bauthor{\binits{K.} \bsnm{Takahashi}},
\bauthor{\binits{R.} \bsnm{Takahashi}},
\bauthor{\binits{R.} \bsnm{Takahashi}},
\bauthor{\binits{T.} \bsnm{Takahashi}},
\bauthor{\binits{H.} \bsnm{Takahashi}},
\bauthor{\binits{T.} \bsnm{Akiteru}},
\bauthor{\binits{T.} \bsnm{Takano}},
\bauthor{\binits{K.} \bsnm{Taniguchi}},
\bauthor{\binits{A.} \bsnm{Taruya}},
\bauthor{\binits{H.} \bsnm{Tashiro}},
\bauthor{\binits{Y.} \bsnm{Torii}},
\bauthor{\binits{M.} \bsnm{Toyoshima}},
\bauthor{\binits{S.} \bsnm{Tsujikawa}},
\bauthor{\binits{Y.} \bsnm{Tsunesada}},
\bauthor{\binits{A.} \bsnm{Ueda}},
\bauthor{\binits{K.-i.} \bsnm{Ueda}},
\bauthor{\binits{M.} \bsnm{Utashima}},
\bauthor{\binits{Y.} \bsnm{Wakabayashi}},
\bauthor{\binits{H.} \bsnm{Yamakawa}},
\bauthor{\binits{K.} \bsnm{Yamamoto}},
\bauthor{\binits{T.} \bsnm{Yamazaki}},
\bauthor{\binits{J.} \bsnm{Yokoyama}},
\bauthor{\binits{C.-M.} \bsnm{Yoo}},
\bauthor{\binits{S.} \bsnm{Yoshida}},
\bauthor{\binits{T.} \bsnm{Yoshino}},
\batitle{The status of {DECIGO}}.
\bjtitle{Journal of Physics: Conference Series}
\bvolume{840},
\bfpage{012010}
(\byear{2017})
\end{barticle}
\endbibitem

\bibitem[\protect\citeauthoryear{{Sch{\"o}del}
  et~al.}{2007}]{2007A&A...469..125S}
\begin{barticle}
\bauthor{\binits{R.} \bsnm{{Sch{\"o}del}}},
\bauthor{\binits{A.} \bsnm{{Eckart}}},
\bauthor{\binits{T.} \bsnm{{Alexander}}},
\bauthor{\binits{D.} \bsnm{{Merritt}}},
\bauthor{\binits{R.} \bsnm{{Genzel}}},
\bauthor{\binits{A.} \bsnm{{Sternberg}}},
\bauthor{\binits{L.} \bsnm{{Meyer}}},
\bauthor{\binits{F.} \bsnm{{Kul}}},
\bauthor{\binits{J.} \bsnm{{Moultaka}}},
\bauthor{\binits{T.} \bsnm{{Ott}}},
\bauthor{\binits{C.} \bsnm{{Straubmeier}}},
\batitle{{The structure of the nuclear stellar cluster of the Milky Way}}.
\bjtitle{\aap}
\bvolume{469}(\bissue{1}),
\bfpage{125}--\blpage{146}
(\byear{2007}).
doi:\doiurl{10.1051/0004-6361:20065089}
\end{barticle}
\endbibitem

\bibitem[\protect\citeauthoryear{{Sesana} et~al.}{2008}]{Sesana+08}
\begin{barticle}
\bauthor{\binits{A.} \bsnm{{Sesana}}},
\bauthor{\binits{A.} \bsnm{{Vecchio}}},
\bauthor{\binits{M.} \bsnm{{Eracleous}}},
\bauthor{\binits{S.} \bsnm{{Sigurdsson}}},
\batitle{{Observing white dwarfs orbiting massive black holes in the
  gravitational wave and electro-magnetic window}}.
\bjtitle{\mnras}
\bvolume{391}(\bissue{2}),
\bfpage{718}--\blpage{726}
(\byear{2008}).
doi:\doiurl{10.1111/j.1365-2966.2008.13904.x}
\end{barticle}
\endbibitem

\bibitem[\protect\citeauthoryear{{Seth} et~al.}{2010}]{2010ApJ...714..713S}
\begin{barticle}
\bauthor{\binits{A.C.} \bsnm{{Seth}}},
\bauthor{\binits{M.} \bsnm{{Cappellari}}},
\bauthor{\binits{N.} \bsnm{{Neumayer}}},
\bauthor{\binits{N.} \bsnm{{Caldwell}}},
\bauthor{\binits{N.} \bsnm{{Bastian}}},
\bauthor{\binits{K.} \bsnm{{Olsen}}},
\bauthor{\binits{R.D.} \bsnm{{Blum}}},
\bauthor{\binits{V.P.} \bsnm{{Debattista}}},
\bauthor{\binits{R.} \bsnm{{McDermid}}},
\bauthor{\binits{T.} \bsnm{{Puzia}}},
\batitle{{The NGC 404 Nucleus: Star Cluster and Possible Intermediate-mass
  Black Hole}}.
\bjtitle{\apj}
\bvolume{714}(\bissue{1}),
\bfpage{713}--\blpage{731}
(\byear{2010}).
doi:\doiurl{10.1088/0004-637X/714/1/713}
\end{barticle}
\endbibitem

\bibitem[\protect\citeauthoryear{{Seto} and {Muto}}{2010}]{Seto&Muto10}
\begin{barticle}
\bauthor{\binits{N.} \bsnm{{Seto}}},
\bauthor{\binits{T.} \bsnm{{Muto}}},
\batitle{{Relativistic astrophysics with resonant multiple inspirals}}.
\bjtitle{\prd}
\bvolume{81}(\bissue{10}),
\bfpage{103004}
(\byear{2010}).
doi:\doiurl{10.1103/PhysRevD.81.103004}
\end{barticle}
\endbibitem

\bibitem[\protect\citeauthoryear{{Seto} and {Muto}}{2011}]{Seto&Muto11}
\begin{barticle}
\bauthor{\binits{N.} \bsnm{{Seto}}},
\bauthor{\binits{T.} \bsnm{{Muto}}},
\batitle{{Resonant trapping of stars by merging massive black hole binaries}}.
\bjtitle{\mnras}
\bvolume{415}(\bissue{4}),
\bfpage{3824}--\blpage{3830}
(\byear{2011}).
doi:\doiurl{10.1111/j.1365-2966.2011.18988.x}
\end{barticle}
\endbibitem

\bibitem[\protect\citeauthoryear{{Steinberg} et~al.}{2019}]{steinberg+19}
\begin{barticle}
\bauthor{\binits{E.} \bsnm{{Steinberg}}},
\bauthor{\binits{E.R.} \bsnm{{Coughlin}}},
\bauthor{\binits{N.C.} \bsnm{{Stone}}},
\bauthor{\binits{B.D.} \bsnm{{Metzger}}},
\batitle{{Thawing the frozen-in approximation: implications for self-gravity in
  deeply plunging tidal disruption events}}.
\bjtitle{\mnras}
\bvolume{485}(\bissue{1}),
\bfpage{146}--\blpage{150}
(\byear{2019}).
doi:\doiurl{10.1093/mnrasl/slz048}
\end{barticle}
\endbibitem

\bibitem[\protect\citeauthoryear{{Stone} et~al.}{2013}]{Stone+13}
\begin{barticle}
\bauthor{\binits{N.} \bsnm{{Stone}}},
\bauthor{\binits{R.} \bsnm{{Sari}}},
\bauthor{\binits{A.} \bsnm{{Loeb}}},
\batitle{{Consequences of strong compression in tidal disruption events}}.
\bjtitle{\mnras}
\bvolume{435},
\bfpage{1809}--\blpage{1824}
(\byear{2013}).
doi:\doiurl{10.1093/mnras/stt1270}
\end{barticle}
\endbibitem

\bibitem[\protect\citeauthoryear{{Stone} and {Metzger}}{2016}]{StoneMetzger16}
\begin{barticle}
\bauthor{\binits{N.C.} \bsnm{{Stone}}},
\bauthor{\binits{B.D.} \bsnm{{Metzger}}},
\batitle{{Rates of stellar tidal disruption as probes of the supermassive black
  hole mass function}}.
\bjtitle{\mnras}
\bvolume{455},
\bfpage{859}--\blpage{883}
(\byear{2016}).
doi:\doiurl{10.1093/mnras/stv2281}
\end{barticle}
\endbibitem

\bibitem[\protect\citeauthoryear{{Stone} and {Loeb}}{2011}]{Stone&Loeb11}
\begin{barticle}
\bauthor{\binits{N.} \bsnm{{Stone}}},
\bauthor{\binits{A.} \bsnm{{Loeb}}},
\batitle{{Prompt tidal disruption of stars as an electromagnetic signature of
  supermassive black hole coalescence}}.
\bjtitle{\mnras}
\bvolume{412}(\bissue{1}),
\bfpage{75}--\blpage{80}
(\byear{2011}).
doi:\doiurl{10.1111/j.1365-2966.2010.17880.x}
\end{barticle}
\endbibitem

\bibitem[\protect\citeauthoryear{{Stone} et~al.}{2013}]{stone+2013}
\begin{barticle}
\bauthor{\binits{N.} \bsnm{{Stone}}},
\bauthor{\binits{R.} \bsnm{{Sari}}},
\bauthor{\binits{A.} \bsnm{{Loeb}}},
\batitle{{Consequences of strong compression in tidal disruption events}}.
\bjtitle{\mnras}
\bvolume{435}(\bissue{3}),
\bfpage{1809}--\blpage{1824}
(\byear{2013}).
doi:\doiurl{10.1093/mnras/stt1270}
\end{barticle}
\endbibitem

\bibitem[\protect\citeauthoryear{{Stone} et~al.}{2019}]{Stone+2019}
\begin{barticle}
\bauthor{\binits{N.C.} \bsnm{{Stone}}},
\bauthor{\binits{M.} \bsnm{{Kesden}}},
\bauthor{\binits{R.M.} \bsnm{{Cheng}}},
\bauthor{\binits{S.} \bsnm{{van Velzen}}},
\batitle{{Stellar tidal disruption events in general relativity}}.
\bjtitle{General Relativity and Gravitation}
\bvolume{51}(\bissue{2}),
\bfpage{30}
(\byear{2019}).
doi:\doiurl{10.1007/s10714-019-2510-9}
\end{barticle}
\endbibitem

\bibitem[\protect\citeauthoryear{{Strubbe} and
  {Quataert}}{2009}]{StrubbeQuataert09}
\begin{barticle}
\bauthor{\binits{L.E.} \bsnm{{Strubbe}}},
\bauthor{\binits{E.} \bsnm{{Quataert}}},
\batitle{{Optical flares from the tidal disruption of stars by massive black
  holes}}.
\bjtitle{\mnras}
\bvolume{400},
\bfpage{2070}--\blpage{2084}
(\byear{2009}).
doi:\doiurl{10.1111/j.1365-2966.2009.15599.x}
\end{barticle}
\endbibitem

\bibitem[\protect\citeauthoryear{{Syer} and {Ulmer}}{1999}]{SyerUlmer98}
\begin{barticle}
\bauthor{\binits{D.} \bsnm{{Syer}}},
\bauthor{\binits{A.} \bsnm{{Ulmer}}},
\batitle{{Tidal disruption rates of stars in observed galaxies}}.
\bjtitle{\mnras}
\bvolume{306},
\bfpage{35}--\blpage{42}
(\byear{1999}).
doi:\doiurl{10.1046/j.1365-8711.1999.02445.x}
\end{barticle}
\endbibitem

\bibitem[\protect\citeauthoryear{{Tanikawa}}{2018}]{Tanikawa18}
\begin{barticle}
\bauthor{\binits{A.} \bsnm{{Tanikawa}}},
\batitle{{High-resolution Hydrodynamic Simulation of Tidal Detonation of a
  Helium White Dwarf by an Intermediate Mass Black Hole}}.
\bjtitle{\apj}
\bvolume{858}(\bissue{1}),
\bfpage{26}
(\byear{2018}).
doi:\doiurl{10.3847/1538-4357/aaba79}
\end{barticle}
\endbibitem

\bibitem[\protect\citeauthoryear{{Tanikawa} et~al.}{2017}]{Tanikawa+17}
\begin{barticle}
\bauthor{\binits{A.} \bsnm{{Tanikawa}}},
\bauthor{\binits{Y.} \bsnm{{Sato}}},
\bauthor{\binits{K.} \bsnm{{Nomoto}}},
\bauthor{\binits{K.} \bsnm{{Maeda}}},
\bauthor{\binits{N.} \bsnm{{Nakasato}}},
\bauthor{\binits{I.} \bsnm{{Hachisu}}},
\batitle{{Does Explosive Nuclear Burning Occur in Tidal Disruption Events of
  White Dwarfs by Intermediate-mass Black Holes?}}
\bjtitle{\apj}
\bvolume{839}(\bissue{2}),
\bfpage{81}
(\byear{2017}).
doi:\doiurl{10.3847/1538-4357/aa697d}
\end{barticle}
\endbibitem

\bibitem[\protect\citeauthoryear{{Tejeda} et~al.}{2017}]{Tejeda+17}
\begin{barticle}
\bauthor{\binits{E.} \bsnm{{Tejeda}}},
\bauthor{\binits{E.} \bsnm{{Gafton}}},
\bauthor{\binits{S.} \bsnm{{Rosswog}}},
\bauthor{\binits{J.C.} \bsnm{{Miller}}},
\batitle{{Tidal disruptions by rotating black holes: relativistic hydrodynamics
  with Newtonian codes}}.
\bjtitle{\mnras}
\bvolume{469}(\bissue{4}),
\bfpage{4483}--\blpage{4503}
(\byear{2017}).
doi:\doiurl{10.1093/mnras/stx1089}
\end{barticle}
\endbibitem

\bibitem[\protect\citeauthoryear{{Thorne}}{1998}]{Thorne98}
\begin{bbook}
\bauthor{\binits{K.S.} \bsnm{{Thorne}}},
\bbtitle{{Probing Black Holes and Relativistic Stars with Gravitational Waves}}
(\bpublisher{University of Chicago Press}, \blocation{???}, \byear{1998}),
p. \bfpage{41}
\end{bbook}
\endbibitem

\bibitem[\protect\citeauthoryear{{Toscani} et~al.}{2019}]{toscani+2019}
\begin{barticle}
\bauthor{\binits{M.} \bsnm{{Toscani}}},
\bauthor{\binits{G.} \bsnm{{Lodato}}},
\bauthor{\binits{R.} \bsnm{{Nealon}}},
\batitle{{Gravitational wave emission from unstable accretion discs in tidal
  disruption events}}.
\bjtitle{\mnras}
\bvolume{489}(\bissue{1}),
\bfpage{699}--\blpage{706}
(\byear{2019}).
doi:\doiurl{10.1093/mnras/stz2201}
\end{barticle}
\endbibitem

\bibitem[\protect\citeauthoryear{{Vick} et~al.}{2017}]{2017MNRAS.468.2296V}
\begin{barticle}
\bauthor{\binits{M.} \bsnm{{Vick}}},
\bauthor{\binits{D.} \bsnm{{Lai}}},
\bauthor{\binits{J.} \bsnm{{Fuller}}},
\batitle{{Tidal dissipation and evolution of white dwarfs around massive black
  holes: an eccentric path to tidal disruption}}.
\bjtitle{\mnras}
\bvolume{468}(\bissue{2}),
\bfpage{2296}--\blpage{2310}
(\byear{2017}).
doi:\doiurl{10.1093/mnras/stx539}
\end{barticle}
\endbibitem

\bibitem[\protect\citeauthoryear{{Vigneron} et~al.}{2018}]{Vigneron18}
\begin{barticle}
\bauthor{\binits{Q.} \bsnm{{Vigneron}}},
\bauthor{\binits{G.} \bsnm{{Lodato}}},
\bauthor{\binits{A.} \bsnm{{Guidarelli}}},
\batitle{{Tidal disruption of stars in a supermassive black hole binary system:
  the influence of orbital properties on fallback and accretion rates}}.
\bjtitle{\mnras}
\bvolume{476}(\bissue{4}),
\bfpage{5312}--\blpage{5322}
(\byear{2018}).
doi:\doiurl{10.1093/mnras/sty585}
\end{barticle}
\endbibitem

\bibitem[\protect\citeauthoryear{{Webbink} et~al.}{1983}]{Webbink83}
\begin{barticle}
\bauthor{\binits{R.F.} \bsnm{{Webbink}}},
\bauthor{\binits{S.} \bsnm{{Rappaport}}},
\bauthor{\binits{G.J.} \bsnm{{Savonije}}},
\batitle{{On the evolutionary status of bright, low-mass X-ray sources.}}
\bjtitle{\apj}
\bvolume{270},
\bfpage{678}--\blpage{693}
(\byear{1983}).
doi:\doiurl{10.1086/161159}
\end{barticle}
\endbibitem

\bibitem[\protect\citeauthoryear{{Wu}}{2018}]{2018AJ....155..118W}
\begin{barticle}
\bauthor{\binits{Y.} \bsnm{{Wu}}},
\batitle{{Diffusive Tidal Evolution for Migrating Hot Jupiters}}.
\bjtitle{\aj}
\bvolume{155}(\bissue{3}),
\bfpage{118}
(\byear{2018}).
doi:\doiurl{10.3847/1538-3881/aaa970}
\end{barticle}
\endbibitem

\bibitem[\protect\citeauthoryear{{Yalinewich} et~al.}{2019}]{yalinewich+19}
\begin{botherref}
\oauthor{\binits{A.} \bsnm{{Yalinewich}}},
\oauthor{\binits{E.} \bsnm{{Steinberg}}},
\oauthor{\binits{T.} \bsnm{{Piran}}},
\oauthor{\binits{J.H.} \bsnm{{Krolik}}},
{Radio Emission from the unbound Debris of Tidal Disruption Events}.
arXiv e-prints,
1903--02575
(2019)
\end{botherref}
\endbibitem

\bibitem[\protect\citeauthoryear{{Zalamea} et~al.}{2010}]{2010MNRAS.409L..25Z}
\begin{barticle}
\bauthor{\binits{I.} \bsnm{{Zalamea}}},
\bauthor{\binits{K.} \bsnm{{Menou}}},
\bauthor{\binits{A.M.} \bsnm{{Beloborodov}}},
\batitle{{White dwarfs stripped by massive black holes: sources of coincident
  gravitational and electromagnetic radiation}}.
\bjtitle{\mnras}
\bvolume{409}(\bissue{1}),
\bfpage{25}--\blpage{29}
(\byear{2010}).
doi:\doiurl{10.1111/j.1745-3933.2010.00930.x}
\end{barticle}
\endbibitem

\bibitem[\protect\citeauthoryear{{Zhong} et~al.}{2014}]{Zhong14}
\begin{barticle}
\bauthor{\binits{S.} \bsnm{{Zhong}}},
\bauthor{\binits{P.} \bsnm{{Berczik}}},
\bauthor{\binits{R.} \bsnm{{Spurzem}}},
\batitle{{Super Massive Black Hole in Galactic Nuclei with Tidal Disruption of
  Stars}}.
\bjtitle{\apj}
\bvolume{792}(\bissue{2}),
\bfpage{137}
(\byear{2014}).
doi:\doiurl{10.1088/0004-637X/792/2/137}
\end{barticle}
\endbibitem

\end{thebibliography}
\bibliographystyle{aps-nameyear}          

\end{document}